\documentclass[twocolumn,twocolappendix]{aastex701}	

\shorttitle{Fisher Forecasts for Roman HLIS}
\shortauthors{K. Cao et al.}
\usepackage{CJKutf8}

\usepackage{graphicx}	
\usepackage{amsmath}	
\usepackage{amssymb}	
\usepackage{multirow}

\begin{document}

\title{Fisher Forecasts for Cosmological Yields from $3\!\times\!2$pt Analysis of the Roman Space Telescope High Latitude Imaging Survey}

\author[orcid=0000-0002-1699-6944]{Kaili Cao (\begin{CJK*}{UTF8}{gbsn}曹开力\end{CJK*})}
\affiliation{Center for Cosmology and AstroParticle Physics (CCAPP), The Ohio State University, 191 West Woodruff Ave, Columbus, OH 43210, USA}
\affiliation{Department of Physics, The Ohio State University, 191 West Woodruff Ave, Columbus, OH 43210, USA}
\email[show]{cao.1191@osu.edu}

\author[orcid=0000-0001-7775-7261]{David H. Weinberg}
\affiliation{Center for Cosmology and AstroParticle Physics (CCAPP), The Ohio State University, 191 West Woodruff Ave, Columbus, OH 43210, USA}
\affiliation{Department of Astronomy, The Ohio State University, 140 West 18th Avenue, Columbus, OH 43210, USA}
\email{weinberg.21@osu.edu}

\author[orcid=0000-0003-4776-0333]{Vivian Miranda}
\affiliation{C. N. Yang Institute for Theoretical Physics, Stony Brook University, Stony Brook, NY 11794, USA}
\email{vivian.miranda@stonybrook.edu}

\author[orcid=0000-0003-2177-9407]{Nihar Dalal}
\affiliation{Center for Cosmology and AstroParticle Physics (CCAPP), The Ohio State University, 191 West Woodruff Ave, Columbus, OH 43210, USA}
\affiliation{Department of Physics, The Ohio State University, 191 West Woodruff Ave, Columbus, OH 43210, USA}
\email{dalal.64@osu.edu}

\author[orcid=0000-0002-1894-3301]{Tim Eifler}
\affiliation{Department of Astronomy and Steward Observatory, University of Arizona, 933 North Cherry Avenue, Tucson, AZ 85721, USA}
\affiliation{Department of Physics, University of Arizona, 1118 E. Fourth Street, Tucson, AZ 85721, USA}
\email{timeifler@arizona.edu}

\author[orcid=0000-0003-0871-8941]{Jiachuan Xu}
\affiliation{Department of Astronomy and Steward Observatory, University of Arizona, 933 North Cherry Avenue, Tucson, AZ 85721, USA}
\affiliation{Department of Physics, Northeastern University, Boston, MA 02115, USA}
\email{jiachuanxu@arizona.edu}

\author[orcid=0009-0006-1916-8550]{Haley Bowden}
\affiliation{Department of Astronomy and Steward Observatory, University of Arizona, 933 North Cherry Avenue, Tucson, AZ 85721, USA}
\email{hbowden@arizona.edu}

\author[orcid=0000-0001-6764-073X]{Katarina Markovi\v{c}}
\affiliation{Jet Propulsion Laboratory, California Institute of Technology, 4800 Oak Grove Drive, Pasadena, CA 91109, USA}
\email{katarina.markovic@jpl.nasa.gov}

\collaboration{all}{Roman HLIS Cosmology PIT}

\begin{abstract}

The High Latitude Imaging Survey (HLIS) of NASA's Nancy Grace Roman Space Telescope will provide powerful tests of cosmological models through sensitive measurements of cosmic shear, galaxy-galaxy lensing (GGL), and galaxy clustering. As part of the HLIS Project Infrastructure Team's Data Challenge 1 (DC1), we carry out Fisher forecasts of cosmological parameter constraints from combinations of these probes, focusing on inverse-variance figures of merit (FoMs) for the parameters $\sigma_8$ and $\Omega_{\rm{m}}$, which scale the amplitude of weak lensing signals. We find good agreement between Fisher analysis and Markov chain Monte Carlo (MCMC) analysis of the DC1 baseline data vector, and we investigate varied priors on cosmological parameters and on nuisance parameters describing unknown biases in photometric redshifts or shear measurements. The DC1 benchmark modeling assumes a ``lens $=$ source'' analysis with linear galaxy bias and no marginalization over baryonic physics. Under these assumptions, the forecast constraints from GGL+clustering are substantially stronger than those from cosmic shear, with the combination of all three probes (``$3\!\times\!2$pt'') providing moderate further improvement. Adding tight external priors on the power spectrum shape parameters $n_{\rm{s}}$, $\Omega_{\rm{b}}$, and $h_0$ can improve the $(\sigma_8, \Omega_{\rm{m}})$ FoMs by factors of $1.2$--$3.5$. The smallest scale angular bins provide much more information than the largest scale bins, and the highest redshift tomographic bins provide more information than the lowest redshift bins. Factor-of-two changes in the priors on photo-$z$ and shear biases, relative to the benchmark values based on anticipated calibration accuracy, produce changes of $\lesssim 20\%$ in FoMs.

\end{abstract}


\section{Introduction} \label{sec:intro}

Since its inception in the Astro2010 Decadal Survey \citep{2010nwnh.book......}, the Nancy Grace Roman Space Telescope has had measurement of cosmic structure through weak gravitational lensing \citep[see, e.g.,][for some reviews]{2001PhR...340..291B, 2013PhR...530...87W, 2015RPPh...78h6901K, 2018ARA&A..56..393M} as a core science goal. The currently scheduled date of launch is September 2026. The $2.4$-m mirror and stable space-based observing platform allow excellent image quality, while the $300$-megapixel near-IR wide field camera enables large area surveys \citep{2019arXiv190205569A}. Weak lensing cosmology will be achieved mainly through the High Latitude Wide Area Survey (HLWAS), which will observe $2415 \,{\rm deg}^2$ in Y106, J129, H158 imaging and grism spectroscopy, and an additional $2702 \,{\rm deg}^2$ in H-band imaging only.\footnote{See the report of the Roman Observations Time Allocation Committee, \citet{2025arXiv250510574O}, with details in their Appendix C.1.} We refer to the imaging component of this survey as the High Latitude Imaging Survey (HLIS), and in this paper we examine cosmological performance forecasts for the $2415 \,{\rm deg}^2$ ``Medium Tier,'' which is projected to have an effective source density $n_{\rm eff} \approx 41.3 \,{\rm arcmin}^{-2}$ and a total of $360$M source galaxy shape measurements.

Large weak lensing surveys have enabled the first high-precision measurements of matter clustering at redshifts $z<1$, especially the ``Stage III'' surveys \citep[in the parlance of][]{2006astro.ph..9591A}: KiDS \citep[the Kilo Degree Survey;][]{2022A&A...664A.170V, 2023A&A...669A..69B, 2023A&A...675A.189D, 2023A&A...679A.133L, 2025A&A...703A.158W}, DES \citep[the Dark Energy Survey;][]{2022PhRvD.105b3514A, 2022PhRvD.105b3515S, 2026arXiv260114559D, 2026arXiv260210065D, 2026PhRvD.113j3503G}, and HSC \citep[the Hyper Suprime Cam;][]{2019PASJ...71...43H, 2020PASJ...72...16H, 2023PhRvD.108l3517M, 2023PhRvD.108l3518L, 2023PhRvD.108l3519D, 2023PhRvD.108l3520M, 2023PhRvD.108l3521S}. Matter clustering can be inferred directly from cosmic shear, the correlation of shape distortions induced by foreground matter, or from the combination of galaxy clustering and galaxy-galaxy lensing (GGL), which measures the galaxy-matter cross-correlation around foreground lens galaxies using the mean tangential shear of background source galaxies. Joint analyses of the shear-shear, shear-galaxy, and galaxy-galaxy correlation functions, commonly referred to as $3\!\times\!2$pt, allow cross-checks, breaking of parameter degeneracies, and higher precision. Weak lensing surveys enable other statistical approaches that sharpen their cosmological sensitivity, such as higher-order statistics \citep[e.g.,][]{2004MNRAS.348..897T, 2025PhRvD.112l3514G, 2025PhRvD.112l3515G}, cluster weak lensing \citep[e.g.,][]{2024PhRvD.110h3511S, 2025PhRvD.112h3535A, 2025A&A...703A..25L, 2025arXiv251025706S}, and non-linear analyses that exploit GGL information on small scales \citep[e.g.,][]{2006ApJ...652...26Y, 2009MNRAS.394..929C, 2017MNRAS.467.3024L, 2020MNRAS.491...51S, 2020MNRAS.492.2872W, 2022MNRAS.510.5376S, 2025arXiv251215962L}. However, in this paper we will focus on $3\!\times\!2$pt analyses where linear perturbation theory is expected to provide accurate predictions.

Performance forecasts play many important roles in cosmological experiments, which include motivating the experiments in the first place, defining science requirements, refining experimental design, supporting the construction of analysis and inference pipelines, devising strategies for combining results from multiple experiments and probes, and identifying which sources of systematic uncertainty can have the largest impact on the results. Comprehensive performance forecasts for Stage IV weak lensing experiments include \citet{2023A&A...675A.120E} for Euclid \citep{2011arXiv1110.3193L, 2022A&A...662A.112E, 2025A&A...697A...1E}, \citet{2022MNRAS.513.1210M} for the Vera C. Rubin Observatory's Legacy Survey of Space and Time \citep[LSST;][]{2012arXiv1211.0310L, 2019ApJ...873..111I}, and \citet{2021MNRAS.507.1746E} for Roman, with \citet{2021MNRAS.507.1514E} focusing specifically on the synergies between LSST and Roman. The absolute values of ``figures of merit'' (FoMs) computed from such forecasts are sensitive to assumptions about systematic uncertainties, survey performance, external information, and the underlying cosmological model space, but within any forecast one can vary these assumptions to quantify their impact. This paper and its companion (J. Xu et al. 2026, in preparation; hereafter XuDC1) focus on forecasts for the Medium Tier of the Roman HLIS.

This paper represents a collective effort of the Roman Project Infrastructure Team (PIT) ``Maximizing Cosmological Science with the Roman High Latitude Imaging Survey'' (PI: O. Dor\'e),\footnote{\url{https://roman-hlis-cosmology.caltech.edu/}} and specifically the Cosmological Parameters Inference Pipeline (CPIP) group. CPIP's development focus is the Cobaya-CosmoLike Joint Architecture pipeline ({\sc CoCoA};\footnote{Pronunciation: co-CO-ah.} V. Miranda et al. 2026, in preparation), which builds on the CosmoLike software tools for predicting galaxy clustering and weak lensing observables \citep{2014MNRAS.440.1379E, 2017MNRAS.470.2100K, 2020JCAP...05..010F} and the Cobaya platform \citep{2021JCAP...05..057T} for cosmological inference, which includes convenient interfaces to CAMB \citep{2011ascl.soft02026L} and CLASS \citep{2011arXiv1104.2932L} Boltzmann codes. In addition to creating data vectors, CosmoLike's spin-off code CosmoCov \citep{2020JCAP...05..010F} uses analytic methods to compute covariance matrices given assumptions about survey properties.

The CPIP team has recently conducted its first internal Data Challenge (DC1), described in detail by XuDC1. In brief, DC1 used {\sc CoCoA} to create mock weak lensing and galaxy clustering data vectors and CosmoCov to compute corresponding covariance matrices for a variety of (blinded) parameter choices. Different CPIP subgroups have then attempted to recover these parameters. In its basic form, because the creation of data vectors and inference of parameters are both performed with {\sc CoCoA}, this exercise does not test for theoretical systematics associated with imperfect predictions, and it implicitly assumes that the ``nuisance'' parameters used by {\sc CoCoA} provide an adequate description of systematics associated with photometric redshifts, shear calibration, and intrinsic alignments. However, DC1 provides a useful test of different analysis choices and computational implementations, and it lays the groundwork for more sophisticated data challenges in the future.

Most of the responses to DC1 use the ``industry standard'' Markov chain Monte Carlo technique \citep[MCMC; e.g.,][]{1953JChPh..21.1087M, 1970Bimka..57...97H, 2010CAMCS...5...65G, 2013PASP..125..306F}, which stochastically samples the posterior distribution of parameter values. In this paper, we instead use Fisher information analysis, with the maximum likelihood or maximum posterior probability found by coordinate descent and the distribution of parameter values around this maximum found by approximating the likelihood or posterior as a multi-variate Gaussian. \citet{1997PhRvL..79.3806T} and \citet{1997ApJ...480...22T} introduced Fisher forecasting methods in cosmology, and the technical appendix of the Dark Energy Task Force report \citep[DETF;][]{2006astro.ph..9591A} provides a pedagogical summary of the method. While the assumptions of Fisher information analysis are more restrictive than those of MCMC analysis, it is informative to compare the results of these methods, and the computational efficiency of the Fisher approach makes it easy to consider many variants of the standard analysis. In this paper, we take advantage of this computational efficiency to examine the impact on the cosmological constraining power of different priors on nuisance parameters and cosmological parameters and to assess the relative contribution of different observables, tomographic redshift bins, and physical scales to the cosmological constraints.

In addition to five cosmological parameters, DC1 incorporates nuisance parameters describing galaxy clustering bias, photometric redshift bias, and multiplicative shear calibration bias in each of eight tomographic bins, plus a $2$-parameter description of intrinsic alignments. We focus our attention on the baseline DC1 data vectors, which were generated with specific (blinded) choices of these parameters. These data vectors do not include measurement noise, so the maximum likelihood (ML) parameter values should correspond to the true input values. However, the values of nuisance parameter are (deliberately) not centered within their priors, so maximum a posteriori (MAP) parameter values do not correspond to true input values. The widths of the priors on nuisance parameters are based on the performance goals of the measurement pipelines. Achieving these goals is technically challenging, and the PIT's Shear and Clustering Measurement (SCM) group has made significant progress in developing algorithms and codes that can meet Roman's stringent requirements \citep[see, e.g.,][for some recent efforts]{2024PASP..136l4506L, 2025ApJS..277...55C, 2025MNRAS.542..608B}.

This paper is structured as follows. In Section~\ref{sec:meth}, we describe methods involved in this work, including {\sc CoCoA}, DC1, search for best-fit parameters, and Fisher analysis. Some further details about DC1 are provided in Appendix~\ref{app:dc1}. Section~\ref{sec:dcbase} presents our results for baseline data vectors in DC1 and compares them to MCMC results. Extended corner plots can be found in Appendix~\ref{app:corner}. In Section~\ref{sec:dissect} and Section~\ref{sec:prior}, we investigate the breakdown of information and the impact of assumed priors, respectively. We summarize and discuss our main conclusions in Section~\ref{sec:conclu}. Appendix~\ref{app:math} addresses several mathematical points, most importantly demonstrating the accuracy of the Fisher approximation in Section~\ref{ss:logp}. We briefly address the impact of cosmology-dependent covariance on the constraining power in Appendix~\ref{app:varcov}. Readers seeking a fast route to our main results can jump to Figures~\ref{fig:corsub_fourier} to \ref{fig:probe}, read the summary in Section~\ref{sec:conclu}, then loop back to Sections~\ref{sec:dissect} and \ref{sec:prior} for more complete explanations.

\section{Methods} \label{sec:meth}

This section describes methods involved in this work. Sections~\ref{ss:cocoa} and \ref{ss:dc1} present the Cobaya-CosmoLike Joint Architecture ({\sc CoCoA}; V. Miranda et al. 2026, in preparation) and the Roman HLIS Cosmology PIT Cosmological Parameters Inference Pipeline (CPIP) Data Challenge 1 (DC1; XuDC1), respectively. Section~\ref{ss:optim} introduces our modified coordinate descent algorithm for finding best-fit parameter values. Section~\ref{ss:fisher} details the mathematical formalism of our Fisher information analysis. Details of the DC1 data vectors --- in particular the choice of angular and redshift bins --- are given in Appendix~\ref{app:dc1}.

\subsection{Cobaya-CosmoLike Joint Architecture ({\sc CoCoA})} \label{ss:cocoa}

As its name indicates, {\sc CoCoA} is an integration of CosmoLike \citep{2014MNRAS.440.1379E, 2017MNRAS.470.2100K, 2020JCAP...05..010F} and Cobaya \citep{2021JCAP...05..057T}. For a given set of cosmological and ``nuisance'' parameters, CosmoLike predicts two-point correlation functions (2PCFs) in real space and/or the corresponding power spectra in Fourier space based on matter power spectra computed with CAMB \citep{2011ascl.soft02026L} and cosmological perturbation theory. In the context of $3\!\times\!2$pt analysis, these are:
\begin{itemize}
    \item Cosmic shear: autocorrelation of cosmic shear; $\xi_\pm (\theta)$ in real space and $C_{\rm ss} (\ell)$ in Fourier space.
    \item Galaxy-galaxy lensing (GGL): cross-correlation of cosmic shear and galaxy distribution; $\gamma_{\rm t} (\theta)$ in real space and $C_{\rm gs} (\ell)$ in Fourier space. The subscript ``t'' stands for ``tangential.''
    \item Galaxy clustering: autocorrelation of galaxy distribution; $w (\theta)$ in real space and $C_{\rm gg} (\ell)$ in Fourier space.
\end{itemize}
Cobaya samples the parameter space, by default via the Metropolis--Hastings algorithm \citep{1953JChPh..21.1087M, 1970Bimka..57...97H}. We refer the readers to V. Miranda et al. (2026, in preparation) for a thorough description of the {\sc CoCoA} software and highlight some important aspects below.

To capture redshift-dependent information, e.g., the growth of large-scale structure, galaxy samples are usually divided into tomographic bins according to photometric redshifts. In DC1, all detected galaxies are divided into $n_{\rm tomo} = 8$ bins, each containing approximately equal numbers of galaxies. 2PCFs and power spectra are then measured for pairs of tomographic bins, which can include a tomographic bin and itself. For cosmic shear, the two bins are interchangeable, hence the total number of functions is $n_{\rm tomo} (n_{\rm tomo}+1) / 2$. For GGL, one bin serves as the ``source'' sample for shear while the other serves as the ``lens'' sample for galaxy positions. In DC1, we use the same galaxy samples for shear and position measurements. This choice is sometimes referred to as ``lens $=$ source.'' The ``lens $=$ source'' approach has not been the default in Stage III analyses, in part because of the difficulty in controlling galaxy clustering systematics when operating near the magnitude limit of the photometric survey. It remains to be see whether the greater uniformity and control of space-based imaging will make ``lens $=$ source'' feasible for Roman. This approach is beneficial because it allows self-calibration of photo-$z$ biases \citep{2020JCAP...12..001S, 2024arXiv241104088E} and because the high lens density reduces shape noise and shot noise in the GGL and clustering measurements. In Appendix~\ref{app:qtrlens}, we examine the impact of lowering the lens density in all bins by a factor of $4$, while holding the source density fixed. We find that this $\times 4$ reduction leads to a factor $\sim 2$ reduction in FoM for data combinations that include clustering. In principle, the signal is only non-zero when the source sample is behind the lens sample; however, due to the redshift uncertainties, it can be non-zero even if the source sample has a lower centroid redshift than the lens sample. Therefore, the maximum number of functions is $n_{\rm tomo}^2$. For galaxy clustering, since galaxy positions in different tomographic bins are not expected to be correlated, only the autocorrelation in each bin is studied. Cross-correlations from different bins can be a useful diagnostic of photometric redshifts error distributions, but we do not examine this possibility here.

While {\sc CoCoA} is able to take additional parameters into account, in this work we focus on $31$ parameters studied in the DC1 baseline case. The $31$ parameters are:
\begin{itemize}
    \item $5$ cosmological parameters: matter density $\Omega_{\rm m}$, fluctuation amplitude $\sigma_8$, spectral index $n_{\rm s}$, baryonic density $\Omega_{\rm b}$, and Hubble constant $h_0 \equiv H_0 / (100 \,{\rm km} \,{\rm s}^{-1} \,{\rm Mpc}^{-1})$. Their respective definitions are covered in standard cosmology textbooks \citep[e.g.,][]{2023cctp.book.....H} and omitted here. We only consider the flat $\Lambda$CDM ($\Lambda$ denotes the cosmological constant; CDM stands for cold dark matter) model in the baseline case.
    \item $8$ linear galaxy bias parameters, one for each tomographic bin, denoted as $b_i$ ($i = 1, 2, \ldots, 8$). The galaxy bias originates from using (detectable) galaxies as tracers of matter distribution. The galaxy-galaxy lensing signal is proportional to galaxy bias, and the galaxy clustering signal is proportional to galaxy bias squared. Therefore, neither of these two probes is expected to have much constraining power by itself, but the degeneracy can be lifted by combining them. We note that the small scales studied in DC1 (see Appendix~\ref{app:dc1}) are well into the regime where non-linear and baryonic effects matter. Meanwhile, DC1 is based upon an optimistic assumption that theoretical models will be sufficiently robust to exploit those scales by the time we are analyzing the Roman HLIS.
    \item $8$ photometric redshift (photo-$z$) bias parameters, $\Delta_{z}^i$ ($i = 1, 2, \ldots, 8$). Since spectroscopic redshifts are only available for a small fraction of galaxies, we need to infer redshifts from multi-band photometry for the vast majority. See \citet{2026OJAp....958200R} for how we plan to measure photometric redshifts from Roman data.
    \item $8$ multiplicative shear bias parameters, $m_i$ ($i = 1, 2, \ldots, 8$). Since we can only statistically measure distortions from galaxy shapes, it is difficult to completely eliminate this bias via calibration \citep{2017arXiv170202600H}.
    \item $2$ parameters for modeling intrinsic alignments (IAs), an overall amplitude $A_{\rm IA}$ and a redshift scaling index $\eta_{\rm IA}$. We adopt the non-linear alignment model \citep{2004PhRvD..70f3526H, 2007NJPh....9..444B} for this purpose.
\end{itemize}
Throughout this paper, we use the vector ${\boldsymbol \theta}$ to denote the collection of parameters and use Greek letters (e.g., $\alpha$ and $\beta$) to index them.

Relative to other cosmological probes such as type Ia supernovae and baryon acoustic oscillations (BAO), a distinctive strength of weak lensing is its sensitivity to the amplitude of matter clustering, characterized in our parameter set by $\sigma_8$, and to the matter density parameter $\Omega_{\rm m}$. We therefore focus most of our attention in this paper on the ($\sigma_8$, $\Omega_{\rm m}$) constraints, including the impact of different choices of priors on nuisance parameters and other cosmological parameters. Within the HLIS data, the scale dependence of galaxy clustering, and to a lesser extent cosmic shear and GGL, provides constraints on $n_{\rm s}$, $\Omega_{\rm b}$, and $h_0$, but the constraints from external data may well be better. Departures from a $\Lambda$ expansion history, such as those suggested by the DESI DR2 analysis \citep{desicollaboration2025desidr2resultsii}, have a small but measurable effect on weak lensing for the same values of ($n_{\rm s}$, $\sigma_8$, $\Omega_{\rm b}$, $h_0$, $\Omega_{\rm m}$) because they change the distance-redshift relation. We defer investigation of such models to future work.

\subsection{CPIP Data Challenge 1 (DC1)} \label{ss:dc1}

In DC1, blinded data vectors, suggested masks, and covariance matrices are provided, and participants are invited to infer parameters and uncertainties from them. In each space, there is a baseline data vector and several alternate data vectors. The baseline data vectors are prepared using $31$ parameters in the default inference pipeline (see Section~\ref{ss:cocoa}); the alternate vectors include more complicated effects that are not captured by the default parameter set. Readers are referred to Appendix~\ref{app:dc1} of this paper and the DC1 GitHub repository\footnote{\url{https://github.com/CosmoLike/roman_cpip_data_challenge}} for further information about DC1.

In DC1, the real-space (2PCF) domain is $\theta \in [2.5, 250] \,{\rm arcmin}$, while the Fourier-space (power spectrum) domain is $\ell \in [30, 4000]$. In each space, the domain is divided into $15$ logarithmic angular scale bins. Functions for different probes and tomographic bin pairs are concatenated into data vectors to simplify array operations. To summarize, in DC1, the total length of a real-space data vector is
\begin{equation}
    \underbrace{15 \times [8 (8+1) / 2] \times 2}_{\rm Shear} + \underbrace{15 \times (8^2 - 3)}_{\rm GGL} + \underbrace{15 \times 8}_{\rm Clus.} = 2115,
    \label{eq:dvlen_real}
\end{equation}
where $\times 2$ comes from the fact that $\xi_\pm (\theta)$ denotes two different functions, while the length of a Fourier-space data vector is
\begin{equation}
    \underbrace{15 \times [8 (8+1) / 2]}_{\rm Shear} + \underbrace{15 \times (8^2 - 2)}_{\rm GGL} + \underbrace{15 \times 8}_{\rm Clus.} = 1590.
    \label{eq:dvlen_fourier}
\end{equation}

The detailed choice of angular and tomographic bins is given in Appendix~\ref{app:dc1}. Also, some bins or bin combinations are masked in the analyses, because the signal is too close to zero to be informative and may be susceptible to numerical noise. For example, some tomographic bin pairs with ${\bar z}_{\rm lens} > {\bar z}_{\rm source}$ are excluded from GGL if the expected overlap of redshift distributions is negligible. One angular bin is masked out from the GGL data vectors in Fourier space, and at least two and up to four angular bins are masked out from the GGL and galaxy clustering data vectors in real space; all angular bins are retained for cosmic shear in both spaces.

\begin{figure*}
    \epsscale{1.1}
    \plotone{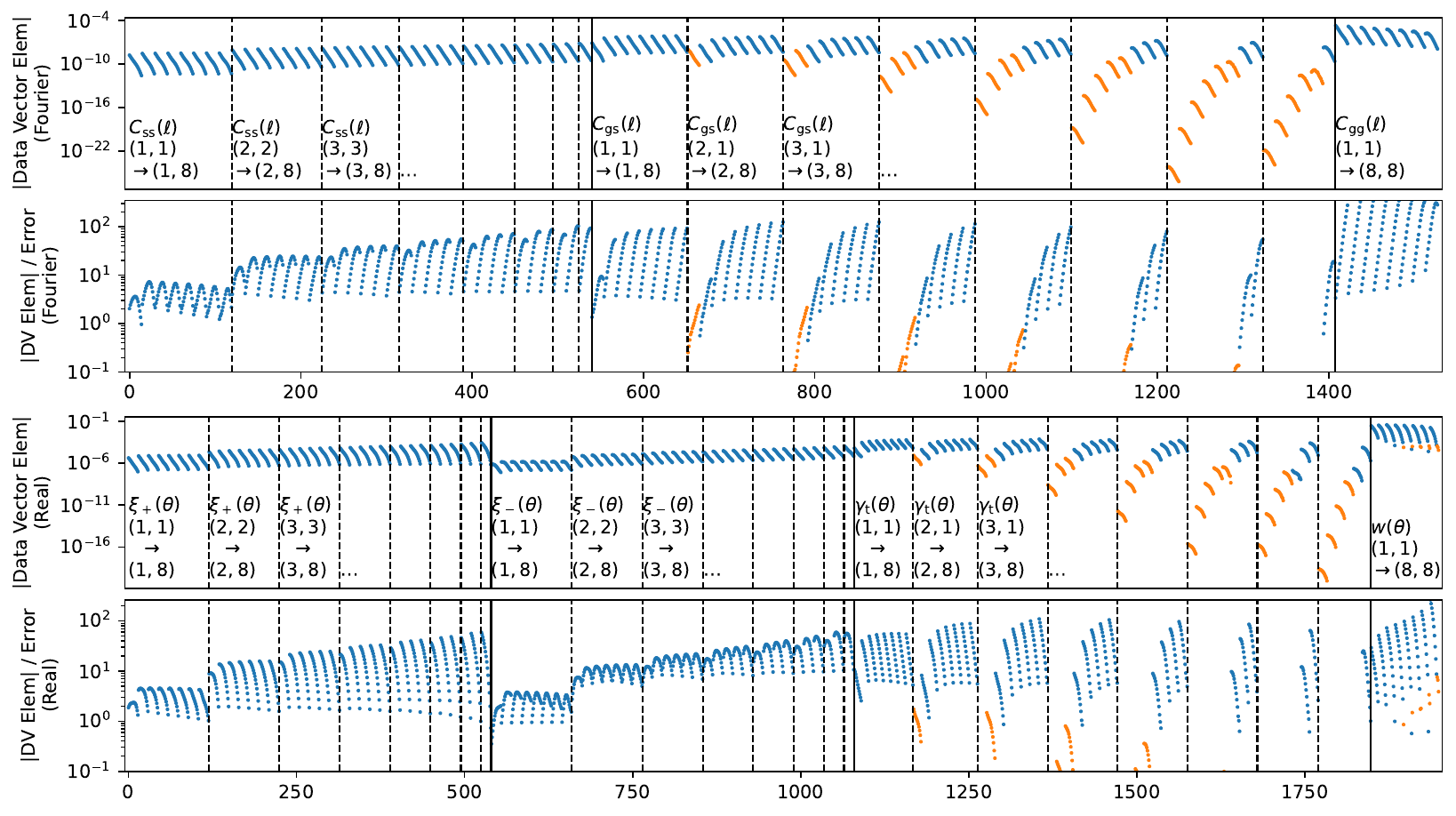}
    \caption{\label{fig:dv_layout}Layouts of baseline data vectors in Fourier space (first two rows) and real space (last two rows). The first and third rows present the absolute values of the data vector elements, with positive elements shown in blue and negative elements shown in orange. Most negative elements correspond to GGL measurements in which the lens tomographic bin lies behind the source tomographic bin. The second and fourth rows present the ratios between the absolute values of data vector elements and the corresponding errors. The errors are the square roots of the diagonal elements of the covariance matrices, which are shown later in Section~\ref{ss:cocoa}. The ordering of sub-blocks is labeled in the first and third rows. $1$ denotes the lowest redshift tomographic bin and $8$ the highest. For GGL, pairs such as $(1, 2)$ and $(2, 1)$ both appear, with the first index denoting the lens bin and the second, the source bin. In real space, all values of $\xi_+$ appear first, then all values of $\xi_-$. The segment for each pair of bins goes from small $\ell$ to large $\ell$ in Fourier space and from small $\theta$ to large $\theta$ in real space.}
\end{figure*}

Figure~\ref{fig:dv_layout} presents the layout of the baseline data vector in each space. Each short streak of points represents the range of angular scales for a particular pair of tomographic bins. In Fourier space, these are ordered by increasing $\ell$, hence decreasing angular scale, while in real space they are ordered by increasing $\theta$. Most data vector elements are positive (blue); for GGL, they are negative (orange) when the source sample is mostly behind the lens sample. The ratios between data vector elements and the corresponding errors can be viewed as single-element signal-to-noise ratios (SNRs). For cosmic shear, the maximum is reached at medium angular scales, as larger scales are subject to greater cosmic variance and small scales are more affected by shape noise. For GGL and clustering, however, these ratios monotonically increase with smaller scales. Across all three probes, we see the consistent trends that the ``SNRs'' increase with redshift, indicating that higher redshift tomographic bins contain more information. Galaxy clustering is the only probe for which single-element SNRs exceed $10^2$. However, in linear theory there is perfect degeneracy between the amplitude of galaxy clustering and the unknown galaxy bias factor, so galaxy clustering only provides information about the amplitude of matter clustering when it is combined with GGL.

Calculating the likelihood or posterior probability for parameter estimates (see Section~\ref{ss:optim} below) requires a covariance matrix. For DC1, we use a covariance matrix computed from analytic formulae by CosmoCov,\footnote{\url{https://github.com/CosmoLike/CosmoCov}} as described by V. Miranda et al. (2026, in preparation). A covariance matrix produced by CosmoCov has two major components: i) the Gaussian component, which captures the uncertainties due to finite number of modes and measurement noise, and ii) the non-Gaussian component, which is dominated by the super-sample covariance \citep[SSC;][]{2013PhRvD..87l3504T, 2018JCAP...06..015B}. SSC results from the fact that modes on scales comparable to or larger than the survey volume perturb the mean density of the volume relative to the true cosmic mean. Because of non-linear coupling between large-scale and small-scale modes, this uncertainty on the scale of the survey volume propagates into correlated uncertainties on smaller scales. We discuss the potential gain by mitigating SSC \citep[e.g.,][]{2019JCAP...10..004D} in Section~\ref{ss:ssc}.

\begin{figure*}
    \plotone{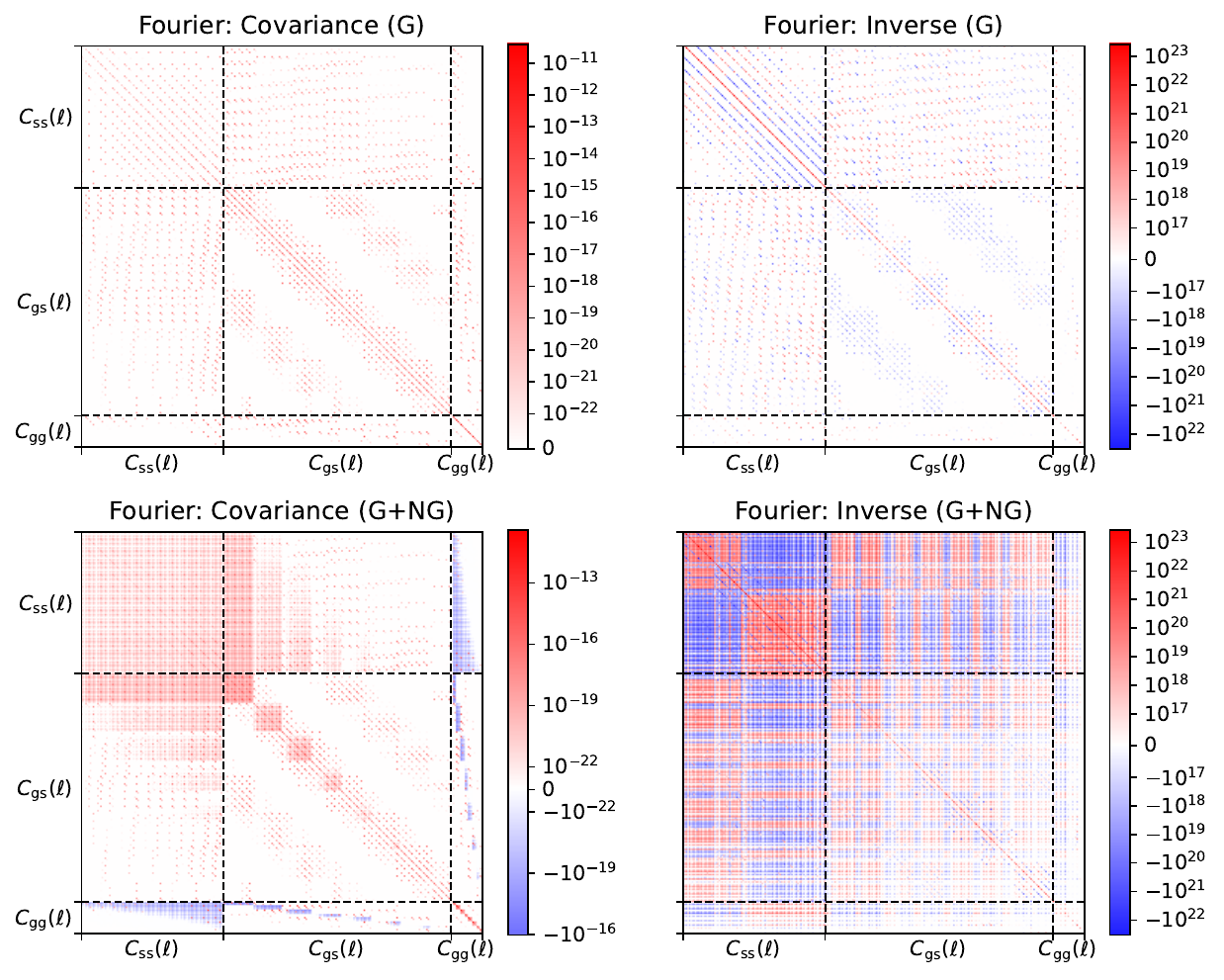}
    \caption{\label{fig:covinv_fourier}Covariance matrices (left column) and their inverses (right column) in Fourier space. The upper row only includes the Gaussian (``G'') component of the covariance matrix, while the lower row includes both Gaussian and non-Gaussian (``NG'') components. In each panel, a symmetric logarithmic scale is used to better present the structure of the matrix, and boundaries between different segments of the data vector (see Section~\ref{ss:cocoa}) are marked with black dashed lines. Within each observable, data elements loop first over scale (inner loop) and then over tomographic redshift bin pair (outer loop). Each square block in the $C_{\rm gs} (\ell)$ cells corresponds to a single redshift bin of lens galaxies.}
\end{figure*}

\begin{figure*}
    \plotone{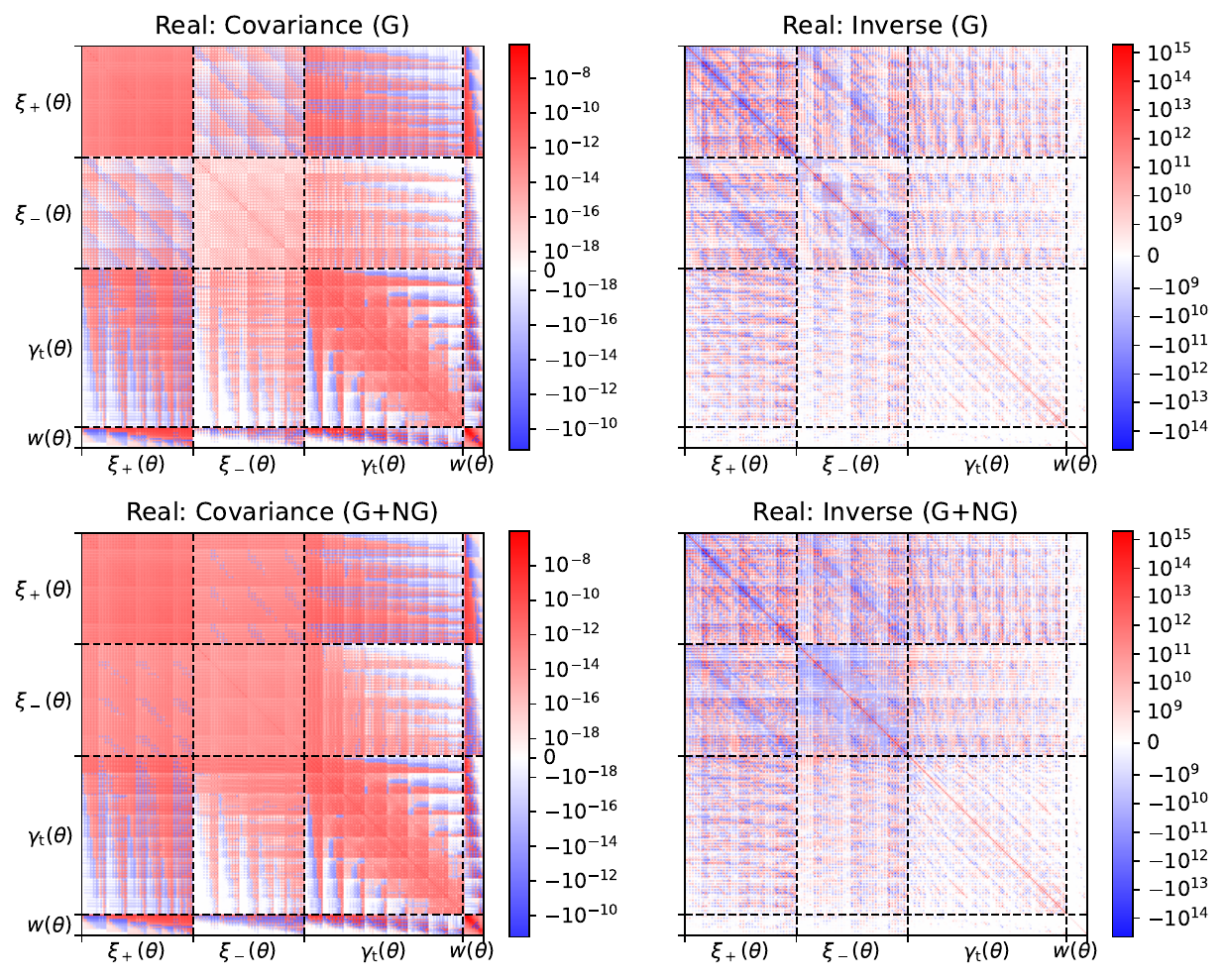}
    \caption{\label{fig:covinv_real}Same as Figure~\ref{fig:covinv_fourier}, but in real space.}
\end{figure*}

For both MCMC and Fisher analyses, inverses of covariance matrices are necessary for computing likelihoods or posterior probabilities of model data vectors. Figures~\ref{fig:covinv_fourier} and \ref{fig:covinv_real} present DC1 covariance matrices and their inverses in Fourier space and real space, respectively; see Appendix~\ref{ss:invert} for how we invert covariance matrices produced by CosmoCov. In each space, the structure of these matrices follows that of the concatenated data vector (see Figure~\ref{fig:dv_layout}).

In Fourier space (Figure~\ref{fig:covinv_fourier}), we see that each block of the Gaussian component of the covariance matrix (upper left panel) is diagonal, because only error bars at the same angular scale are correlated. Its inverse (upper right panel) has a similar structure. With the non-Gaussian component, the full covariance matrix (lower left panel) includes correlations across different angular scales and some negative correlations between galaxy clustering and the other two probes, but the overall structure and the scaling of diagonal elements are barely affected. In its inverse (lower right panel), we see more non-zero elements, as expected. The situation in real space (Figure~\ref{fig:covinv_real}) is different, as the Gaussian component of the covariance matrix and its inverse (upper row) already include correlations across different angular scales. This is understandable, since angular scales in real space can be considered as linear combinations of angular scales in Fourier space, hence the former are correlated even if the latter are not. Therefore, the full covariance and its inverse (lower row) are less obviously different from the Gaussian component-only case (upper row). We address the cosmology dependence of the covariance matrix in Appendix~\ref{app:varcov}, arguing that it has minimal impact on cosmological forecasts.

On small scales, uncertainties in baryonic effects can have an important impact on cosmological inference \citep[e.g.,][]{2011MNRAS.415.3649V, 2019OJAp....2E...4C}. These effects can be summarized via principal component analysis of hydrodynamical simulations \citep[e.g.,][]{2015MNRAS.454.2451E}. To account for baryonic physics uncertainties, one can introduce free parameters that scale these principal components and marginalize over them when inferring other parameters. Alternatively, one can use the Sherman-Morrison formula to modify the covariance matrix, which is equivalent to marginalizing over wide priors on the baryonic effects. In DC1 and this work, we do not directly address the baryonic effects in either way, so we effectively assume that baryonic effects are perfectly known on the angular scales included in the analysis. The impact of modeling and marginalizing over baryonic feedback and other non-linear effects remains an important topic for future investigations.

\subsection{Finding Best-fit Parameter Values} \label{ss:optim}

We refer readers to XuDC1 \citep[also see][]{2024arXiv241104088E} for detailed descriptions of the mapping from parameters ${\boldsymbol \theta}$ to the model-predicted data vector ${\boldsymbol m}$ \citep[also see][for precursor studies using {\sc CoCoA}]{2023PhRvD.107l3529Z, 2024PhRvD.110f3532X}. Given this mapping, we can try to find the best-fit parameter values by maximizing either the likelihood ${\cal L} ({\boldsymbol \theta})$ or the posterior probability ${\cal P} ({\boldsymbol \theta})$. For our purposes, the (logarithmic) likelihood is defined as
\begin{equation}
    \ln {\cal L} ({\boldsymbol \theta}) \equiv -\frac12 \chi^2 ({\boldsymbol m})
    = -\frac12 ({\boldsymbol m} - {\boldsymbol d})^{\rm T} {\bf C}^{-1} ({\boldsymbol m} - {\boldsymbol d}),
    \label{eq:def_like}
\end{equation}
where ${\boldsymbol d}$ is the DC1 ``observed'' data vector, and ${\boldsymbol m}$ is a function of ${\boldsymbol \theta}$. The (logarithmic) posterior probability is defined as
\begin{equation}
    \ln {\cal P} ({\boldsymbol \theta}) \equiv \ln {\cal L} ({\boldsymbol \theta}) + \ln {\rm Prior} ({\boldsymbol \theta}),
    \label{eq:def_post}
\end{equation}
where ${\rm Prior} ({\boldsymbol \theta})$ is the prior probability of ${\boldsymbol \theta}$. Note that these formulae implicitly assume that the errors in data vector elements are Gaussian, which is not strictly true \citep{2009MNRAS.395.2065T, 2020MNRAS.499.2977L} but may be a reasonable approximation for these statistics on these scales.

DC1 assumes uncorrelated Gaussian priors on photo-$z$ biases and shear biases centered at zero; the priors on cosmological, intrinsic alignments, and galaxy bias parameters are flat and wide.\footnote{For MCMC, ``wide'' means a sufficiently wide domain covering all plausible values; for Fisher analysis, ``wide'' means infinitely wide.} Denoting the standard deviations (often referred to as ``widths'') as $\Delta (\Delta_{z}^i)$ and $\Delta (m_i)$, the global prior probability can be written as
\begin{equation}
    \ln {\rm Prior} ({\boldsymbol \theta}) = -\frac12 \sum_{i=1}^8 \left[ \frac{\Delta_{z}^i}{\Delta (\Delta_{z}^i)} \right]^2 -\frac12 \sum_{i=1}^8 \left[ \frac{m_i}{\Delta (m_i)} \right]^2.
    \label{eq:prior}
\end{equation}
Note that it is normalized so that $\ln {\rm Prior} ({\boldsymbol \theta}) = 0$ when $\Delta_{z}^i = 0$ and $m_i = 0$ for $i = 1, 2, \ldots, 8$. In DC1, $\Delta (\Delta_{z}^i) = 0.002$ in Fourier space, $0.003$ in real space,\footnote{The difference was unintended.} and $\Delta (m_i) = 0.005$ in both spaces; the assumed widths do not depend on the redshift. Thanks to its Gaussianity, the prior probability can be written in terms of a prior precision matrix\footnote{A precision matrix is the inverse of a covariance matrix. To avoid confusion with the covariance matrix of data vector elements ${\bf C}$, we use ${\bf \Sigma}$ to denote the covariance matrix of parameters throughout this work.} ${\bf \Sigma}_{\rm Prior}^{-1}$ (see Section~\ref{ss:fisher} for its usage). When the prior probability is Equation~(\ref{eq:prior}), ${\bf \Sigma}_{\rm Prior}^{-1}$ is just a diagonal matrix with elements $1 / \Delta^2 (\Delta_{z}^i)$, $1 / \Delta^2 (m_i)$, and $0$. Zero elements on the diagonal of ${\bf \Sigma}_{\rm Prior}^{-1}$ correspond to parameters with flat and wide priors. In other words, ${\bf \Sigma}_{\rm Prior}^{-1}$ does not include any information about these parameters, and the corresponding elements in the prior covariance matrix ${\bf \Sigma}_{\rm Prior}$ are infinity. This is acceptable, since ${\bf \Sigma}_{\rm Prior}$ is never directly used.

The best-fit parameter values are those leading to either maximum likelihood (ML) or maximum posterior probability (known as maximum a posteriori, MAP). Given the large number of parameters, a grid search is an impractical way of finding the maximum. Furthermore, computing numerical (partial) derivatives is a challenging task and is unaffordable at every step. Therefore, we implement a modified version of the coordinate descent algorithm \citep[e.g.,][]{wright2015coordinate} to perform the maximization. Specifically, we start from an initial guess and vary each parameter in turn while keeping others fixed. For each parameter, we try integer multiples of some step size away from the current value, and pick the maximum likelihood or posterior probability among these points. We loop over all the parameters, shrink step sizes when no new progress is made, and repeat this process until the first $\gtrsim 6$ decimal places of $\ln {\cal L}$ or $\ln {\cal P}$ no longer change. Since a zero gradient is a necessary but not sufficient condition for a maximum, sometimes our search stagnates at a zero-gradient, non-maximum point. To account for this scenario, we switch back and forth between the coordinate directions (i.e., varying one parameter while keeping others fixed) and linear combinations of the parameters, so that some of the partial derivatives with respect to these linear combinations are non-zero at such points. Finally, we verify our best-fit parameter values using the {\sc SciPy} implementation of the downhill simplex method.\footnote{\url{https://docs.scipy.org/doc/scipy/reference/generated/scipy.optimize.fmin.html}}

Due to the unsmoothness of CAMB predictions, our modified coordinate descent algorithm failed to find the global maxima of likelihood and posterior probability in real space during DC1. After DC1, we reran the maximization with different starting points and successfully recovered the truth parameter values (with small discrepancies due to differences in computing facilities and software versions). One lesson we learned is that to find the global maxima of unsmooth functions, it is advisable to try different starting points, find a collection of local maxima, and pick the global maxima.

Using the same computing facility, namely the Cardinal cluster\footnote{\url{https://www.osc.edu/resources/technical_support/supercomputers/cardinal}} at the Ohio Supercomputer Center \citep{OhioSupercomputerCenter1987}, our modified coordinate descent algorithm only needs about $1\,{\rm hours} \times 8\,{\rm cores}$ to converge to the fifth significant figure in $\ln {\cal L}$ or $\ln {\cal P}$ for a run with a single starting point, while the {\sc CoCoA} implementation of MCMC needs about $5\,{\rm days} \times 8\,{\rm cores} \times 8\,{\rm chains}$ to reach an $R - 1$ value ($R$ is the Gelman-Rubin statistic) of $\sim 0.02$. We note that machine learning emulators for either matter power spectra \citep[e.g.,][]{2010ApJ...713.1322L, 2025RASTI...4....2J} or data vectors \citep[e.g.,][]{2025PhRvD.111l3519Z, 2025PhRvD.111l3520S} can substantially reduce the computational cost of each data vector evaluation for both methods and thus alleviate this discrepancy in total core-hour consumption. See XuDC1 for proof-of-principle.

\subsection{Fisher Information Analysis} \label{ss:fisher}

For our purposes, the Fisher matrix is defined as the precision matrix of parameters and computed as
\begin{equation}
    {\bf \Sigma}_{\cal L}^{-1} \equiv F_{\alpha \beta} = {\boldsymbol m}_{,\alpha}^{\rm T} {\bf C}^{-1} {\boldsymbol m}_{,\beta},
    \label{eq:fisher}
\end{equation}
where the subscript $_{,\alpha}$ denotes partial differentiation with respect to parameter $\theta_\alpha$. It describes the amount of information about the parameters ${\boldsymbol \theta}$. Intuitively, the precision matrix of data vector elements ${\bf C}^{-1}$ is the amount of information from the observational data, and the partial derivatives are how this information is translated to parameters of interest. Because of the high dimensionality and finite precision in our scenario, numerical (partial) derivatives need to be carefully verified. We do this by tuning the step size for each parameter so that two different expressions of numerical derivatives ($5$-point stencil and $5$-point linear regression) lead to consistent results, with fractional differences in the norms of derivative vectors at the ${\cal O} (10^{-4})$ level for the cosmological parameters and several orders of magnitude better for the others.\footnote{We note that such fractional differences are not monotonic functions of step sizes. In other words, as we reduce the step size, the fractional difference may first decrease and then increase. ${\cal O} (10^{-4})$ is the level of fractional differences at optimal step sizes for derivatives with respect to cosmological parameters; it cannot be further reduced by simply tuning step sizes.} For $A_{\rm s}$ and $n_{\rm s}$, this strategy fails due to the unsmoothness of CAMB predictions, hence we use a linear fit to each element of a {\sc CoCoA} data vector within a relatively large domain of $A_{\rm s}$ or $n_{\rm s}$ to find the first derivative. In Section~\ref{ss:logp}, we demonstrate that a Gaussian posterior computed from our Fisher matrix describes the directly computed posterior in the neighborhood of the maximum, providing an end-to-end test of our Fisher matrix computation including the numerical derivatives.

The inverse of the Fisher matrix is the covariance matrix of parameters ${\bf \Sigma}_{\cal L}$, which encodes the uncertainties in parameters from the observational data alone. In this case, the partial derivatives in Equation~(\ref{eq:fisher}) should be computed at the ML parameter values ${\boldsymbol {\hat \theta}}_{\cal L}$. To incorporate the prior precision matrix ${\bf \Sigma}_{\rm Prior}^{-1}$, thanks to the Gaussianity of Equation~(\ref{eq:prior}), we have
\begin{equation}
    {\bf \Sigma}_{\cal P}^{-1} = {\bf \Sigma}_{\cal L}^{-1} + {\bf \Sigma}_{\rm Prior}^{-1},
    \label{eq:sigma_add}
\end{equation}
where ${\bf \Sigma}_{\rm Prior}^{-1}$ is the prior precision matrix defined in Section~\ref{ss:optim}; its inverse, ${\bf \Sigma}_{\cal P}$, encodes the uncertainties in parameters from the combination of observational data and our prior knowledge. In principle, the likelihood part Equation~(\ref{eq:fisher}) should now be based on partial derivatives computed at the MAP parameter values ${\boldsymbol {\hat \theta}}_{\cal P}$ instead of the ML values ${\boldsymbol {\hat \theta}}_{\cal L}$. However, the fractional differences between the two sets of numerical derivatives are only at the ${\cal O} (10^{-4})$ level, so we consistently use derivatives computed at ${\boldsymbol {\hat \theta}}_{\cal P}$ in Sections~\ref{sec:dissect} and \ref{sec:prior}. See Appendix~\ref{app:corner} for further justification.

\section{Forecasts for the DC1 Baseline Case} \label{sec:dcbase}

\begin{figure*}
    \plotone{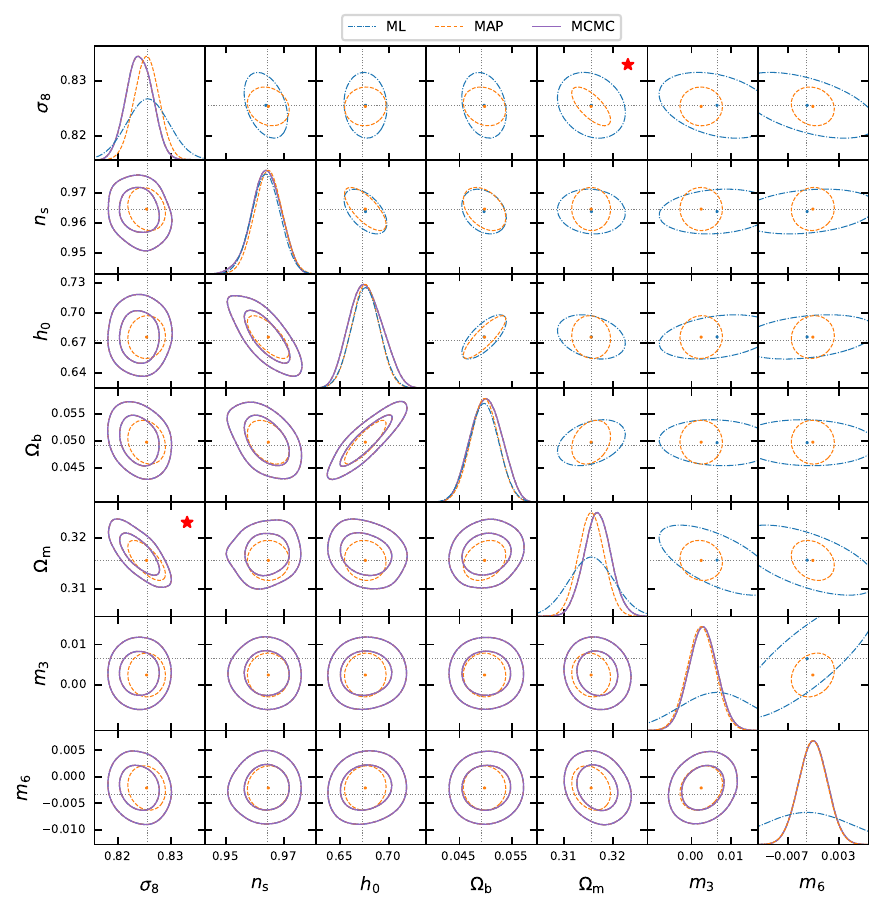}
    \caption{\label{fig:corsub_fourier}Corner plots for representative parameters in Fourier space. $7$ parameters are shown in this figure: all cosmological parameters studied in this work ($\sigma_8$, $n_{\rm s}$, $h_0$, $\Omega_{\rm b}$, and $\Omega_{\rm m}$), as well as multiplicative shear biases ($m_3$ and $m_6$) in the $3^{\rm rd}$ and $6^{\rm th}$ tomographic bins. In the panels above the diagonal, the maximum likelihood results (``ML''; blue) and maximum a posteriori results (``MAP''; orange) are compared. Peak values are shown as dots, and boundaries of $1\sigma$ credible regions are shown as ellipses. In the panels below the diagonal, the MAP results are compared to MCMC results, which are shown in purple, with both $1\sigma$ and $2\sigma$ regions. The diagonal panels show the 1D marginalized distributions from MCMC results and Fisher results, which are normalized so that peaks of MAP results have the same heights as those of MCMC results. The truth parameter values (i.e., those used to make the ``observed'' data vector ${\boldsymbol d}$) are marked as gray dotted horizontal and vertical lines. We mark the panels corresponding to ($\sigma_8$, $\Omega_{\rm m}$) constraints, which are the focus of this paper, with a star ($\star$). An extended version with $6$ additional parameters can be found in Appendix~\ref{app:corner}.}
\end{figure*}

\begin{figure*}
    \plotone{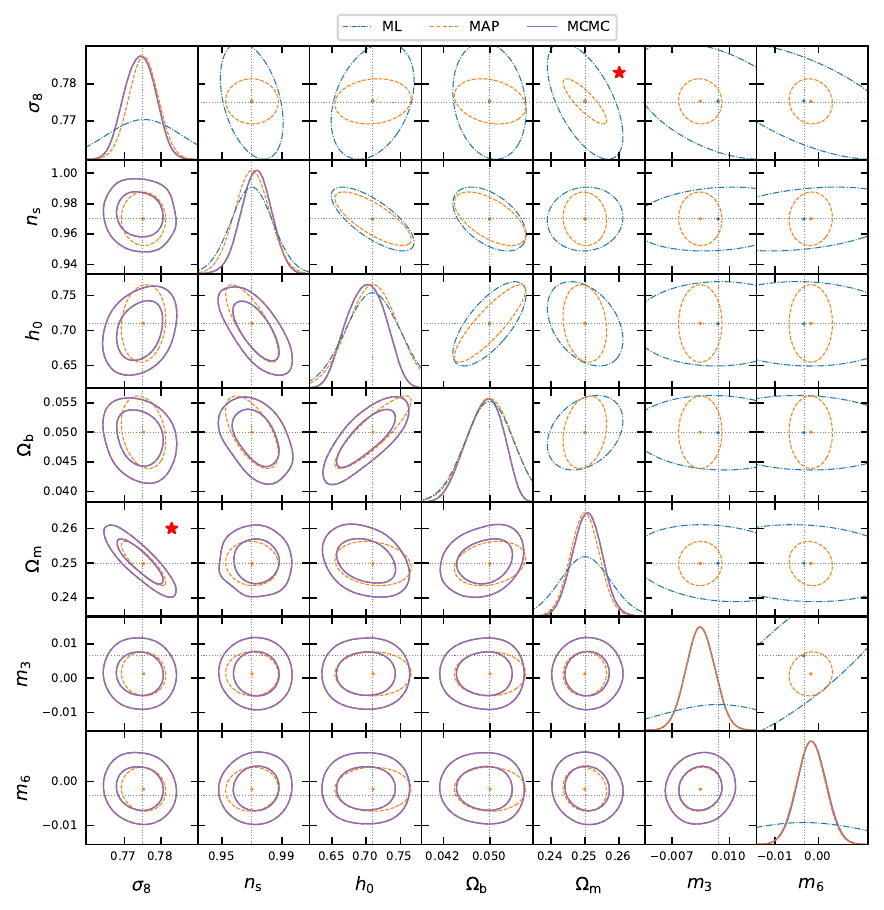}
    \caption{\label{fig:corsub_real}Same as Figure~\ref{fig:corsub_fourier}, but in real space. An extended version with $6$ additional parameters can be found in Appendix~\ref{app:corner}.}
\end{figure*}

Figures~\ref{fig:corsub_fourier} and \ref{fig:corsub_real} are corner plots for representative parameters in Fourier and real spaces, respectively. In the above-diagonal panels, blue dots and ellipses represent the maximum likelihood (ML) parameter values ${\boldsymbol {\hat \theta}}_{\cal L}$ and $1\sigma$ credible regions according to ${\bf \Sigma}_{\cal L}$, while orange dots and ellipses represent the maximum a posteriori parameter values ${\boldsymbol {\hat \theta}}_{\cal P}$ and $1\sigma$ credible regions according to ${\bf \Sigma}_{\cal P}$. The ellipses are visualized following \citet{2009arXiv0906.4123C}; they correspond to $\Delta \chi^2 \equiv \chi^2 ({\boldsymbol \theta}) - \chi^2 ({\boldsymbol {\hat \theta}}) \simeq 2.3$ and enclose $\sim 68\%$ of the marginal probability in each 2D subspace for a pair of parameters. Panels below the diagonal compare the posterior parameter constraints from the Fisher analysis to those derived from MCMC (purple contours). Specifically, we use the MCMC analysis titled ``Scarlet''\footnote{In DC1, Team Scarlet consisted of five students and a postdoctoral researcher at The Ohio State University: Nihar Dalal, Kaili Cao, Michael Gabe, Emily Macbeth, Charuhas Shiveshwarkar, and Anthony Harbo Torres. The main ``Scarlet'' results presented in XuDC1 were obtained by running {\sc CoCoA} following DC1 instructions at the Ohio Supercomputer Center.} in XuDC1 and plot contours containing $68\%$ and $95\%$ of weighted chain points in the 2D projection for each pair of parameters. The orange Fisher ellipses should be compared to the inner purple contours, as they represent $1\sigma$ credible regions in each case.

In both spaces, the ML parameter values ${\boldsymbol {\hat \theta}}_{\cal L}$ (blue dots) almost perfectly agree with the truth parameters (gray dotted lines). This is expected: Since the baseline data vector ${\boldsymbol d}$ is noiseless in each space, we should be able to retrieve the truth parameters from it and obtain a logarithmic likelihood of $\ln {\cal L} ({\boldsymbol {\hat \theta}}_{\cal L}) = 0$; the actual value is $\sim -0.032$ in Fourier space and $\sim -0.005$ in real space, because the computing facilities and software versions for producing the ``observed'' data vector ${\boldsymbol d}$ and the model data vector ${\boldsymbol m} ({\boldsymbol {\hat \theta}})$ are different. When priors are included, the $1\sigma$ credible regions shrink substantially: The orange ellipses are usually enclosed by blue ones, with only a few exceptions (e.g., the $m_3$--$m_6$ panel).

The MAP parameter values ${\boldsymbol {\hat \theta}}_{\cal P}$ sometimes noticeably deviate from the truth values, though the latter lie within the $1\sigma$ credible region. These deviations arise because the DC1 priors on photo-$z$ biases and shear biases are all centered at zero, but the truth values of these biases are not set to zero in DC1. The MAP values for cosmological parameters remain almost exactly equal to the true values. Of course, if the data vector were not noiseless, we would expect deviations at the $\sim 1\sigma$ level for cosmological parameters, in both ML and MAP. The comparison of ML and MAP values in an observational analysis is valuable, as substantial differences could be a sign of inappropriate priors or other problems. Meanwhile, the agreement between MAP and MCMC results (shown in the below diagonal panels) is promising. In some cases, especially for the nuisance parameters (e.g., the $m_3$--$m_6$ panel), the boundaries of $1\sigma$ credible regions almost perfectly overlap. The agreement is not as good in the subspace of cosmological parameters, especially in real space, but the MAP and MCMC results are still consistent with each other, and the areas of $1\sigma$ credible regions are similar.

We suspect that much if not all of the difference between Fisher and MCMC in these panels arises because the Fisher analysis effectively assumes unbounded flat priors on cosmological parameters while the MCMC analysis imposes bounded flat priors. For example, the MCMC analysis limits $0.55 \leq h_0 \leq 0.80$, which is narrow enough to distort the shape of contours. The DC1 data vector constrains $h_0$, $\Omega_{\rm b}$, and $n_{\rm s}$ through the shape of the power spectrum, and because these parameters have largely degenerate impact on the shape, the constraints on the individual parameters are weak. By contrast, $\sigma_8$ and $\Omega_{\rm m}$ influence the amplitude of the weak lensing signal. While non-Gaussianity could also contribute to the difference of contour shapes, we show in Appendix~\ref{ss:logp} that the multivariate Gaussian approximation to the posterior probability is quite accurate, even over a $\ln {\cal P}$ range much larger than that corresponding to $95\%$ or $99\%$ confidence regions. An alternative explanation is that, since the $h_0$ parameter has more non-Gaussianity in its posterior (see Section~\ref{ss:logp}), we would expect the Fisher approximation to perform worst there.

\begin{figure*}
    \plotone{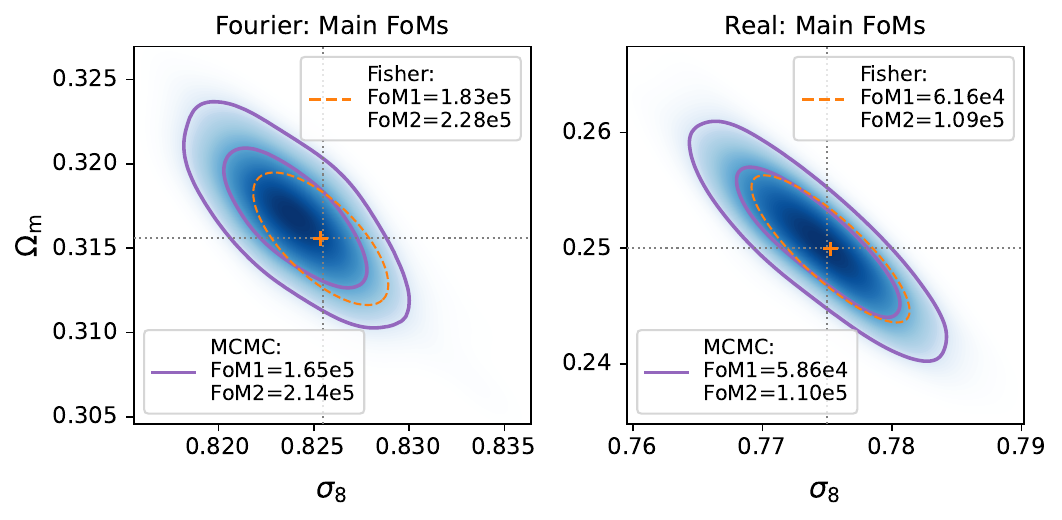}
    \caption{\label{fig:fisher_mcmc}The subspace of $\sigma_8$ and $\Omega_{\rm m}$, in which our two figures of merit, FoM1 = $1/\Delta^2(\sigma_8)$ and FoM2 $= {\det}^{-1/2}({\rm Cov} (\sigma_8, \Omega_{\rm m}))$, are defined. The two panels present results in Fourier space and real space, respectively. In addition to elements from Figures~\ref{fig:corsub_fourier} and \ref{fig:corsub_real}, the (weighted) densities of parameter values sampled by MCMC are visualized in blue.}
\end{figure*}

To facilitate comparisons in the next two sections, we define two figures of merit (FoMs):
\begin{equation}
    {\rm FoM1} = \frac{1}{\Delta^2(\sigma_8)},
    \label{eq:fom1}
\end{equation}
the reciprocal of the square of the 1D marginalized error bar for $\sigma_8$, and
\begin{equation}
    {\rm FoM2} = {\det}^{-1/2}({\rm Cov} (\sigma_8, \Omega_{\rm m})),
    \label{eq:fom2}
\end{equation}
the reciprocal of the square root of the determinant of the 2D marginalized covariance matrix for $\sigma_8$ and $\Omega_{\rm m}$. Thus, FoM1 is a 1-parameter figure of merit and FoM2 is a 2-parameter figure of merit, though both are defined to scale as an inverse variance. The signal-to-noise ratio, a model-independent figure of merit, is discussed in Appendix~\ref{ss:snr2}. Figure~\ref{fig:fisher_mcmc} shows the subspace in which the two FoMs are defined. Fisher analysis often overestimates both FoMs by $\sim 5\%$ relative to MCMC. This difference is understandable, since according to the Cram{\'e}r--Rao inequality \citep{rao1945information, cramer1999mathematical}, FoMs based on a Fisher analysis should be the upper limits and thus larger than FoMs based on MCMC. However, the difference could also be a consequence of the bounded priors of the MCMC analysis, potentially causing non-Gaussian behavior, as discussed above.

We note that an initial comparison of our Fisher matrix contours to MCMC contours showed larger differences than seen here, which we eventually traced to a mistake in the calculation of numerical derivatives with respect to $A_{\rm s}$ and $n_{\rm s}$. Thus, the error bars and FoMs in this paper differ from those reported for the blinded Fisher analysis in XuDC1, titled ``Lachesis.'' This example highlights the value of data challenges and comparison of independent calculations, even in idealized cases where results ``should'' agree.

\section{Dissecting the Information Content of the Observables} \label{sec:dissect}

In this section, we explore the contribution of different components of the data and priors to the cosmological constraining power. In Section~\ref{ss:probe}, we discuss how marginalization choices affect the FoMs and study the constraining power from different subsets of the three probes. In Sections~\ref{ss:tomo} and \ref{ss:scale}, we investigate the information from different tomographic bins and angular scales, respectively. Finally in Section~\ref{ss:ssc}, we estimate the potential gain of information by mitigating the super-sample covariance.

\subsection{Marginalization Choices and Different Probes} \label{ss:probe}

\begin{figure*}
    \plotone{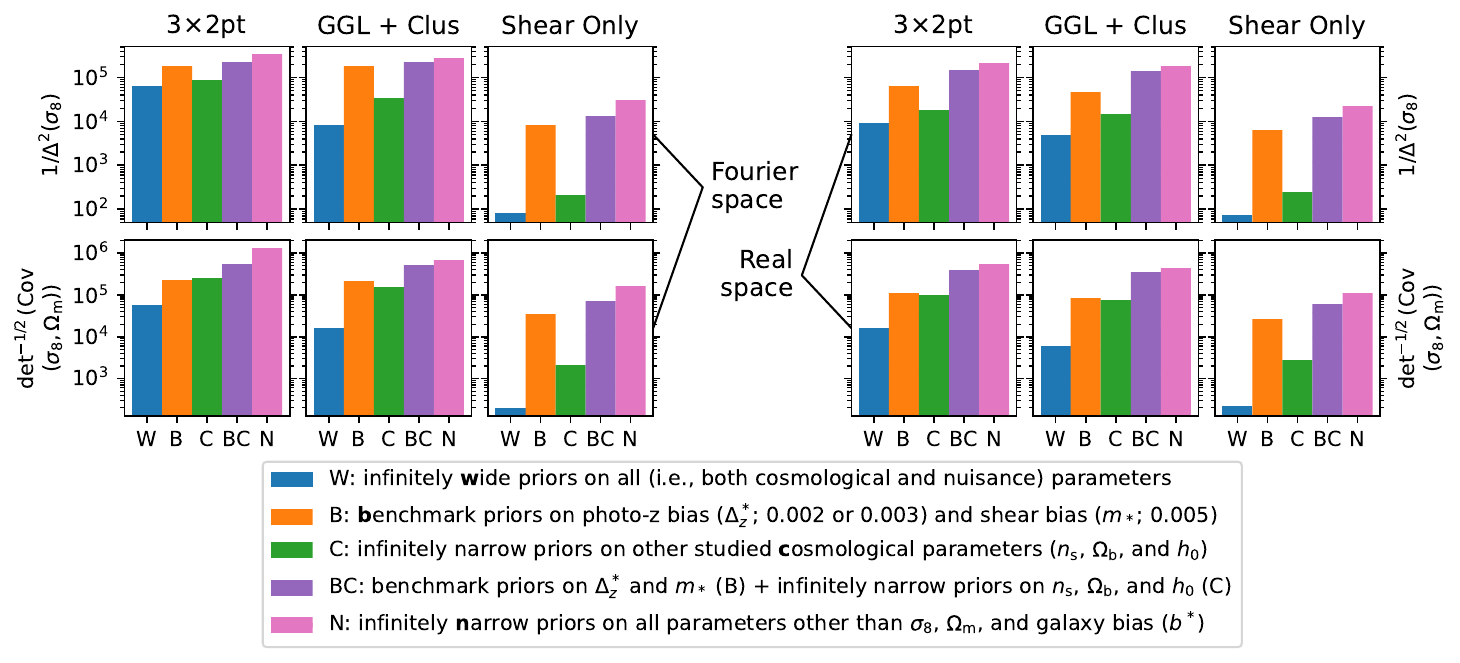}
    \caption{\label{fig:probe}Constraining power of different combinations of probes: the combination of all three probes (``$3\!\times\!2$pt''), the combination of galaxy-galaxy lensing (``GGL'') and galaxy clustering (``Clus''), and cosmic shear (``Shear'') only. The two rows present our two figures of merit, $1/\Delta^2(\sigma_8)$ and ${\det}^{-1/2}({\rm Cov} (\sigma_8, \Omega_{\rm m}))$, respectively. The left panels show results in real space, and the right panels show results in Fourier space. In each panel, different colors correspond to different marginalization choices, as briefly explained in the legend at the bottom of the figure; see the text for detailed explanations. Specific values are tabulated in Table~\ref{tab:probe}.}
\end{figure*}

\begin{table*}[]
    \centering
    \begin{tabular}{cc|r@{.}lr@{.}lr@{.}lr@{.}lr@{.}lr@{.}lr@{.}l}
    \hline
        FoM (Space) & Choice & \multicolumn{2}{c}{$3\!\times\!2$pt} & \multicolumn{2}{c}{GGL+Clus} & \multicolumn{2}{c}{Shear Only} & \multicolumn{2}{c}{Shear+GGL} & \multicolumn{2}{c}{GGL Only} & \multicolumn{2}{c}{Shear+Clus} & \multicolumn{2}{c}{Clus Only} \\
    \hline
        \multirow{5}{5em}{$1/\Delta^2(\sigma_8)$$\times 10^{-3}$ (Fourier)}
        & W & 63&92 & 8&123 & 0&0798 & 5&462 & 1&665 & 1&179 & 0&7769 \\
        & B & 182&7 & 177&2 & 8&151 & 43&06 & 16&35 & 135&8 & 114&9 \\
        & C & 88&00 & 34&07 & 0&2056 & 6&788 & 3&215 & 11&13 & 8&014 \\
        & BC & 222&7 & 218&3 & 13&01 & 51&29 & 16&80 & 183&0 & 146&2 \\
        & N & 336&7 & 278&3 & 29&60 & 75&64 & 18&59 & 244&4 & 156&0 \\
    \hline
        \multirow{5}{5em}{$1/\Delta^2(\sigma_8)$$\times 10^{-3}$ (Real)}
        & W & 9&249 & 5&003 & 0&0755 & 3&117 & 0&6226 & 0&3989 & 0&2129 \\
        & B & 61&61 & 46&61 & 6&186 & 23&81 & 3&356 & 21&85 & 3&317 \\
        & C & 17&56 & 14&75 & 0&2446 & 4&714 & 1&686 & 4&956 & 1&724 \\
        & BC & 149&1 & 136&7 & 12&09 & 43&13 & 9&280 & 89&68 & 3&523 \\
        & N & 205&5 & 178&6 & 22&04 & 54&24 & 10&19 & 143&1 & 3&545 \\
    \hline
        \multirow{5}{5em}{$\det^{-1/2}({\rm Cov}$ $(\sigma_8, \Omega_{\rm m}))$$\times 10^{-3}$ (Fourier)}
        & W & 56&77 & 16&18 & 0&1964 & 6&202 & 2&023 & 4&015 & 2&437 \\
        & B & 227&9 & 208&7 & 34&63 & 94&49 & 7&203 & 178&1 & 40&24 \\
        & C & 250&8 & 149&7 & 2&071 & 20&55 & 13&09 & 80&45 & 67&42 \\
        & BC & 527&9 & 507&0 & 68&52 & 154&9 & 30&80 & 451&7 & 328&4 \\
        & N & 1325& & 671&0 & 159&1 & 300&5 & 32&87 & 647&3 & 340&8 \\
    \hline
        \multirow{5}{5em}{$\det^{-1/2}({\rm Cov}$ $(\sigma_8, \Omega_{\rm m}))$$\times 10^{-3}$ (Real)}
        & W & 16&51 & 6&018 & 0&2176 & 4&067 & 0&9077 & 1&771 & 0&6840 \\
        & B & 109&2 & 84&43 & 26&32 & 62&35 & 2&674 & 56&23 & 4&129 \\
        & C & 99&90 & 72&23 & 2&638 & 22&34 & 10&93 & 33&94 & 17&05 \\
        & BC & 383&4 & 337&8 & 60&06 & 140&5 & 26&09 & 272&7 & 43&07 \\
        & N & 524&3 & 429&3 & 106&0 & 207&1 & 27&54 & 393&8 & 43&29 \\
    \hline
    \end{tabular}
    \caption{\label{tab:probe}Constraining power of different combinations of probes. The third column corresponds to the combination of all three probes (``$3\!\times\!2$pt''), and the fourth to ninth columns correspond to the three subsets of two and the three individual probes. The first and third (horizontal) blocks present our first figure of merit (FoM), $1/\Delta^2(\sigma_8)$, while the second and fourth blocks present our second FoM, ${\det}^{-1/2}({\rm Cov} (\sigma_8, \Omega_{\rm m}))$. Both FoMs are scaled by a factor of $10^{-3}$ to enhance readability. Within each block, different rows correspond to different marginalization choices explained in the text. See Figure~\ref{fig:probe} for a visualization of the third to fifth columns.}
\end{table*}

Figure~\ref{fig:probe} shows the two FoMs for three combinations of probes --- the combination of all three probes, the combination of galaxy-galaxy lensing (GGL) and galaxy clustering, and cosmic shear only --- with different marginalization choices in both real and Fourier spaces.  These are as follows:
\begin{itemize}
    \item Choice ``W'': The blue bars present the constraining power from observational data alone, i.e., with flat and wide priors on all parameters. This amounts to directly inverting ${\bf \Sigma}_{\cal L}^{-1}$ calculated using Equation~(\ref{eq:fisher}). In other words, we superimpose an all-zero prior precision matrix ${\bf \Sigma}_{\rm Prior}^{-1}$, which corresponds to infinitely wide prior on all parameters.
    \item Choice ``B'': This is the benchmark case most closely matched to the Data Challenge 1 (DC1), with Gaussian priors on photo-$z$ biases and shear biases. ${\bf \Sigma}_{\rm Prior}^{-1}$ corresponds to Equation~(\ref{eq:prior}), with $\Delta (\Delta_{z}^i)$ and $\Delta (m_i)$ set following DC1.
    \item Choice ``C'': This mimics the situation where we have prior knowledge about cosmology from probes other than weak lensing. Like choice ``W,'' this case uses an all-zero ${\bf \Sigma}_{\rm Prior}^{-1}$; however, it assumes that other cosmological parameters studied in this work ($n_{\rm s}$, $\Omega_{\rm b}$, and $h_0$) have infinitely narrow priors. In reality, the constraints on these parameters from external data like cosmic microwave background \citep[CMB; e.g.,][]{2020A&A...641A...6P} and baryon acoustic oscillations \citep[BAO; e.g.,][]{desicollaboration2025desidr2resultsii} are of course not infinitely tight, and the results shown here should be interpreted as upper limits. Nevertheless, the CMB and BAO constraints on these parameters are much stronger than those from weak lensing observations, and this choice isolates the impact of these power spectrum shape parameters on the weak lensing FoMs for ($\sigma_8$, $\Omega_{\rm m}$).
    \item Choice ``BC'': This is the combination of choices ``B'' and ``C,'' i.e., with priors on some bias parameters and external cosmological knowledge. Mathematically, this means both ${\bf \Sigma}_{\rm Prior}^{-1}$ from DC1 and infinitely narrow priors on $n_{\rm s}$, $\Omega_{\rm b}$, and $h_0$. In the future, if cosmological parameters based on Roman HLIS data are in reasonable agreement with CMB and BAO, they will be combined to attain tighter constraints. Choice ``BC'' estimates the upper limits for such combinations.
    \item Choice ``N'': This case is more extreme than choice ``BC.'' We assume infinitely narrow priors on all parameters other than $\sigma_8$, $\Omega_{\rm m}$ and galaxy biases; in other words, compared to choice ``BC,'' not only do we take the limits of $\Delta (\Delta_{z}^i) \to 0$ and $\Delta (m_i) \to 0$, but we also assume that intrinsic alignments are perfectly known from astrophysical measurements. Note that we still marginalize over galaxy bias parameters $b_i$, as otherwise constraints from GGL or clustering alone would be unrealistically tight.
\end{itemize}
FoMs for all non-empty subsets of the three probes are tabulated in Table~\ref{tab:probe}.

We focus first on the combination of all three probes (``$3\!\times\!2$pt''), shown by the left panels in each half of Figure~\ref{fig:probe}. In all cases (real and Fourier space, both FoMs), adding the benchmark priors (``B'') on photo-$z$ and shear biases improves constraining power substantially (factors of $2.9$--$6.7$) compared to the wide priors (``W''). Substantial improvements also come from adding cosmological priors on $n_{\rm s}$, $\Omega_{\rm b}$, and $h_0$ (``C''), even without the benchmark priors. Of greatest practical impact, adding the cosmological priors to the benchmark priors (``BC'') produces nearly half an order of magnitude gain in FoM2 $= {\det}^{-1/2}({\rm Cov} (\sigma_8, \Omega_{\rm m}))$ relative to benchmark priors alone, though the gain is smaller for FoM1 $= 1/\Delta^2(\sigma_8)$. This difference demonstrates the value of using external constraints on the shape of the matter power spectrum rather than relying on Roman clustering data alone. The further gain from infinitely narrow priors on nuisance parameters (``N'') is smaller, though still significant in Fourier space. We examine the impact of priors in more detail in Section~\ref{sec:prior}.

For any choice of prior assumptions, the real space FoM is significantly lower than the Fourier space FoM. We believe that this difference reflects the scale cuts applied in each case. For both analyses, small scales are masked, mitigating sensitivity to baryonic effects and other non-linearities not captured by the {\sc CoCoA} model. However, Figure~\ref{fig:probe} suggests that the cut at $\ell_{\max} \simeq 2452$ retains more small-scale information than the cut at $\theta_{\min} \simeq 5.49 \,{\rm arcmin}$, which corresponds roughly to $\ell_{\max} \simeq \pi / \theta_{\min} \simeq 1967$. Further investigation will be needed to see whether this extra information loses its cosmological constraining power once baryonic effects and other theoretical uncertainties in the non-linear regime are accounted for. We discuss the impact of scale cuts further in Section~\ref{ss:scale}.

The other columns of Figure~\ref{fig:probe} and, more comprehensively, Table~\ref{tab:probe} show FoM results for different subsets of the three probes. The most striking result is that shear alone is always much less constraining than GGL+clustering, while the constraints for GGL+clustering are close to those for the full $3\!\times\!2$pt. In recent analysis of, e.g., DES weak lensing, the constraints from cosmic shear and from GGL+galaxy clustering show comparable constraining power \citep{2022PhRvD.105b3520A}. At least within the assumptions made for DC1 (which, importantly, include ``lens $=$ source''), the Roman constraints are expected to be dominated by GGL+clustering (see Figure~\ref{fig:dv_layout}). Such expectation is also supported by our MCMC results for shear only and GGL+clustering (not shown in figures). A more complex bias model might degrade the constraints from GGL+clustering, if linear bias is not an adequate description of HLIS precision on these scales.

The shear-only constraints with wide priors (``W'') are particularly weak, though adding either the benchmark priors on nuisance parameters or strong constraints on ($n_{\rm s}$, $\Omega_{\rm b}$, $h_0$) improves them considerably. Table~\ref{tab:probe} also shows that GGL alone or clustering alone gives weak constraints. This is as expected: Unknown galaxy bias factors are a severe degeneracy for either of these observables on its own, but they are calibrated by the combination because GGL scales (in the linear regime) as $b_g$ while the galaxy auto-correlation scales as $b_g^2$.

\subsection{Tomographic Bins} \label{ss:tomo}

\begin{figure*}
    \plotone{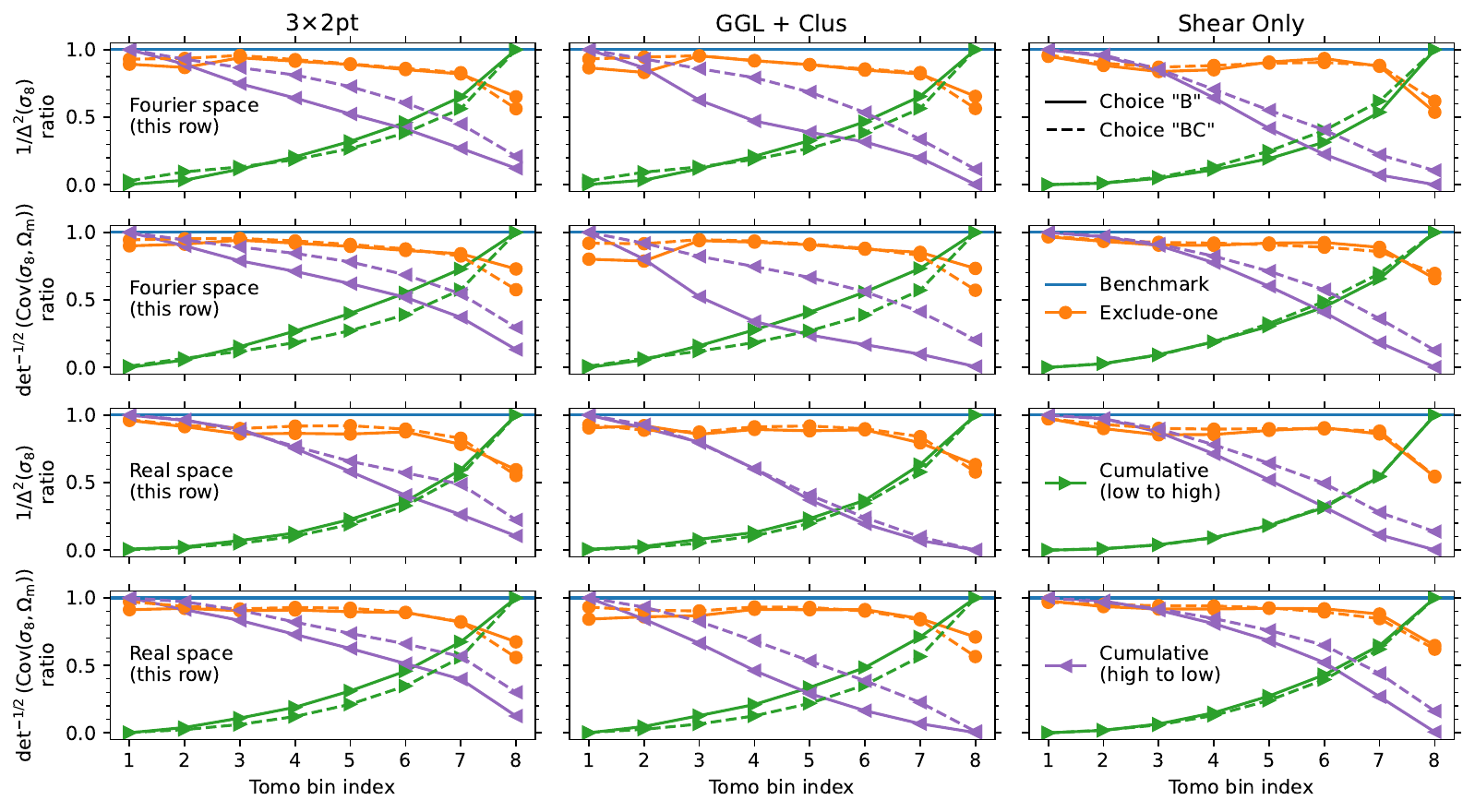}
    \caption{\label{fig:tomo}Breakdown of contributions from tomographic bins. From left to right, the three columns correspond to the combination of all three probes (``$3\!\times\!2$pt''), the combination of galaxy-galaxy lensing (``GGL'') and galaxy clustering (``Clus''), and cosmic shear (``Shear'') only, respectively. Like in Figure~\ref{fig:probe}, the first (last) two rows show results in real (Fourier) space; the first and third rows present FoM1 $= 1/\Delta^2(\sigma_8)$, while the second and fourth rows present FoM2 $= {\det}^{-1/2}({\rm Cov} (\sigma_8, \Omega_{\rm m}))$. The ratios are computed by dividing figures of merit in variant cases (with a subset of tomographic bins) by the corresponding ones with all tomographic bins included. In all panels, solid (dashed) curves correspond to marginalization choice ``B'' (``BC''), as defined in Figure~\ref{fig:probe} and explained in Section~\ref{sec:dissect}. Blue represents the benchmark case, orange represents the variant cases with all-but-one tomographic bins, and green (purple) represents the combination of the labeled bin and those on its left (right).}
\end{figure*}

Figure~\ref{fig:tomo} shows the breakdown of information from different tomographic bins. The four rows correspond to the two FoMs in real and Fourier spaces, and the three columns are three combinations of probes. Interestingly, the overall behavior is similar for all cases. It is clear that higher-redshift bins contribute more than lower-redshift bins. For $3\!\times\!2$pt, the combinations of i) the three highest-redshift bins and ii) the six lowest-redshift bins have basically the same constraining power. For GGL+clustering or cosmic shear only, the numbers of bins are slightly different, but the quantitative conclusions are similar. Remarkably, the highest redshift tomographic bin contributes $25$--$45\%$ of the FoM in most cases, while excluding even the three lowest redshift bins makes only a relatively small difference ($20$--$40\%$) to the FoM in most cases. It is thus clear that Roman weak lensing cosmology benefits a lot from Roman's ability to reach unprecedented depths, partially thanks to its highly sensitive Wide Field Instrument \citep[WFI;][]{2020JATIS...6d6001M}. Meanwhile, note that redshift calibration gets hardest at high redshifts \citep[see, e.g.,][]{2023MNRAS.524.5109R} in a real analysis, and there are trade-offs associated with the importance of the highest redshift bins.

\subsection{Angular Scales} \label{ss:scale}

\begin{figure*}
    \plotone{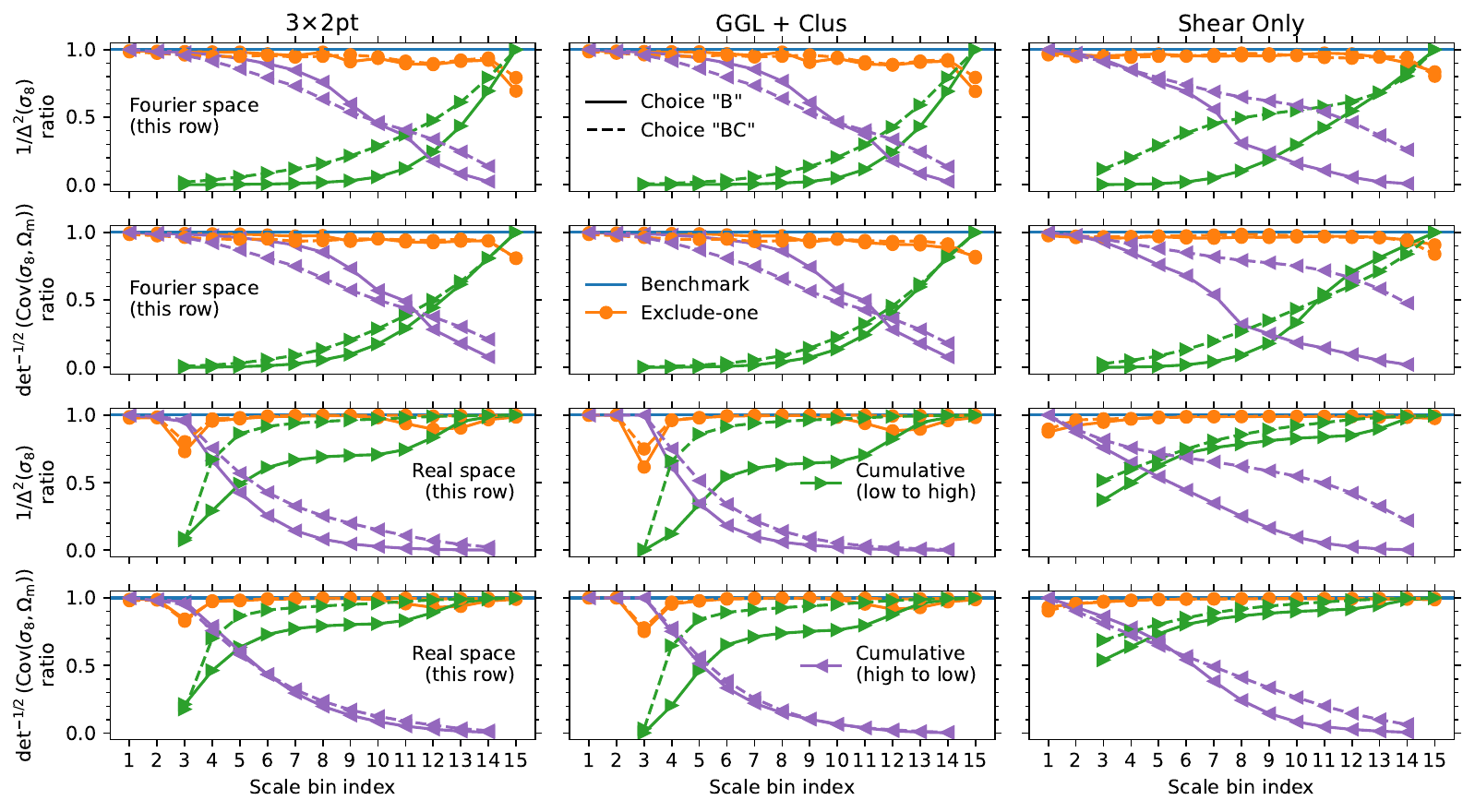}
    \caption{\label{fig:scale}Breakdown of contributions from angular scales. Similar to Figure~\ref{fig:tomo}, but for angular scales. Note that in real space, the angular scale bin index runs from small scale ($\theta \simeq 2.97 \,{\rm arcmin}$ for the $1$st bin) to large scale ($\theta \simeq 219 \,{\rm arcmin}$ for the $15$th bin), while in Fourier space, it runs from large scale ($\ell \simeq 35.31$ for the $1$st bin) to small scale ($\ell \simeq 3398$ for the $15$th bin).}
\end{figure*}

In Fourier space, the number of modes grows rapidly toward small scales (i.e., large $\ell$ values), but theoretical systematics are larger there due to non-linear growth and may limit ability to use the information in practice. At large scales, variations in zodiacal background, dust extinction, and stellar contamination could introduce observational systematics, but it should be possible to calibrate these effects, and we do not introduce any low-$\ell$ cut. Figure~\ref{fig:scale} shows the breakdown of information from different angular scales. As highlighted in the caption, from left to right, the index runs from small scales to large scales in real space and does the opposite in Fourier space. 

We see the clear trend that small-scale bins contribute more than large-scale bins in both spaces. For $3\!\times\!2$pt in Fourier (real) space, the combinations of i) the four (five) smallest-scale bins and ii) the twelve (eleven) largest-scale bins have basically the same constraining power. The other combinations of probes show consistent results. Here we reiterate that in DC1, angular scale cuts are implemented as masks in both spaces. For example, in real space, typically the two smallest-scale bins are masked out for GGL and clustering (but not for shear).\footnote{This explains the ``troughs'' at bin $3$ in exclude-one (i.e., including all-but-one angular scale bins; orange) curves for $3\!\times\!2$pt and ``GGL+Clus'' in real space. A smaller-scale bin contributes more information, hence excluding bin $3$ causes more loss of constraining power than excluding any of bin $4$ and above. Meanwhile, bin $1$ and bin $2$ are already masked out for GGL and clustering, and excluding them only affects cosmic shear. Therefore, the exclude-one curves for the shear-only case in real space are still monotonic.} In Fourier space, we see a similar non-monotonicity in some cases (e.g., FoM1 $= 1/\Delta^2(\sigma_8)$ from cosmic shear only), but the contrast is not as large as in real space.

From Figure~\ref{fig:probe}, it seems like the constraining power in Fourier space is larger than that in real space. Figure~\ref{fig:scale} emphasizes the familiar point that constraining power is sensitive to the minimum scale used in the analysis. A sharp cut in $\ell$ does not correspond to a sharp cut in $\theta$, nor vice versa, but we can approximately match scales through $\theta \approx \pi/\ell$. In DC1, the maximum $\ell$ for Fourier space inference from GGL and clustering is $\ell_{\max} \simeq 2452$, while the minimum $\theta$ for real space inference, $\theta_{\min} \simeq 5.49 \,{\rm arcmin}$, corresponds to $\ell_{\max} \simeq 1967$. Thus, the higher FoMs from Fourier space plausibly come from adopting an effectively smaller minimum scale. In terms of comoving wavenumbers, we have: $\ell_{\max} \simeq 2452$ corresponds to $k_{\max} \simeq 1.07 \,h \,{\rm Mpc}^{-1}$ and $\ell_{\max} \simeq 1967$, $k_{\max} \simeq 0.685 \,h \,{\rm Mpc}^{-1}$ at $z \simeq 1$; $\ell_{\max} \simeq 2452$ corresponds to $k_{\max} \simeq 0.1249 \,h \,{\rm Mpc}^{-1}$ and $\ell_{\max} \simeq 1967$, $k_{\max} \simeq 0.550 \,h \,{\rm Mpc}^{-1}$ at $z \simeq 2$. For comprehensive listings of scales, see tables in Appendix~\ref{app:dc1}. Here we take $k_{\max} = \ell_{\max} / D_{\rm C}$, where $D_{\rm C}$ is the comoving angular diameter distance, but the relation between Fourier wavenumber and harmonic wavenumber is only approximate. Since $k_{\max} \sim 0.1 \,h \,{\rm Mpc}^{-1}$ is the typical scale where simple perturbation theories break down, this comparison indicates that advanced theoretical models like effective field theory of large scale structure \citep[EFTofLSS;][]{2012JCAP...07..051B, 2012JHEP...09..082C} are important for enhancing constraining power. {\sc CoCoA} uses CosmoLike \citep{2014MNRAS.440.1379E, 2017MNRAS.470.2100K, 2020JCAP...05..010F} to compute matter and galaxy clustering in the non-linear regime, but we do not marginalize over any uncertainties in this description.

\subsection{Super-Sample Covariance} \label{ss:ssc}

As mentioned in Section~\ref{ss:cocoa}, super-sample covariance \citep[SSC;][]{2013PhRvD..87l3504T, 2018JCAP...06..015B} is the dominant component of the non-Gaussian covariance. Mitigating SSC \citep[e.g.,][]{2019JCAP...10..004D} can reduce the covariance of the data vector and thus enhance the constraining power on the parameters of interest. SSC results from the fact that every weak lensing survey has a finite volume and modes on scales larger than this volume can couple non-linearly to smaller scale modes. For Roman HLIS, it will be possible to use the larger-area LSST \citep{2012arXiv1211.0310L, 2019ApJ...873..111I} to measure large scale modes of the ``galaxy overdensity'' and mitigate SSC by estimating the corresponding modes of the matter density within the enclosed HLIS footprint. Since it is mitigation, not elimination, it makes sense to write the resulting covariance matrix as a linear combination of its Gaussian (``G'') and non-Gaussian (``NG''; mainly SSC) components
\begin{equation}
    {\rm Cov} = {\rm Cov}^{\rm G} + \lambda {\rm Cov}^{\rm NG},
    \label{eq:ssc}
\end{equation}
where $\lambda \in [0, 1]$ is an undetermined coefficient. For all cases studied in this work, except those in this section, $\lambda$ is always $1$.

\begin{figure*}
    \plotone{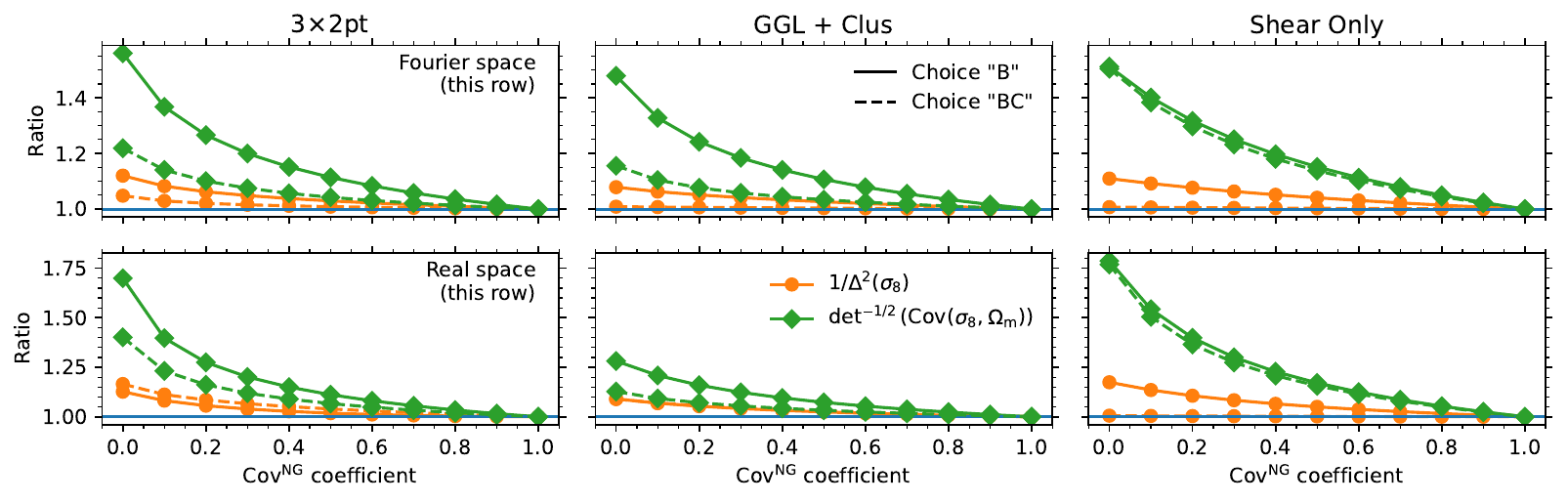}
    \caption{\label{fig:ssc}Gain of information by mitigating super-sample covariance. The three columns are the same as those in Figures~\ref{fig:tomo} and \ref{fig:scale}. The two rows show results in real and Fourier space, respectively. The $x$-axis is the coefficient $\lambda$; for each data point, the covariance matrix is a linear combination of the Gaussian and non-Gaussian components: ${\rm Cov} = {\rm Cov}^{\rm G} + \lambda {\rm Cov}^{\rm NG}$. Perfect SSC mitigation corresponds to $\lambda = 0$ and no mitigation to $\lambda = 1$. The $y$-axis represents FoM ratios, and solid (dashed) curves correspond to marginalization choice ``B'' (``BC''). The benchmark case is shown as blue horizontal lines, while our two figures of merit are shown in orange and green, respectively.}
\end{figure*}

Figure~\ref{fig:ssc} shows how our FoMs vary with this $\lambda$. In most cases, the FoMs are monotonic functions of $\lambda$, as expected. For FoM1 $= 1/\Delta^2(\sigma_8)$, even perfect mitigation of SSC ($\lambda = 0$) produces fairly modest gains. However, the impact is larger for FoM2 $= {\det}^{-1/2}({\rm Cov} (\sigma_8, \Omega_{\rm m}))$, with a factor of $\sim 1.6$ gain for $3\!\times\!2$pt with choice ``B''; the gain is less for choice ``BC''. This suggests that much of the improvement with choice ``B'' is coming from better constraints on the shape of the power spectrum, which provides information about $\Omega_{\rm m}$ but not about $\sigma_8$. However, for shear only, the trends are basically the same, regardless of whether external cosmological knowledge is included. Figure~\ref{fig:ssc} implies that SSC mitigation must be at least $50\%$ effective in removing the non-Gaussian covariance (i.e., $\lambda \leq 0.5$) to have much impact on cosmological inference. However, with strong mitigation the potential gains are significant, depending on the observables used and the strength of external priors.

\section{Impact of Priors} \label{sec:prior}

As shown in Section~\ref{sec:dcbase}, the assumed priors on ``nuisance'' parameters (namely photo-$z$ biases and shear biases) significantly enhance the constraining power from observational data alone. In DC1, the widths of priors are empirically chosen based on previous surveys. Furthermore, priors of the same type are assumed to be uncorrelated across different tomographic bins. In this section, we study how assumptions about priors change our figures of merit. Sections~\ref{ss:photoz} and \ref{ss:shear} address priors on photometric redshift bias and multiplicative shear bias, respectively. For each type of prior, we investigate scaling and correlation separately.

Realistic covariance matrices for these biases will ultimately come from simulations and data. Here, to develop some intuitive understanding, we look at two specific forms of injected correlations: i) all bin pairs being correlated with the same coefficient
\begin{equation}
    {\rm Cov} (\rho) = \sigma^2 \begin{pmatrix}
    1 & \rho & \rho & \cdots & \rho & \rho \\
    \rho & 1 & \rho & \cdots & \rho & \rho \\
    \rho & \rho & 1 & \cdots & \rho & \rho \\
    \vdots & \vdots & \vdots & \ddots & \vdots & \vdots \\
    \rho & \rho & \rho & \cdots & 1 & \rho \\
    \rho & \rho & \rho & \cdots & \rho & 1 \\
    \end{pmatrix},
    \label{eq:corr_all}
\end{equation}
and ii) only adjacent pairs being correlated with the same coefficient
\begin{equation}
    {\rm Cov} (\rho) = \sigma^2 \begin{pmatrix}
    1 & \rho & 0 & \cdots & 0 & 0 \\
    \rho & 1 & \rho & \cdots & 0 & 0 \\
    0 & \rho & 1 & \cdots & 0 & 0 \\
    \vdots & \vdots & \vdots & \ddots & \vdots & \vdots \\
    0 & 0 & 0 & \cdots & 1 & \rho \\
    0 & 0 & 0 & \cdots & \rho & 1 \\
    \end{pmatrix}.
    \label{eq:corr_pair}
\end{equation}
In both cases, $\rho$ denotes the correlation coefficient. Since a covariance matrix must be positive (semi)definite, we study $\rho \in [-0.1, 0.9]$ for Equation~(\ref{eq:corr_all}) and $\rho \in [-0.5, 0.5]$ for Equation~(\ref{eq:corr_pair}). Intuitively, if there were only two tomographic bins, the two forms would be identical, and both positive and negative values of $\rho$ would reduce the determinant of the covariance matrix (which is $1 - \rho^2$). However, correlated priors only affect the constraining power via the sum ${\bf \Sigma}_{\cal L}^{-1} + {\bf \Sigma}_{\rm Prior}^{-1}$, and the actual relationship can be more complicated.

\subsection{Photometric Redshift Bias} \label{ss:photoz}

\begin{figure*}
    \plotone{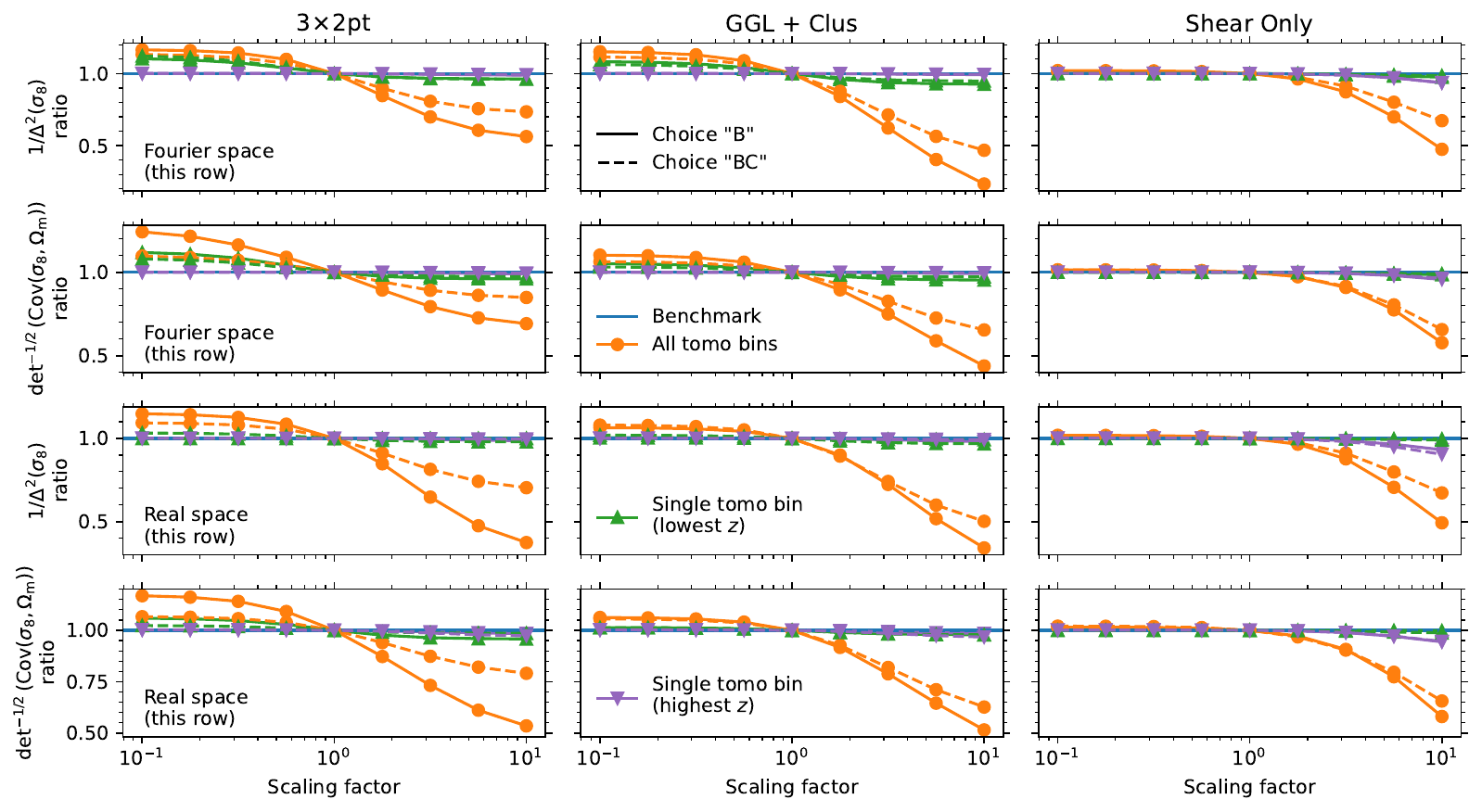}
    \caption{\label{fig:photoz_scale}Impact of scaling of priors on photo-$z$ biases. The panel layout is the same as in Figures~\ref{fig:tomo} and \ref{fig:scale}. In all panels, solid (dashed) curves correspond to marginalization choice ``B'' (``BC''), as defined in Figure~\ref{fig:probe} and explained in Section~\ref{sec:dissect}. The benchmark case, with all priors set following DC1, is shown as blue horizontal lines. For orange curves, we rescale all priors on photo-$z$ biases simultaneously; for green (purple) curves, we only rescale the prior on photo-$z$ bias in the lowest-$z$ (highest-$z$) bin while keeping others fixed. Note that in DC1, priors on photo-$z$ biases have width $0.002$ in Fourier space and $0.003$ real space, regardless of the redshift.}
\end{figure*}

Figure~\ref{fig:photoz_scale} shows how our figures of merit change when assumed prior widths for photo-$z$ biases are rescaled by a factor between $0.1$ and $10$. In DC1, the default widths in Fourier and real spaces are $0.002$ and $0.003$, respectively. The FoMs are monotonic functions of the scaling factor, as expected. A closer look reveals that the dependence is non-linear. For $3\!\times\!2$pt (left column), both gain and loss of information seem to saturate at certain smaller or larger widths. Nevertheless, the dynamic ranges are already significant, from $\sim +10\%$ when the errors are $10$ times smaller to $\sim -30\%$ or even less when they are $10$ times larger, implying the importance of control over photo-$z$ systematics. When photo-$z$ priors in all tomographic bins are rescaled by the same factor, for a factor of $\sim 2$ range around the fiducial prior, the change in the $3\!\times\!2$pt FoM is roughly $\pm (6$--$15)\%$ for $1/\Delta^2(\sigma_8)$ and $\pm (4$--$13)\%$ for ${\det}^{-1/2}({\rm Cov} (\sigma_8, \Omega_{\rm m}))$.

For GGL+clustering (middle column) or cosmic shear only (right column), we see that the gain of information due to smaller $\Delta_{z}^i$ is limited (up to $\sim +5\%$), while the loss due to larger values is substantial, at the $\sim -40\%$ level when no external cosmological information is included (solid curves). When the control over photo-$z$ bias is weaker than the DC1 assumption (scaling factor $> 1$), the photo-$z$ prior of the highest redshift tomographic bin has the largest impact, and we show the impact of varying only the prior for this bin with the purple curves. When the control is stronger than the DC1 assumption (scaling factor $< 1$), especially for $3\!\times\!2$pt (left) and when external cosmological information is included (dashed curves), the lowest redshift tomographic bin is more influential, and we show the impact of varying only the prior for this bin with the green curves.

\begin{figure*}
    \plotone{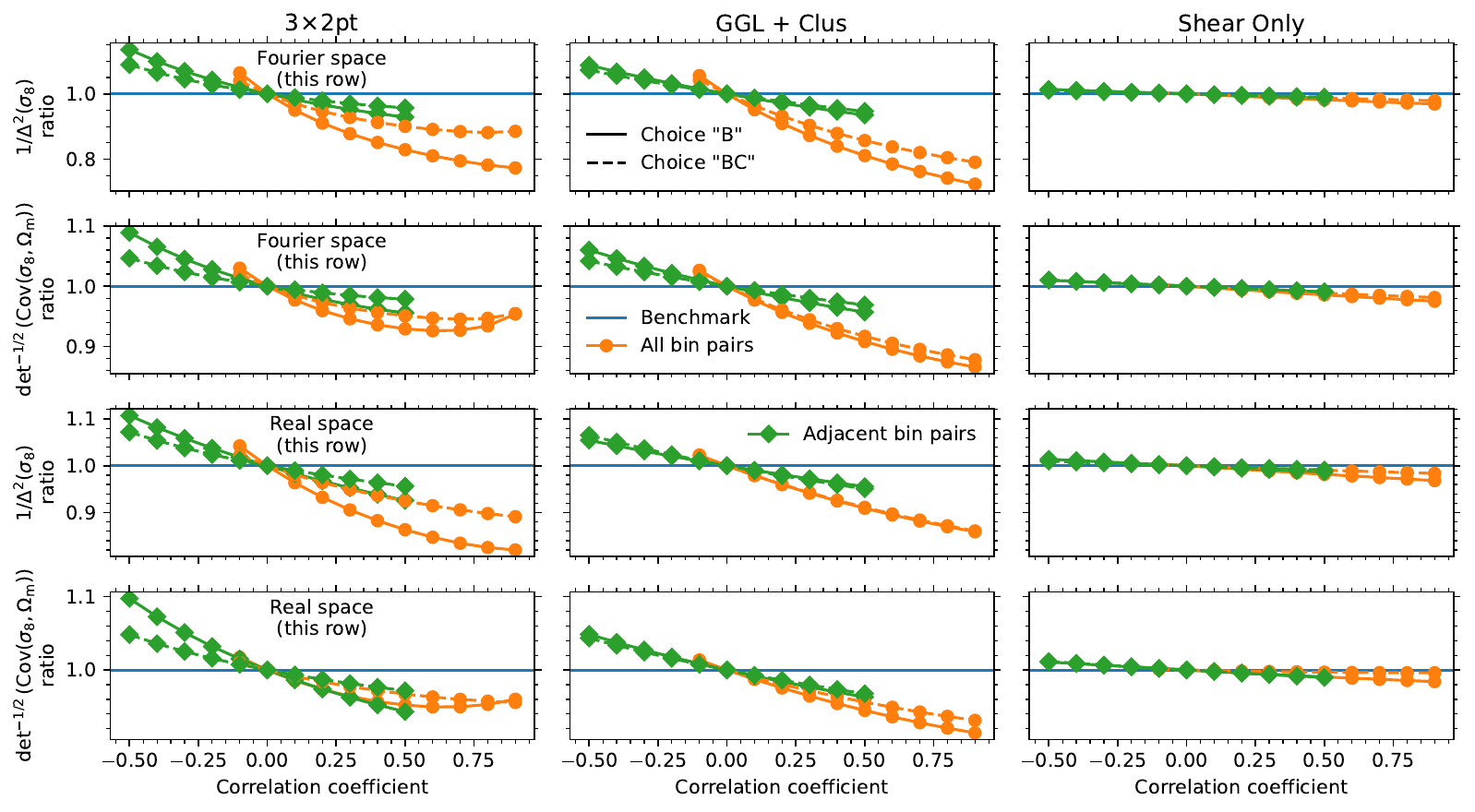}
    \caption{\label{fig:photoz_corr}Impact of correlation of priors on photo-$z$ biases. The panel layout and line styles are the same as in Figure~\ref{fig:photoz_scale}. Again, the benchmark case is shown as blue horizontal lines. For orange curves, we inject correlations to all bin pairs using Equation~(\ref{eq:corr_all}); for green curves, we only inject correlations to adjacent bin pairs using Equation~(\ref{eq:corr_pair}).}
\end{figure*}

Training and calibration methods for photometric redshifts could plausibly lead to biases that are correlated from bin to bin (because of common errors) or anti-correlated (because of galaxies shifting from one bin to a neighboring bin). Figure~\ref{fig:photoz_corr} shows how the constraining power varies with the correlation coefficient $\rho$ in either Equation~(\ref{eq:corr_all}) or Equation~(\ref{eq:corr_pair}). In most cases, the FoMs monotonically decrease with an increasing correlation coefficient. However, even with extreme correlation coefficients, the impact is only at the $\sim 10\%$ level. Therefore, we refrain from discussing the trends in detail. Realistic covariance matrices for photo-$z$ biases might lead to different conclusions, but Figures~\ref{fig:photoz_scale} and \ref{fig:photoz_corr} suggest that the magnitude of photo-$z$ priors is more important to control than the bin-to-bin correlation.

Photometric redshifts in the HLIS will be based on a combination of Roman and LSST photometry, and calibrated with spectroscopic surveys and, potentially, with clustering-based redshifts and with galaxy spectral energy distribution (SED) models. Figure~\ref{fig:photoz_scale} implies that if this calibration can be achieved even within a factor of $2$--$3$ of the level adopted for DC1, the impact of photo-$z$ bias uncertainties on cosmology will be limited. However, these are still gains to be made if the photo-$z$ calibration can be even better than that assumed in DC1. Furthermore, analyses that use complementary weak lensing and clustering statistics or extend to more non-linear scales may achieve substantially higher FoMs. In this case, photo-$z$ bias uncertainties could become a limiting factor, though it is also possible that these alternative measures will themselves constrain the photo-$z$ biases.

\subsection{Multiplicative Shear Bias} \label{ss:shear}

\begin{figure*}
    \plotone{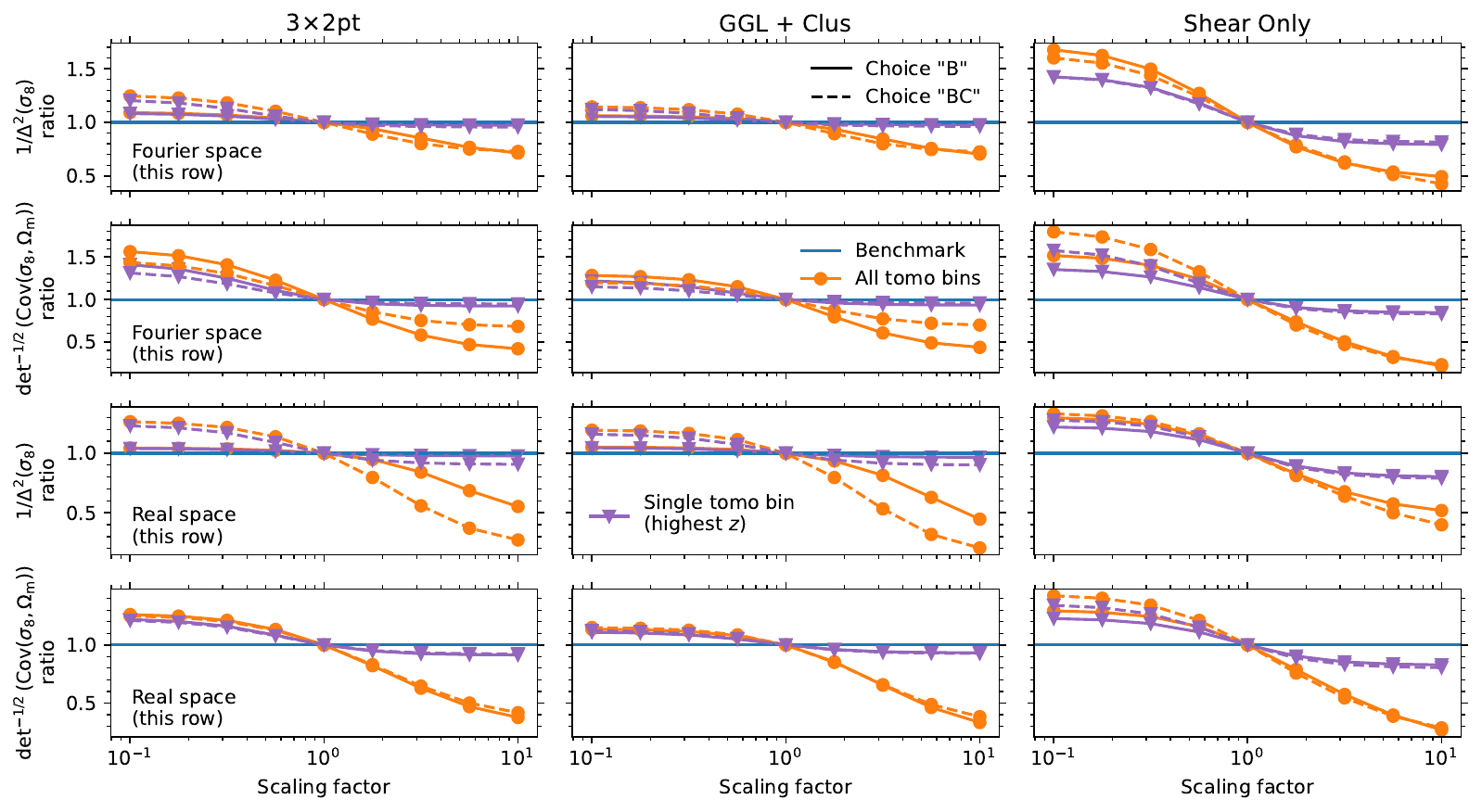}
    \caption{\label{fig:shear_scale}Impact of scaling of priors on shear biases. Same as Figure~\ref{fig:photoz_scale}, but for shear biases. The green curves (only rescaling the prior on shear bias in the lowest-$z$ bin) are omitted as they are mostly flat. Note that in DC1, all priors on shear biases have width $0.005$ in both real and Fourier spaces.}
\end{figure*}

In addition to photo-$z$ biases, DC1 includes priors on multiplicative shear biases as well. Figure~\ref{fig:shear_scale} shows how the FoMs change when these priors have different widths. Compared to Figure~\ref{fig:photoz_scale}, we see significantly larger dynamical ranges with the same domain for the scaling factor. For $3\!\times\!2$pt (left column), when cosmological constraints from external data are included (dashed curves), the gain of information when the errors are $10$ times smaller is up to $\sim +50\%$, and the loss when they are $10$ times larger is up to $\sim -50\%$. A significant fraction of the gain from much smaller shear bias uncertainties comes from the highest-$z$ tomographic bin alone, as one can see by comparing purple and orange curves of the same line type. Factor of $\sim 2$ changes in the shear bias uncertainty (for all bins) produce roughly $\pm 15\%$ changes in FoM2 $= {\det}^{-1/2}({\rm Cov} (\sigma_8, \Omega_{\rm m}))$ for the case with strong cosmological priors (``BC,'' dashed orange) in real or Fourier space, and smaller changes (around $\pm 5\%$) for FoM1 $= 1/\Delta^2(\sigma_8)$ for benchmark (``B,'' solid orange) priors.

For GGL+clustering (middle column), the loss of information is similar to the $3\!\times\!2$pt case when the errors are larger, but the gain is not as significant when they are smaller, only at the $\sim +20\%$ level. However, for cosmic shear only (right column), the fractional gains or losses in the FoM are similar to or larger than those for $3\!\times\!2$pt. These comparison results are understandable, as cosmic shear autocorrelation is fully based on shear measurements, GGL is partially based on them, and galaxy clustering does not involve shape measurements at all. For cosmic shear, changing the shear bias uncertainty by a factor of $\sim 2$ changes FoMs by $\pm (5$--$20)\%$.

\begin{figure*}
    \plotone{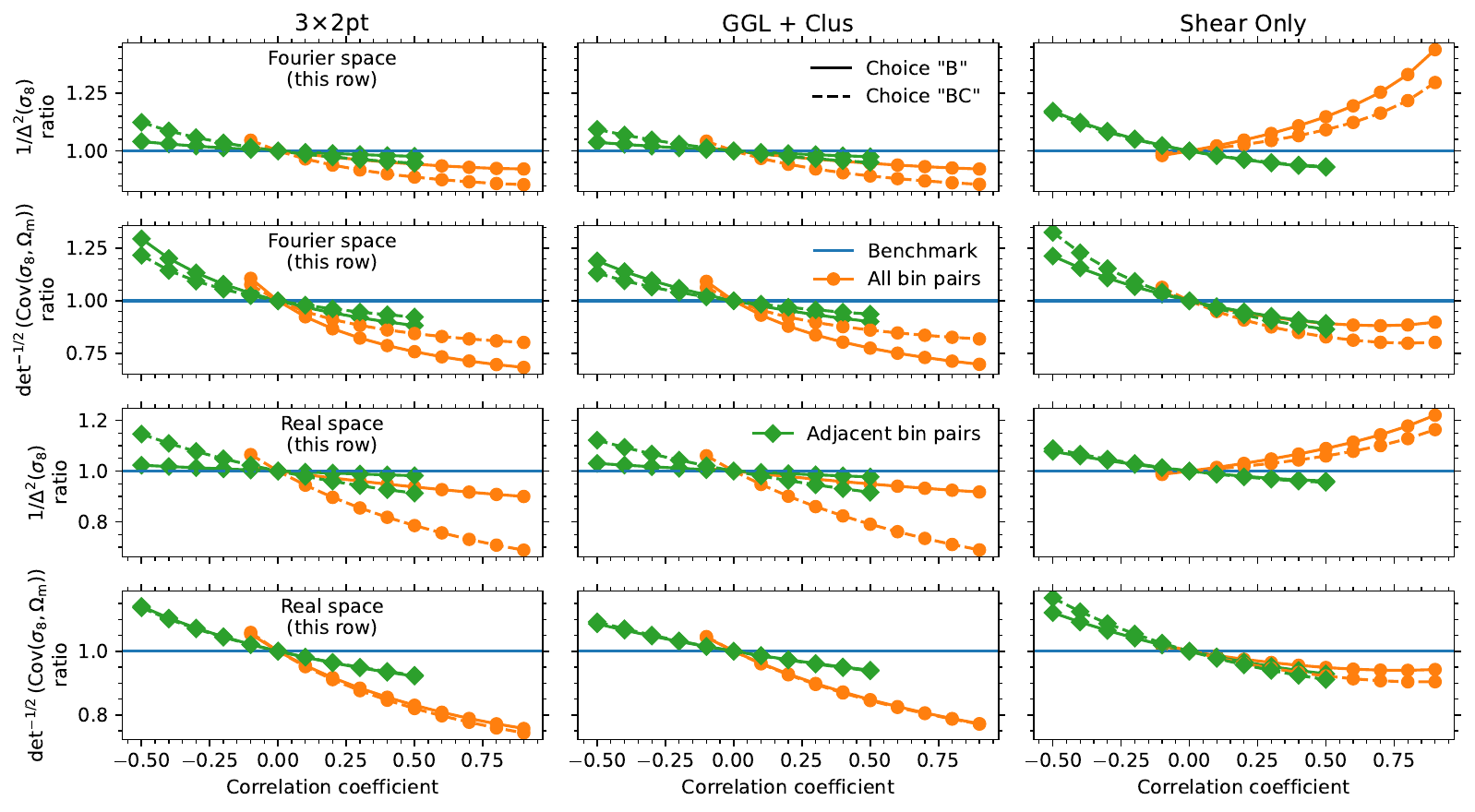}
    \caption{\label{fig:shear_corr}Impact of correlation of priors on shear biases. The structure follows that of Figure~\ref{fig:photoz_corr}, but for shear biases.}
\end{figure*}

Figure~\ref{fig:shear_corr} shows how the constraining power varies with correlated priors on shear biases. Again, the dynamic ranges of FoMs are larger than those for photo-$z$ biases (shown in Figure~\ref{fig:photoz_corr}). For $3\!\times\!2$pt (left column) and GGL+clustering (middle column), we see a consistent trend that FoMs decrease with larger $\rho$, especially when external cosmological knowledge is included (dashed curves). For extreme cases of correlation or anti-correlation, effects can be as large as $\sim 25\%$. Since shear biases are calibrated in similar fashion for all source redshift bins, positive correlations are plausible. In the uncorrelated case, errors in different tomographic bins tend to cancel, but with strong positive correlation the effective ``global'' uncertainty in shear bias is larger, reducing the FoM. Figure~\ref{fig:shear_corr} shows that shear calibration efforts need to characterize the correlation of bias uncertainties across redshift bins in addition to characterizing the magnitude of these uncertainties.

Surprisingly, for cosmic shear, a positive correlation increases the FoM with choice ``B,'' where all cosmological parameters are constrained from weak lensing alone. In this case, locking the shear bias parameters together apparently helps break their degeneracy with cosmological parameters. We note, for skeptical readers, that it is mathematically permissible for the constraining power to increase with correlations of either sign. However, taking advantage of this effect would require knowing the correlation accurately, and it goes away in any case when bringing in external cosmological priors (choice ``BC'') or GGL+clustering. We note that this analysis assumes that we can model GGL with linear galaxy bias, which may not be the case. We leave investigations of more sophisticated models of galaxy bias for future work.

\section{Conclusions} \label{sec:conclu}

In this paper, we have conducted Fisher forecasts for cosmological yields from $3\!\times\!2$pt (cosmic shear, galaxy-galaxy lensing, and galaxy clustering) analysis of the Roman High Latitude Imaging Survey (HLIS). Our Fisher analysis is based upon the Cobaya-CosmoLike Joint Architecture ({\sc CoCoA}; V. Miranda et al. 2026, in preparation) developed by our Cosmological Parameters Inference Pipeline (CPIP) Working Group. The model data vectors and covariance matrices are taken from the CPIP Data Challenge 1 (Section~\ref{ss:cocoa}). Instead of running Markov chain Monte Carlo (MCMC), we have performed maximum likelihood (ML) and maximum a posteriori (MAP) estimation of parameters (Section~\ref{ss:optim}) and used the Fisher formalism to estimate uncertainties (Section~\ref{ss:fisher}). As measures of cosmological performance, we have focused on two figures of merit, FoM1 $= 1/\Delta^2(\sigma_8)$ and FoM2 $= {\det}^{-1/2}({\rm Cov} (\sigma_8, \Omega_{\rm m}))$.

For the benchmark priors adopted in the data challenge, we find good agreement in cosmological performance between our Fisher information forecast and an MCMC forecast, as illustrated in Figures~\ref{fig:corsub_fourier} and \ref{fig:corsub_real}. In our baseline scenarios, we find FoM1 $= 1.83 \times 10^5$ (Fisher) vs. $1.65 \times 10^5$ (MCMC) and FoM2 $= 2.28 \times 10^5$ (Fisher) vs. $2.14 \times 10^5$ (MCMC) in Fourier space; FoM1 $= 6.16 \times 10^4$ (Fisher) vs. $5.86 \times 10^4$ (MCMC) and FoM2 $= 1.09 \times 10^5$ (Fisher) vs. $1.10 \times 10^5$ (MCMC) in real space. The shapes of MCMC contours for some parameter pairs differ from the elliptical form dictated by Fisher analysis, which may reflect the impact of bounded priors or non-Gaussian likelihood or both. We examine the accuracy of the Gaussian likelihood approximation further in Appendix~\ref{ss:logp}. We also note that our coordinate descent maximization did not initially find the correct ML and MAP solutions in real space because data vectors as a function of parameters are not always smooth, a cautionary lesson for analyses where the true answer is not known.

Our Fourier space analyses give consistently higher FoMs than our real space analyses, by factors of $\sim 2$ for the benchmark priors. We attribute this difference to a smaller effective scale cut in the Fourier space analyses (see Section~\ref{ss:scale}). Our analysis does not include any theoretical systematics associated with baryonic effects or non-linear bias, and it is possible that marginalizing over such systematics would remove the difference by reducing the cosmological information from small scales.

By exploiting the speed and flexibility of Fisher analysis, we have investigated many variants of the benchmark analysis. Our key findings are as follows:
\begin{itemize}
    \item Given the benchmark assumptions (including ``lens $=$ source,'' linear galaxy bias, and no marginalization over baryonic physics), the cosmological constraints from GGL+clustering are substantially stronger than those from cosmic shear, by factors of $\sim 5$--$20$ in FoM. Constraints from $3\!\times\!2$pt are similar to those from GGL+clustering, though slightly stronger. This dominance of GGL+clustering over cosmic shear holds for all of the prior combinations that we consider, and it is noticeably different from current weak lensing surveys where these two approaches are comparably powerful. (See Figure~\ref{fig:probe} and Table~\ref{tab:probe}.) Appendix~\ref{app:qtrlens} shows that reducing the lens density by a factor of $4$ lower the $3\!\times\!2$pt FoM by a factor of $\sim 2$, so maintaining the high lens density assumed in DC1 is highly desirable for cosmological analyses with the HLIS. XuDC1 examine the impact of scale cuts and baryonic physics on cosmic shear and $3\!\times\!2$pt constraints.
    \item Adding tight priors on $n_{\rm s}$, $\Omega_{\rm b}$, and $h_0$, which affect the shape of the matter power spectrum, substantially improves the constraints on $\Omega_{\rm m}$ and $\sigma_8$, by factors of $\sim 1.2$ (FoM1) or $\sim 2.3$ (FoM2) in the case of $3\!\times\!2$pt Fourier space analysis with benchmark priors on other nuisance parameters. (See Figure~\ref{fig:probe} and Table~\ref{tab:probe}, comparing choice ``BC'' and choice ``B.'') Since these parameters and the shape of the power spectrum can be constrained by CMB or other galaxy clustering observations, this improved performance may be a realistic expectation for joint analyses.
    \item With these tight $(n_{\rm s}, \Omega_{\rm b}, h_0)$ priors, going from the benchmark priors on photo-$z$ bias and multiplicative shear bias parameters and wide priors on intrinsic alignments parameters to infinitely tight priors gives further improvements of $\sim 1.5$ (FoM1) or $\sim 3.1$ (FoM2) (Fourier space, $3\!\times\!2$pt). This difference shows that gains in cosmological performance are possible if control of systematics can be tightened beyond the level represented in our benchmark priors. In all cases we maintain wide priors on galaxy bias. Note that our IA model is fairly restrictive, with only two free parameters, and a more flexible description might lead to greater degeneracy with cosmological parameters. (See Figure~\ref{fig:probe} and Table~\ref{tab:probe}.)
    \item The high-$z$ tomographic bins contain more information than the low-$z$ bins. Omitting just the highest redshift bin can reduce the FoM by $25$--$45\%$, though the impact is smaller when tight priors on $(n_{\rm s}, \Omega_{\rm b}, h_0)$ are incorporated. (See Figure~\ref{fig:tomo}.) The power in the highest redshift bins demonstrates the value of Roman's deep near-IR imaging and emphasizes the importance of maintaining systematics control at high redshift.
    \item As expected, FoMs are sensitive to the minimum scale considered in the analysis. Excluding the two smallest scale bins noticeably degrades the FoM (e.g., by $40\%$ for FoM2 from $3\!\times\!2$pt in Fourier space), while excluding the four largest scale bins has minimal impact. (See Figure~\ref{fig:scale}.) While scale cuts are frequently used to mitigate sensitivity to uncertain baryonic and non-linear effects, in the long term it is preferable to continue to small scales and marginalize over flexible models of these effects. Developing such models and testing them at the high accuracy needed for Roman analysis is a major challenge.
    \item Mitigating super-sample covariance can noticeably improve the FoM if it can lower the non-Gaussian contribution to the covariance matrix by at least a factor of two. The impact of SSC mitigation is larger for FoM2 than for FoM1, and it is larger for cosmic shear analysis than for $3\!\times\!2$pt. (See Figure~\ref{fig:ssc}.) SSC mitigation is less important than improving performance at small scales and high redshift, but it is worth pursuing.
    \item For $3\!\times\!2$pt analyses, sharpening or expanding the priors on photo-$z$ biases by a factor of $\sim 2$ changes the forecast FoM by $4$--$15\%$, relative to the value for our benchmark priors of $0.002$ (Fourier space) or $0.003$ (real space) in each tomographic bin. With $10 \times$ tighter photo-$z$ priors, which would be difficult to achieve, the FoM can improve by $7$--$24\%$. Photo-$z$ priors $10 \times$ worse than the benchmark would significantly degrade cosmological performance, reducing the FoM by $15$--$60\%$. (See Figure~\ref{fig:photoz_scale}.)
    \item Our baseline analysis assumes that the systematic uncertainty in photo-$z$ bias is uncorrelated from bin to bin, so the covariance matrix of the prior is diagonal. Allowing correlated systematics affects the FoM by $\lesssim 10\%$. (See Figure~\ref{fig:photoz_corr}.)
    \item For $3\!\times\!2$pt analyses, sharpening or expanding the priors on shear multiplicative bias by a factor of $\sim 2$ changes the forecast FoM by up to $\sim 20\%$, relative to the value for our benchmark priors of $0.005$ in each tomographic bin. Sharpening the priors by a factor of ten can improve the FoM by as much as $\sim 50\%$, with most of the improvement associated with the highest redshift bin. Expanding the priors by a factor of ten can degrade the FoM by $30$--$60\%$. Somewhat surprisingly, the fractional impact on the FoM is higher when we assume tight external priors on $(n_{\rm s}, \Omega_{\rm b}, h_0)$; the impact of shear bias uncertainty is lower when these parameters are inferred from the $3\!\times\!2$pt analysis. (See Figure~\ref{fig:shear_scale}.)
    \item The impact of correlated priors on shear biases is larger than for photo-$z$ biases, up to $\sim 25\%$ for maximal correlations, but the impact depends on the specifics of the analysis and the form of correlation assumed. (See Figure~\ref{fig:shear_corr}.)
\end{itemize}

There are many directions for future investigations of cosmological forecasting for Roman, in preparation for the much more exciting challenge of deriving cosmological results from the HLIS weak lensing and clustering measurements. These directions include extension to non-$\Lambda$CDM cosmologies, more sophisticated and flexible models of intrinsic alignments \citep{2019PhRvD.100j3506B}, theoretical models that extend to non-linear scales \citep{2012JCAP...07..051B, 2012JHEP...09..082C}, and additional observables such as cluster weak lensing \citep{2020MNRAS.491.3061S, 2021ApJ...910...28W} and higher order shear statistics \citep{2025PhRvD.112l3514G, 2025PhRvD.112l3515G}. Our results here show that Fisher information analysis is accurate enough to give useful insights, complementing MCMC studies with its speed and flexibility, and helping to focus effort where it is most valuable.

\section*{Acknowledgments}

This paper has undergone internal review in the Roman High Latitude Imaging Survey (HLIS) Cosmology Project Infrastructure Team (PIT). We would like to thank Eric Huff for helpful comments and feedback during the PIT internal review process, which led to substantial improvements of the paper. We also thank Charuhas Shiveshwarkar, Christopher M. Hirata, Chun-Hao To, and Junzhou Zhang for useful discussions regarding cosmology-dependent covariance and the ``lens $=$ source'' assumption. We thank the anonymous reviewer for insightful comments.

This work was supported by the ``Maximizing Cosmological Science with the Roman High Latitude Imaging Survey'' Roman Project Infrastructure Team (NASA grant 22-ROMAN11-0011).

Computations for this project used the Cardinal Cluster at the Ohio Supercomputer Center \citep{Ohio_Supercomputer_Center2024-dl}.

\software{CosmoLike \citep{2014MNRAS.440.1379E, 2017MNRAS.470.2100K, 2020JCAP...05..010F}, Cobaya \citep{2021JCAP...05..057T}; {\sc NumPy} \citep{2020Natur.585..357H}, {\sc SciPy} \citep{2020NatMe..17..261V}, {\sc pandas} \citep{2022zndo...3509134R}, {\sc Matplotlib} \citep{2007CSE.....9...90H}}

\section*{Data Availability}

The {\sc CoCoA} software\footnote{\url{https://github.com/CosmoLike/cocoa}} and the CPIP Data Challenge 1\footnote{\url{https://github.com/CosmoLike/roman_cpip_data_challenge}} are both publicly available on GitHub. Our code for modified coordinate descent and Fisher information analysis is publicly available in the following GitHub repository:

\url{https://github.com/Roman-HLIS-Cosmology-PIT/cpip-fisher-2025.git}

Release v0.1, a frozen version of the code, is available on Zenodo \citep{10.5281/zenodo.20044334} under an open-source Creative Commons Attribution license: \dataset[doi: 10.5281/zenodo.20044334]{https://doi.org/10.5281/zenodo.20044334}.

We caution the readers that the code for this project depends on a specific {\sc CoCoA} installation on a specific computing facility, and the code snippets need to be customized before reuse.

\appendix

\section{More on Data Challenge 1} \label{app:dc1}

\begin{figure}
    \plotone{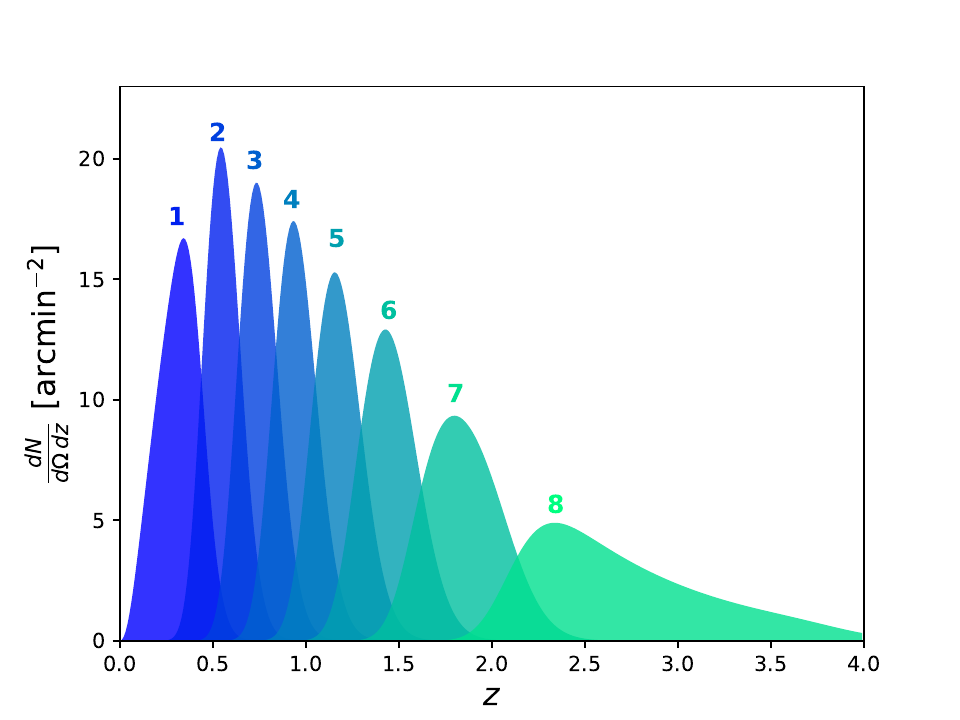}
    \caption{\label{fig:redshift_bins}Redshift distributions in $8$ tomographic bins assumed in DC1. Each bin has an equal number of galaxies, and the total surface density is $n_{\rm eff} = 41.3 \,{\rm arcmin}^{-2}$.}
\end{figure}

\begin{table*}[]
    \centering
    \begin{tabular}{c|ccccccccccccccc}
    \hline
        Bin & $1$* & $2$* & $3$ & $4$ & $5$ & $6$ & $7$ & $8$ & $9$ & $10$ & $11$ & $12$ & $13$ & $14$ & $15$ \\
    \hline
        $\theta_{\min} \,[{\rm arcmin}]$ & $2.500$ & $3.398$ & $4.620$ & $6.280$ & $8.536$ & $11.60$ & $15.77$ & $21.44$ & $29.15$ & $39.62$ & $53.86$ & $73.22$ & $99.53$ & $135.3$ & $183.9$ \\
        $\bar{\theta} \,[{\rm arcmin}]$ & $2.972$ & $4.040$ & $5.492$ & $7.465$ & $10.15$ & $13.79$ & $18.75$ & $25.49$ & $34.65$ & $47.10$ & $64.03$ & $87.04$ & $118.3$ & $160.8$ & $218.6$ \\
        $\ell (\bar{\theta})$ & $3634$ & $2673$ & $1967$ & $1447$ & $1064$ & $782.9$ & $575.9$ & $423.7$ & $311.7$ & $229.3$ & $168.7$ & $124.1$ & $91.28$ & $67.15$ & $49.40$ \\
    \hline
        $\bar{z} = 0.308$ & $0.239$ & $0.325$ & $0.441$ & $0.600$ & $0.815$ & $1.108$ & $1.506$ & $2.048$ & $2.783$ & $3.784$ & $5.143$ & $6.992$ & $9.504$ & $12.92$ & $17.56$ \\
        $\bar{z} = 0.548$ & $0.402$ & $0.546$ & $0.742$ & $1.009$ & $1.371$ & $1.864$ & $2.534$ & $3.444$ & $4.682$ & $6.364$ & $8.651$ & $11.76$ & $15.99$ & $21.73$ & $29.54$ \\
        $\bar{z} = 0.747$ & $0.521$ & $0.709$ & $0.963$ & $1.310$ & $1.780$ & $2.420$ & $3.290$ & $4.472$ & $6.079$ & $8.263$ & $11.23$ & $15.27$ & $20.76$ & $28.22$ & $38.36$ \\
        $\bar{z} = 0.952$ & $0.632$ & $0.859$ & $1.168$ & $1.588$ & $2.159$ & $2.935$ & $3.989$ & $5.423$ & $7.372$ & $10.02$ & $13.62$ & $18.52$ & $25.17$ & $34.22$ & $46.51$ \\
        $\bar{z} = 1.182$ & $0.743$ & $1.009$ & $1.372$ & $1.865$ & $2.535$ & $3.447$ & $4.685$ & $6.369$ & $8.657$ & $11.77$ & $16.00$ & $21.75$ & $29.56$ & $40.18$ & $54.62$ \\
        $\bar{z} = 1.463$ & $0.860$ & $1.170$ & $1.590$ & $2.161$ & $2.938$ & $3.994$ & $5.429$ & $7.379$ & $10.03$ & $13.64$ & $18.54$ & $25.20$ & $34.25$ & $46.56$ & $63.29$ \\
        $\bar{z} = 1.869$ & $1.004$ & $1.364$ & $1.855$ & $2.521$ & $3.427$ & $4.659$ & $6.333$ & $8.609$ & $11.70$ & $15.91$ & $21.62$ & $29.40$ & $39.96$ & $54.32$ & $73.84$ \\
        $\bar{z} = 2.720$ & $1.231$ & $1.673$ & $2.275$ & $3.092$ & $4.203$ & $5.714$ & $7.767$ & $10.56$ & $14.35$ & $19.51$ & $26.52$ & $36.05$ & $49.01$ & $66.62$ & $90.56$ \\
    \hline
    \end{tabular}
    \caption{\label{tab:scale_real}Scales in real space in DC1. The 2nd to 16th columns correspond to the $15$ angular scale bins, with the minimum $\theta$ values in the 2nd row, the central $\theta$ values in the 3rd row, and the $\ell = \pi / \theta$ values corresponding to central $\theta$ values in the 4th row. Bins $1$ and $2$ (labeled with asterisks), and sometimes bins $3$ and $4$ as well (typically at low redshifts), are masked out for GGL and galaxy clustering. The 5th to 12th rows list comoving angular diameter distances $d = D_{\rm C} ({\bar z}) \cdot {\bar \theta}$ (in $h^{-1} \,{\rm Mpc}$), which increase with increasing bin index (larger scales), at mean redshifts in $8$ tomographic bins.}
\end{table*}

\begin{table*}[]
    \centering
    \begin{tabular}{c|ccccccccccccccc}
    \hline
        Bin & $1$ & $2$ & $3$ & $4$ & $5$ & $6$ & $7$ & $8$ & $9$ & $10$ & $11$ & $12$ & $13$ & $14$ & $15$* \\
    \hline
        $\ell_{\max}$ & $41.57$ & $57.60$ & $79.82$ & $110.6$ & $153.3$ & $212.4$ & $294.3$ & $407.8$ & $565.0$ & $783.0$ & $1085$ & $1503$ & $2083$ & $2887$ & $4000$ \\
        $\bar{\ell}$ & $35.31$ & $48.93$ & $67.81$ & $93.96$ & $130.2$ & $180.4$ & $250.0$ & $346.4$ & $480.0$ & $665.1$ & $921.7$ & $1277$ & $1770$ & $2452$ & $3398$ \\
        $\theta (\bar{\ell}) \,[{\rm arcmin}]$ & $305.8$ & $220.7$ & $159.3$ & $114.9$ & $82.95$ & $59.86$ & $43.20$ & $31.18$ & $22.50$ & $16.24$ & $11.72$ & $8.456$ & $6.103$ & $4.404$ & $3.178$ \\
    \hline
        $\bar{z} = 0.308$ & $41.31$ & $57.25$ & $79.33$ & $109.9$ & $152.3$ & $211.1$ & $292.5$ & $405.3$ & $561.6$ & $778.2$ & $1078$ & $1494$ & $2070$ & $2869$ & $3975$ \\
        $\bar{z} = 0.548$ & $24.82$ & $34.39$ & $47.65$ & $66.03$ & $91.49$ & $126.8$ & $175.7$ & $243.4$ & $337.3$ & $467.4$ & $647.7$ & $897.5$ & $1244$ & $1723$ & $2388$ \\
        $\bar{z} = 0.747$ & $19.25$ & $26.68$ & $36.97$ & $51.22$ & $70.98$ & $98.36$ & $136.3$ & $188.9$ & $261.7$ & $362.6$ & $502.5$ & $696.3$ & $964.8$ & $1337$ & $1853$ \\
        $\bar{z} = 0.952$ & $15.98$ & $22.14$ & $30.68$ & $42.51$ & $58.91$ & $81.63$ & $113.1$ & $156.7$ & $217.2$ & $301.0$ & $417.0$ & $577.9$ & $800.7$ & $1110$ & $1538$ \\
        $\bar{z} = 1.182$ & $13.69$ & $18.97$ & $26.28$ & $36.42$ & $50.46$ & $69.92$ & $96.89$ & $134.3$ & $186.0$ & $257.8$ & $357.2$ & $495.0$ & $685.9$ & $950.5$ & $1317$ \\
        $\bar{z} = 1.463$ & $11.88$ & $16.47$ & $22.82$ & $31.62$ & $43.81$ & $60.70$ & $84.12$ & $116.6$ & $161.5$ & $223.8$ & $310.1$ & $429.7$ & $595.5$ & $825.1$ & $1143$ \\
        $\bar{z} = 1.869$ & $10.25$ & $14.21$ & $19.68$ & $27.28$ & $37.80$ & $52.37$ & $72.57$ & $100.6$ & $139.3$ & $193.1$ & $267.6$ & $370.7$ & $513.7$ & $711.9$ & $986.4$ \\
        $\bar{z} = 2.720$ & $8.432$ & $11.68$ & $16.19$ & $22.43$ & $31.09$ & $43.08$ & $59.69$ & $82.71$ & $114.6$ & $158.8$ & $220.1$ & $304.9$ & $422.5$ & $585.5$ & $811.3$ \\
    \hline
    \end{tabular}
    \caption{\label{tab:scale_fourier}Scales in Fourier space in DC1. The 2nd to 16th columns correspond to the $15$ angular scale bins, with the maximum $\ell$ values in the 2nd row, the central $\ell$ values in the 3rd row, and the $\theta = \pi / \ell$ values corresponding to central $\ell$ values in the 4th row. Bin $15$ (labeled with an asterisk) is masked out for GGL. The 5th to 12th rows present the comoving wavenumbers (in units of $h \,{\rm Gpc}^{-1}$), which increase with increasing bin index (smaller scales), at mean redshifts in $8$ tomographic bins. The boundary of the linear regime at low redshift is frequently taken to be $k_{\rm max} \sim (0.1$--$0.2) h \,{\rm Mpc}^{-1} = (100$--$200) h \,{\rm Gpc}^{-1}$, though in detail this boundary depends on the statistic under consideration and the level of accuracy required. Note that DC1 used different cosmological parameter values for generating Fourier space and real space data vectors (see Table~\ref{tab:param}), and we use the Fourier space parameters to compute $k$ values in this table.}
\end{table*}

This appendix supplements Section~\ref{ss:dc1} by providing some additional information about CPIP Data Challenge 1 (DC1). DC1 is designed for the Medium Tier ($2415 \,{\rm deg}^2$ in the Y106, J129, H158 bands) of the High Latitude Imaging Survey (HLIS); for further information about the survey design, see \citet{2025arXiv250510574O}.

Figure~\ref{fig:redshift_bins} presents the redshift distributions in $8$ tomographic bins assumed in DC1. The redshifts were generated by applying the Galaxy Survey Exposure Time Calculator,\footnote{\url{https://roman.gsfc.nasa.gov/science/etc14.html}} assuming $5 \times 140 \,{\rm s}$ exposures and the same signal-to-noise cuts as \citet{2021MNRAS.507.1746E}, to the CANDELS catalog \citep{2013ApJS..207...24G}. The resulting overall distribution was then split into $8$ bins of equal number of sources, and the distribution within each bin was convolved with a Gaussian uncertainty of width $0.05$. The assumed shape noise is $\sigma_\epsilon = 0.26$ per component, contributing $\sigma_\epsilon^2 / n_{\rm eff}$ to the shear covariance matrix \citep[e.g.,][]{2004PhRvD..70d3009H}. Tables~\ref{tab:scale_real} and \ref{tab:scale_fourier} tabulate angular scales in DC1, along with comoving wavenumbers at mean redshifts of $8$ tomographic bins. These tables supplement the discussions about angular scale cuts in Sections~\ref{ss:probe} and \ref{ss:scale}.

The following are some notes to complement the description of a ``pixels-to-cosmology'' pipeline in the main text. From an observational point of view, the distortions of galaxy shapes can be measured using shear measurement algorithms like {\sc Metacalibration} and {\sc Metadetection} \citep{2017arXiv170202600H, 2017ApJ...841...24S, 2020ApJ...902..138S}. Then these 2PCFs can be computed from a galaxy catalog via {\sc TreeCorr} \citep{2015ascl.soft08007J}; note that {\sc TreeCorr} is able to produce higher order statistics like three-point correlation functions \citep[3PCFs;][]{2024arXiv240701798S}, which are beyond the scope of this work. While measurement uncertainties and systematics are usually better understood in real space (also known as configuration space), predictions based on cosmological simulations \citep[e.g.,][]{2019ApJS..245...16H, 2019arXiv190608355H} are more straightforward in Fourier space (or more strictly speaking, harmonic space). Although mathematical transformations allow us to switch between real and Fourier spaces, the resulting cosmological parameters are not always consistent. While reconciliation techniques have been proposed \citep[e.g.,][]{2025MNRAS.540.1668P}, it is still worth studying both spaces separately.

\section{Extended Corner Plots} \label{app:corner}

\begin{figure*}
    \epsscale{1.2} \plotone{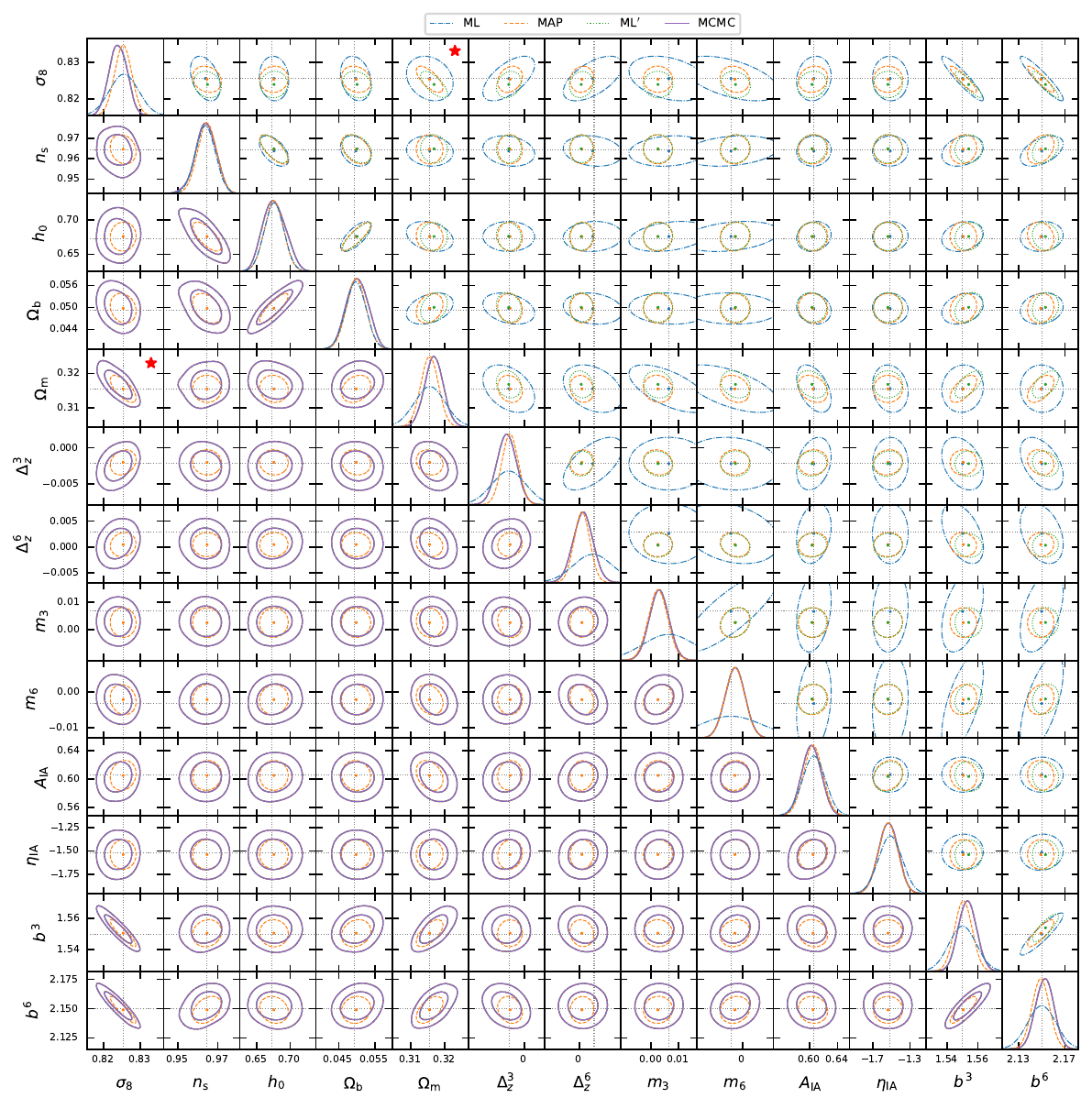}
    \caption{\label{fig:corner_fourier}Extended version of Figure~\ref{fig:corsub_fourier}. $6$ additional parameters are shown in this figure: photo-$z$ biases ($\Delta_{z}^3$ and $\Delta_{z}^6$) and galaxy biases ($b^3$ and $b^6$) in the $3^{\rm rd}$ and $6^{\rm th}$ tomographic bins, as well as the intrinsic alignments parameters ($A_{\rm IA}$ and $\eta_{\rm IA}$). Furthermore, MAP results extrapolated from ML results assuming Gaussianity of the parameter space (``ML$'$'') are shown in green.}
\end{figure*}

\begin{figure*}
    \epsscale{1.2} \plotone{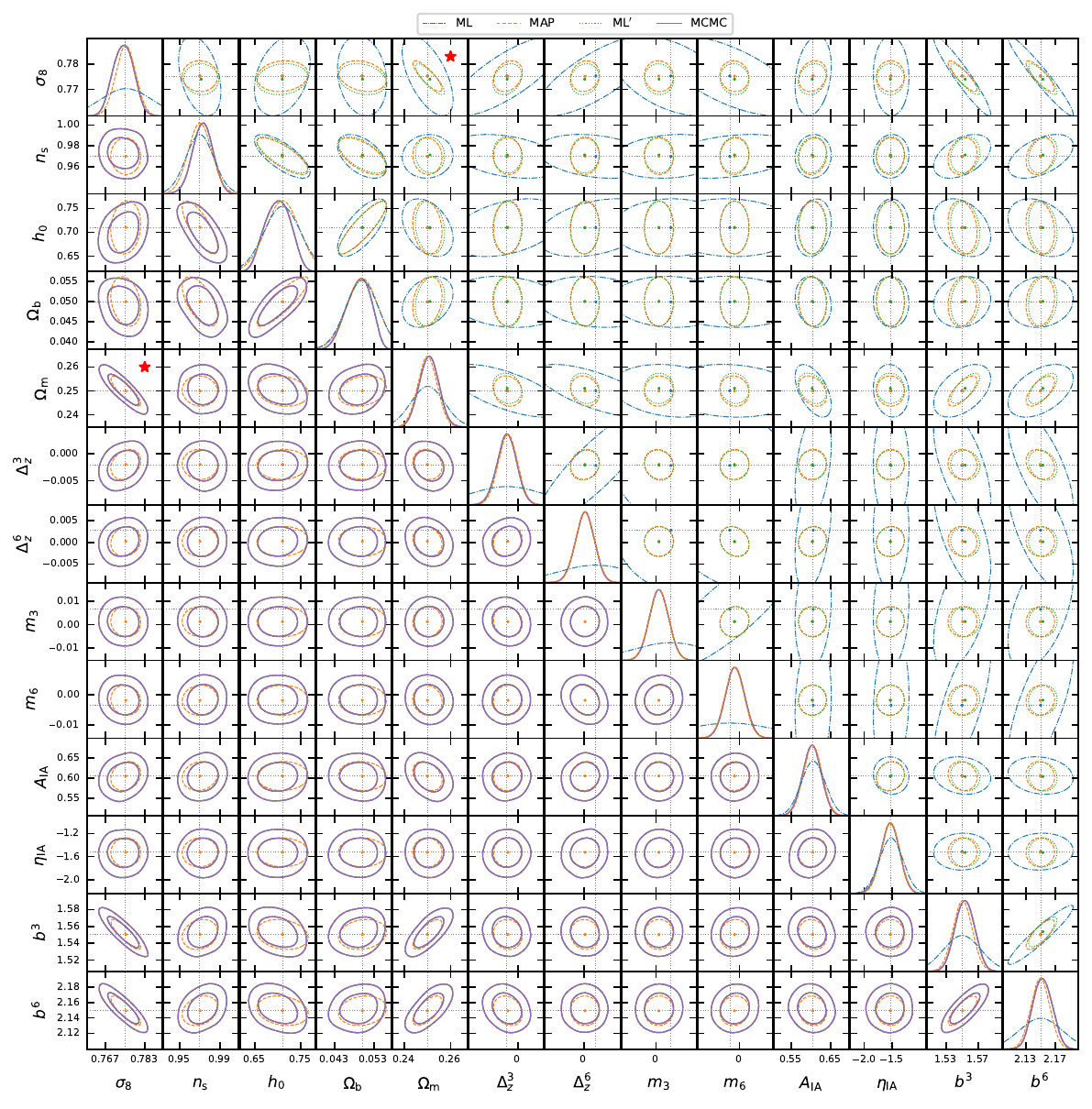}
    \caption{\label{fig:corner_real}Extended version of Figure~\ref{fig:corsub_real}. The additional parameters and green dots and ellipses are the same as those in Figure~\ref{fig:corner_fourier}.}
\end{figure*}

This appendix contains the extended versions of Figures~\ref{fig:corsub_fourier} and \ref{fig:corsub_real}, i.e., Figures~\ref{fig:corner_fourier} and \ref{fig:corner_real}. The extended corner plots support and supplement our observations in Section~\ref{sec:dcbase}, especially regarding $1\sigma$ credible regions in subspaces of nuisance parameters. In addition, these two figures include an additional set of results, shown in green and labeled ``ML$'$,'' which we now explain.

In addition to what is described in Section~\ref{ss:fisher}, there is an alternative way to incorporate the prior distribution. Given that the DC1 priors are Gaussian (see Equation~(\ref{eq:prior})), if we further assume that the likelihood ${\cal L} ({\boldsymbol \theta})$ is a multivariate Gaussian function, then the posterior probability ${\cal P} ({\boldsymbol \theta})$ is also Gaussian, and they are related in a simple way. The addition of precision matrices simply follows Equation~(\ref{eq:sigma_add}); as for the central values of the Gaussian functions, using Equation~(362) of \citet{petersen2008matrix}, we have
\begin{equation}
    {\boldsymbol {\hat \theta}}_{\cal P} = ({\bf \Sigma}_{\cal L}^{-1} + {\bf \Sigma}_{\rm Prior}^{-1})^{-1} ({\bf \Sigma}_{\cal L}^{-1} {\boldsymbol {\hat \theta}}_{\cal L} + {\bf \Sigma}_{\rm Prior}^{-1} {\boldsymbol {\hat \theta}}_{\rm Prior}),
    \label{eq:theta_add}
\end{equation}
where ${\boldsymbol {\hat \theta}}$ is the vector of parameters and ${\bf \Sigma}^{-1}$ denotes a precision matrix of parameters; the subscripts ${\cal P}$, ${\cal L}$, and ${\rm Prior}$ stand for maximum a posteriori, maximum likelihood, and prior information, respectively. Note that ${\bf \Sigma}_{\rm Prior}^{-1}$ comes from the prior probability and ${\boldsymbol {\hat \theta}}_{\rm Prior}$ is an all-zero vector in our case (the values for parameters with flat and wide priors do not matter). In Figures~\ref{fig:corner_fourier} and \ref{fig:corner_real}, the green dots and ellipses represent MAP parameter values ${\boldsymbol {\hat \theta}}_{\cal P}$ obtained via Equation~(\ref{eq:theta_add}) and the corresponding $1\sigma$ credible regions. Unlike for the orange ellipses, the ${\bf \Sigma}_{\cal L}^{-1}$ part of Equation~(\ref{eq:sigma_add}) is based on partial derivatives taken at the ML parameter values ${\boldsymbol {\hat \theta}}_{\cal L}$.

Despite the difference in where derivatives are evaluated, the agreement between the orange and green ellipses is remarkable: They often have the same shapes and orientations, indicating that the Gaussianity of the likelihood is a reasonable assumption. The discrepancies between the two sets of numerical derivatives are at the same level as the numerical uncertainties in those derivatives (${\cal O} (10^{-4})$). Therefore, we conclude that the differences between ML and MAP parameter values have almost no effect on the estimation of constraining power. This justifies our choice of using the same sets of numerical derivatives in Sections~\ref{sec:dissect} and \ref{sec:prior}. The agreement between directly found MAP parameter values ${\boldsymbol {\hat \theta}}_{\cal P}$ (see Section~\ref{ss:optim}) and those from Equation~(\ref{eq:theta_add}) is good for photo-$z$ and shear biases but not as good for cosmological and other ``nuisance'' parameters. Arguably, the level of agreement or disagreement can be viewed as a measure of Gaussianity: For photo-$z$ and shear biases, the Gaussian priors of DC1 dominate, thus the agreement is good; for other parameters, the posterior distributions are not necessarily close to Gaussian, thus the disagreement is significant. Therefore, the usage of Equation~(\ref{eq:theta_add}) is limited.

\begin{figure*}
    \plotone{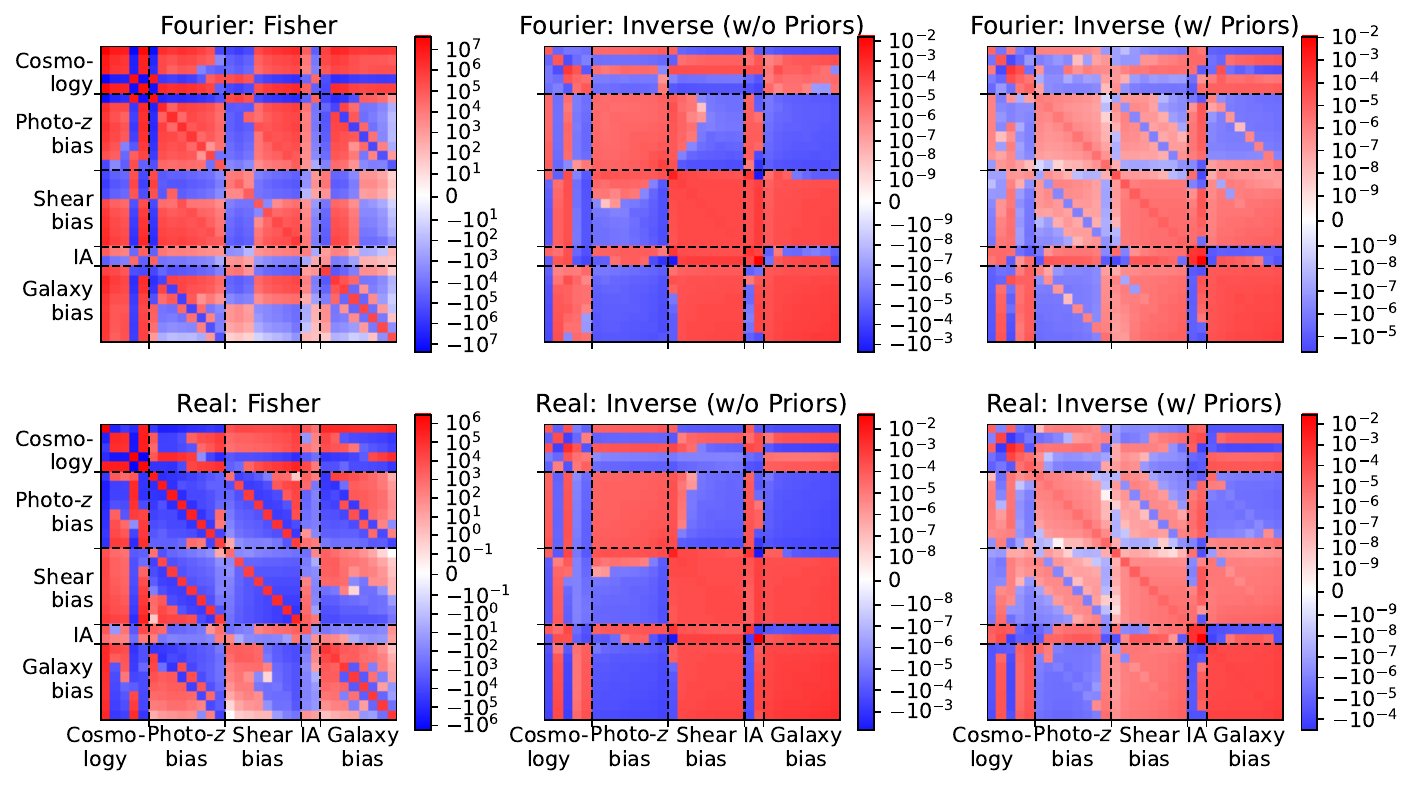}
    \caption{\label{fig:fisher}Fisher matrices and their inverses. The upper (lower) row corresponds to Fourier (real) space. In each row, the first panel shows the Fisher matrix, the second panel shows its inverse, and the third panel shows the inverse of the sum of the Fisher matrix and the precision matrix corresponding to the DC1 priors (on photo-$z$ bias and shear bias). The boundaries between different groups of parameters are marked with black dashed lines. The ordering of cosmological parameters in this figure is: $\sigma_8$, $n_{\rm s}$, $h_0$, $\Omega_{\rm b}$, and $\Omega_{\rm m}$.}
\end{figure*}

Figure~\ref{fig:fisher} presents the Fisher matrices from $3\!\times\!2$pt analysis in both Fourier and real spaces (${\bf \Sigma}_{\cal L}^{-1}$; left column) and their inverses without (${\bf \Sigma}_{\cal L}$; middle column) and with (${\bf \Sigma}_{\cal P}$; right column) default DC1 priors. The structure of these matrices follows the organization of the vector of parameters ${\boldsymbol \theta}$: cosmological parameters, and then different types of ``nuisance'' parameters. The constraining power on cosmological parameters is presented in the cosmology block. As for the ``nuisance'' parameters, by comparing the middle and right columns, we see an obvious difference: In ${\bf \Sigma}_{\cal P}$, the diagonal terms of the photo-$z$ bias or shear bias block are much larger than the off-diagonal terms, but in ${\bf \Sigma}_{\cal L}$ diagonal and off-diagonal terms are comparable. It is clear that ``self-calibration'' of bias parameters from the observational data alone leads to correlated uncertainties in their values, but superimposing uncorrelated priors significantly reduces correlation between error bars.

Inverting a Fisher matrix --- or more strictly speaking, computing Schur complements --- amounts to marginalizing over involved parameters. Applying infinitely narrow priors, i.e., assuming that some of the parameters are perfectly known, is implemented by removing the corresponding rows and columns from the Fisher matrix before inverting it. To include correlated priors, we make $n_{\rm tomo} \times n_{\rm tomo}$ covariance matrices for the bias parameters being studied, invert them to obtain $n_{\rm tomo} \times n_{\rm tomo}$ precision matrices, and put them in the right places of the full precision matrix ${\bf \Sigma}_{\rm Prior}^{-1}$, which is then superimposed to the Fisher matrix ${\bf \Sigma}_{\cal L}^{-1}$. This implementation is adopted throughout Section~\ref{sec:prior}.

\begin{table*}[]
    \centering
    \begin{tabular}{l|lll|lll}
    \hline
        Parameter & ML (Fourier) & MAP (Fourier) & MCMC (Fourier) & ML (real) & MAP (real) & MCMC (real) \\
    \hline
        \multicolumn{7}{l}{\bf Cosmology} \\
        $\Omega_{\rm m} \!=\! 0.3156, 0.250$ & $0.3156 \!\pm\! 0.0045$ & $0.3156 \!\pm\! 0.0026$ & $0.3167^{+0.0030}_{-0.0029}$ & $0.2500 \!\pm\! 0.0073$ & $0.2499 \!\pm\! 0.0042$ & $0.2505 \!\pm\! 0.0042$ \\
        $\sigma_8 \!=\! 0.8255, 0.775$ & $0.8256 \!\pm\! 0.0040$ & $0.8254 \!\pm\! 0.0023$ & $0.8241 \!\pm\! 0.0025$ & $0.7754 \!\pm\! 0.0104$ & $0.7753 \!\pm\! 0.0040$ & $0.7744 \!\pm\! 0.0041$ \\
        $n_{\rm s} \!=\! 0.9645, 0.970$ & $0.9638 \!\pm\! 0.0049$ & $0.9647 \!\pm\! 0.0048$ & $0.9654^{+0.0056}_{-0.0059}$ & $0.9700 \!\pm\! 0.0139$ & $0.9701 \!\pm\! 0.0116$ & $0.9730 \!\pm\! 0.0099$ \\
        $\Omega_{\rm b} \!=\! 0.0492, 0.050$ & $0.0497 \!\pm\! 0.0028$ & $0.0498 \!\pm\! 0.0027$ & $0.0494^{+0.0032}_{-0.0027}$ & $0.0499 \!\pm\! 0.0042$ & $0.0500 \!\pm\! 0.0040$ & $0.0491^{+0.0039}_{-0.0028}$ \\
        $h_0 \!=\! 0.6727, 0.710$ & $0.6761 \!\pm\! 0.0145$ & $0.6759 \!\pm\! 0.0141$ & $0.672^{+0.018}_{-0.017}$ & $0.7096 \!\pm\! 0.0399$ & $0.7101 \!\pm\! 0.0367$ & $0.0700 \!\pm\! 0.027$ \\
    \hline
        \multicolumn{7}{l}{\bf Galaxy bias} \\
        $b^1 \!=\! 1.18$ & $1.1796 \!\pm\! 0.0048$ & $1.1782 \!\pm\! 0.0041$ & $1.1805^{+0.0042}_{-0.0043}$ & $1.1800 \!\pm\! 0.0180$ & $1.1772 \!\pm\! 0.0123$ & $1.179 \!\pm\! 0.012$ \\
        $b^2 \!=\! 1.40$ & $1.3998 \!\pm\! 0.0077$ & $1.3977 \!\pm\! 0.0053$ & $1.40010 \!\pm\! 0.0053$ & $1.3999 \!\pm\! 0.0211$ & $1.3976 \!\pm\! 0.0119$ & $1.400 \!\pm\! 0.012$ \\
        $b^3 \!=\! 1.55$ & $1.5500 \!\pm\! 0.0089$ & $1.5505 \!\pm\! 0.0058$ & $1.5529 \!\pm\! 0.0061$ & $1.5499 \!\pm\! 0.0232$ & $1.5504 \!\pm\! 0.0118$ & $1.553 \!\pm\! 0.012$ \\
        $b^4 \!=\! 1.71$ & $1.7096 \!\pm\! 0.0100$ & $1.7092 \!\pm\! 0.0065$ & $1.7124 \!\pm\! 0.0068$ & $1.7099 \!\pm\! 0.0247$ & $1.7093 \!\pm\! 0.0119$ & $1.712 \!\pm\! 0.012$ \\
        $b^5 \!=\! 1.90$ & $1.8996 \!\pm\! 0.0112$ & $1.8981 \!\pm\! 0.0071$ & $1.9023^{+0.0074}_{-0.0075}$ & $1.8999 \!\pm\! 0.0265$ & $1.8989 \!\pm\! 0.0121$ & $1.901 \!\pm\! 0.012$ \\
        $b^6 \!=\! 2.15$ & $2.1499 \!\pm\! 0.0123$ & $2.1490 \!\pm\! 0.0076$ & $2.1525 \!\pm\! 0.0081$ & $2.1498 \!\pm\! 0.0291$ & $2.1493 \!\pm\! 0.0127$ & $2.152 \!\pm\! 0.013$ \\
        $b^7 \!=\! 2.52$ & $2.5197 \!\pm\! 0.0136$ & $2.5198 \!\pm\! 0.0082$ & $2.5233^{+0.0088}_{-0.0091}$ & $2.5198 \!\pm\! 0.0331$ & $2.5197 \!\pm\! 0.0141$ & $2.522 \!\pm\! 0.014$ \\
        $b^8 \!=\! 3.44$ & $3.4430 \!\pm\! 0.0168$ & $3.4454 \!\pm\! 0.0101$ & $3.4487 \!\pm\! 0.011$ & $3.4403 \!\pm\! 0.0435$ & $3.4412 \!\pm\! 0.0186$ & $3.444 \!\pm\! 0.019$ \\
    \hline
        \multicolumn{7}{l}{\bf Photo-$z$} \\
        $\Delta_{z}^1 \!=\! 0.001414$ & $0.0014 \!\pm\! 0.0025$ & $0.0003 \!\pm\! 0.0011$ & $0.0003 \!\pm\! 0.0013$ & $0.0014 \!\pm\! 0.0062$ & $0.0002 \!\pm\! 0.0020$ & $0.0001 \!\pm\! 0.0020$ \\
        $\Delta_{z}^2 \!=\! 0.004298$ & $0.0043 \!\pm\! 0.0023$ & $0.0028 \!\pm\! 0.0011$ & $0.0031 \!\pm\! 0.0013$ & $0.0043 \!\pm\! 0.0068$ & $0.0025 \!\pm\! 0.0019$ & $0.0024 \!\pm\! 0.0018$ \\
        $\Delta_{z}^3 \!=\! -0.002162$ & $-0.0022 \!\pm\! 0.0024$ & $-0.0020 \!\pm\! 0.0012$ & $-0.0025 \!\pm\! 0.0014$ & $-0.0022 \!\pm\! 0.0071$ & $-0.0020 \!\pm\! 0.0018$ & $-0.0022 \!\pm\! 0.0018$ \\
        $\Delta_{z}^4 \!=\! 0.000047$ & $-0.0000 \!\pm\! 0.0028$ & $-0.0009 \!\pm\! 0.0013$ & $-0.0013 \!\pm\! 0.0015$ & $0.0000 \!\pm\! 0.0078$ & $-0.0011 \!\pm\! 0.0019$ & $-0.0013 \!\pm\! 0.0019$ \\
        $\Delta_{z}^5 \!=\! 0.003450$ & $0.0034 \!\pm\! 0.0032$ & $0.0013 \!\pm\! 0.0014$ & $0.0015 \!\pm\! 0.0017$ & $0.0034 \!\pm\! 0.0085$ & $0.0012 \!\pm\! 0.0021$ & $0.0011 \!\pm\! 0.0021$ \\
        $\Delta_{z}^6 \!=\! 0.002860$ & $0.0026 \!\pm\! 0.0038$ & $0.0005 \!\pm\! 0.0015$ & $0.0007 \!\pm\! 0.0019$ & $0.0028 \!\pm\! 0.0092$ & $0.0002 \!\pm\! 0.0023$ & $0.0002 \!\pm\! 0.0023$ \\
        $\Delta_{z}^7 \!=\! 0.002578$ & $0.0021 \!\pm\! 0.0052$ & $0.0004 \!\pm\! 0.0017$ & $0.0007 \!\pm\! 0.0022$ & $0.0026 \!\pm\! 0.0108$ & $0.0004 \!\pm\! 0.0025$ & $0.0004 \!\pm\! 0.0024$ \\
        $\Delta_{z}^8 \!=\! -0.001002$ & $-0.0026 \!\pm\! 0.0102$ & $-0.0003 \!\pm\! 0.0019$ & $-0.0005 \!\pm\! 0.0028$ & $-0.0010 \!\pm\! 0.0173$ & $-0.0003 \!\pm\! 0.0029$ & $-0.0002 \!\pm\! 0.0029$ \\
    \hline
        \multicolumn{7}{l}{\bf Shear calibration} \\
        $m_1 \!=\! 0.00203$ & $0.0014 \!\pm\! 0.0379$ & $-0.0002 \!\pm\! 0.0050$ & $-0.0001 \!\pm\! 0.0049$ & $0.0019 \!\pm\! 0.0651$ & $-0.0001 \!\pm\! 0.0050$ & $-0.0001 \!\pm\! 0.0050$ \\
        $m_2 \!=\! 0.00114$ & $0.0013 \!\pm\! 0.0110$ & $0.0006 \!\pm\! 0.0040$ & $0.0003 \!\pm\! 0.0040$ & $0.0015 \!\pm\! 0.0197$ & $0.0003 \!\pm\! 0.0046$ & $0.0003 \!\pm\! 0.0046$ \\
        $m_3 \!=\! 0.00660$ & $0.0065 \!\pm\! 0.0097$ & $0.0025 \!\pm\! 0.0036$ & $0.0029 \!\pm\! 0.0036$ & $0.0066 \!\pm\! 0.0170$ & $0.0013 \!\pm\! 0.0042$ & $0.0014 \!\pm\! 0.0042$ \\
        $m_4 \!=\! -0.00774$ & $-0.0078 \!\pm\! 0.0090$ & $-0.0058 \!\pm\! 0.0032$ & $-0.0055 \!\pm\! 0.0032$ & $-0.0078 \!\pm\! 0.0158$ & $-0.0038 \!\pm\! 0.0039$ & $-0.0037 \!\pm\! 0.0038$ \\
        $m_5 \!=\! -0.00101$ & $-0.0011 \!\pm\! 0.0090$ & $-0.0002 \!\pm\! 0.0030$ & $-0.0002^{+0.0029}_{-0.0030}$ & $-0.0011 \!\pm\! 0.0156$ & $0.0002 \!\pm\! 0.0035$ & $0.0003 \!\pm\! 0.0035$ \\
        $m_6 \!=\! -0.00328$ & $-0.0032 \!\pm\! 0.0089$ & $-0.0021 \!\pm\! 0.0028$ & $-0.0021 \!\pm\! 0.0028$ & $-0.0034 \!\pm\! 0.0154$ & $-0.0018 \!\pm\! 0.0033$ & $-0.0016 \!\pm\! 0.0033$ \\
        $m_7 \!=\! 0.00097$ & $0.0010 \!\pm\! 0.0090$ & $0.0012 \!\pm\! 0.0026$ & $0.0014 \!\pm\! 0.0026$ & $0.0009 \!\pm\! 0.0156$ & $0.0010 \!\pm\! 0.0031$ & $0.0012 \!\pm\! 0.0031$ \\
        $m_8 \!=\! 0.00278$ & $0.0030 \!\pm\! 0.0086$ & $0.0023 \!\pm\! 0.0025$ & $0.0027 \!\pm\! 0.0026$ & $0.0027 \!\pm\! 0.0153$ & $0.0021 \!\pm\! 0.0030$ & $0.0024 \!\pm\! 0.0030$ \\
    \hline
        \multicolumn{7}{l}{{\bf IA} (NLA)} \\
        $A_{\rm IA} \!=\! 0.6061$ & $0.6061 \!\pm\! 0.0164$ & $0.6049 \!\pm\! 0.0140$ & $0.603 \!\pm\! 0.014$ & $0.6062 \!\pm\! 0.0304$ & $0.6050 \!\pm\! 0.0236$ & $0.603 \!\pm\! 0.023$ \\
        $\eta_{\rm IA} \!=\! -1.515$ & $-1.5133 \!\pm\! 0.1275$ & $-1.5375 \!\pm\! 0.1044$ & $-1.54^{+0.11}_{-0.10}$ & $-1.5133 \!\pm\! 0.2119$ & $-1.5258 \!\pm\! 0.1670$ & $-1.53^{+0.17}_{-0.16}$ \\
    \hline
    \end{tabular}
    \caption{\label{tab:param}Parameter values and 1D marginalized error bars. The first column presents the truth parameter values used to generate DC1 baseline data vectors. The cosmological parameters are different in Fourier and real spaces, but the nuisance parameters are the same. The second and fifth columns show the maximum likelihood (ML) results, the third and sixth columns show the maximum a posteriori (MAP) results, and the fourth and seventh columns show the Markov chain Monte Carlo (MCMC) results.}
\end{table*}

Table~\ref{tab:param} tabulates parameter values and 1D marginalized error bars in DC1.

\section{Mathematical Remarks} \label{app:math}

In this appendix, we make three mathematical remarks related to our Fisher information analysis in this work.

\subsection{Inversion of Covariance Matrices} \label{ss:invert}

\begin{figure*}
    \plotone{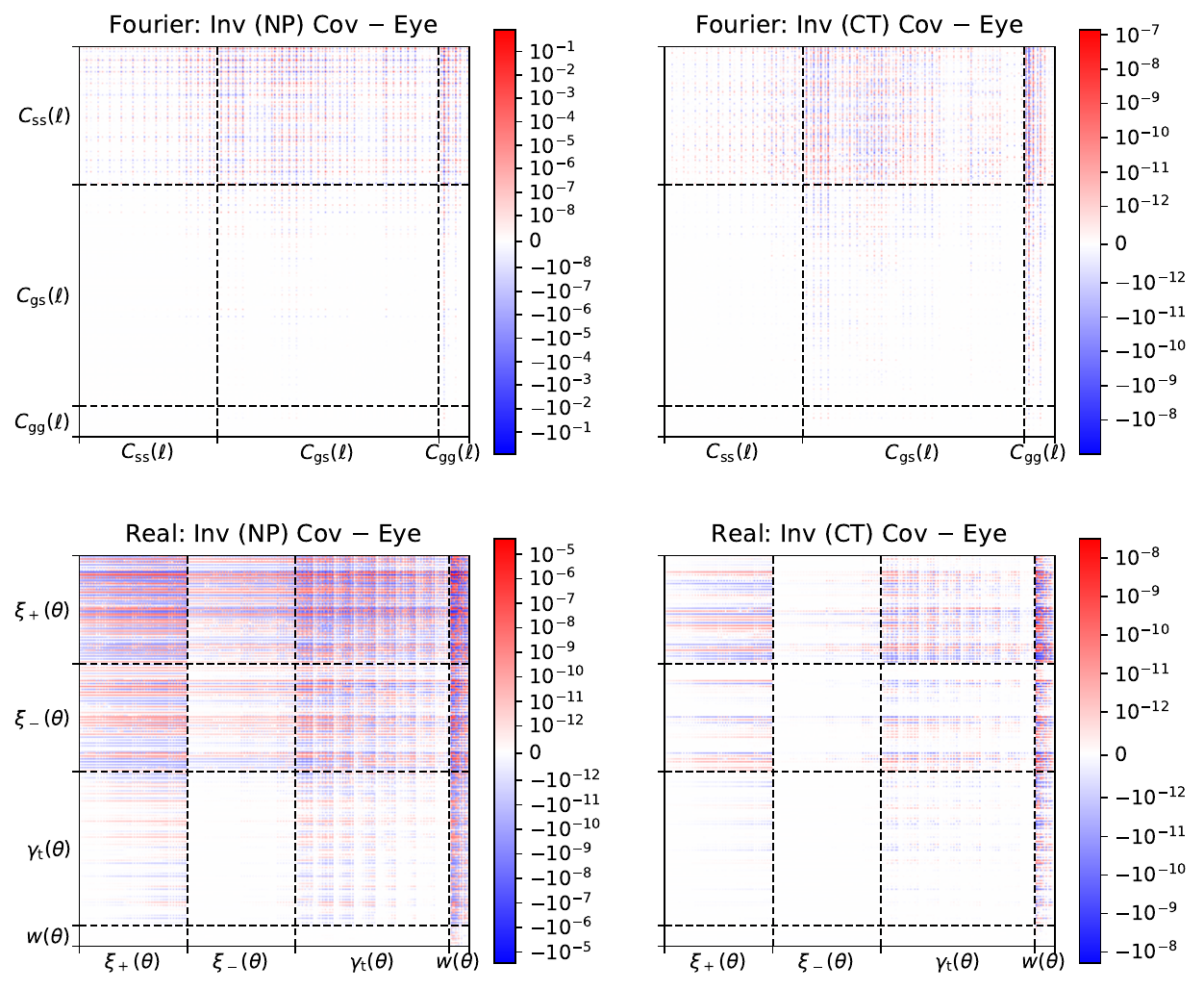}
    \caption{\label{fig:invert}Inversion of covariance matrices. Each panel shows the difference between the product of an inverse matrix and the covariance matrix and the expected identity matrix, which should be zero by definition. The left column naively uses the NumPy (``NP'') routine ({\tt numpy.linalg.inv}), while the right column use the ``correlation trick'' (``CT''). The upper (lower) row presents test results in real (Fourier) space. Like in Figures~\ref{fig:covinv_fourier} and \ref{fig:covinv_real}, a symmetric logarithmic scale is used to better present the structure of the matrix, and boundaries between different segments of the data vector (see Section~\ref{ss:cocoa}) are marked with black dashed lines. Note that the color bar scales are different for different panels.}
\end{figure*}

There are many mature routines for inverting matrices. However, covariance matrices for a large comological data vector are often ill-conditioned --- with a large ratio of maximum and minimum absolute eigenvalues --- and naively applying a regular routine may lead to unsatisfactory results. There is a standard and simple trick to address this issue. Instead of directly inverting the $n \times n$ covariance matrix ${\bf C}$, we compute the corresponding correlation matrix ${\bf \Lambda} {\bf C} {\bf \Lambda}$, where
\begin{equation}
    {\bf \Lambda} \equiv \begin{pmatrix}
    1/\sqrt{C_{11}} & 0 & \cdots & 0 \\
    0 & 1/\sqrt{C_{22}} & \cdots & 0 \\
    \vdots & \vdots & \ddots & \vdots \\
    0 & 0 & \cdots & 1/\sqrt{C_{nn}} \\
    \end{pmatrix},
    \label{eq:invert}
\end{equation}
which has elements that span a much smaller dynamic range and is typically much better conditioned. Then we can use the mathematical identity ${\bf C}^{-1} = {\bf \Lambda} ({\bf \Lambda} {\bf C} {\bf \Lambda})^{-1} {\bf \Lambda}$ to compute the inverse of ${\bf C}$. Despite its simplicity, this method performs very well on our covariance matrices, as shown in Figure~\ref{fig:invert}. Therefore, all inverse covariance matrices involved in this work are computed in this way.

\subsection{On Signal-to-Noise Ratio} \label{ss:snr2}

As mentioned in Section~\ref{sec:dissect}, the signal-to-noise ratio (squared) is a model-independent figure of merit. For the data vector ${\boldsymbol d}$ and the covariance matrix ${\bf C}$, it is simply defined as
\begin{equation}
    {\rm SNR}^2 = {\boldsymbol d}^{\rm T} {\bf C}^{-1} {\boldsymbol d}.
\end{equation}
We choose not to include it in the text, because we have found that its relationship with figures of merit for the cosmological constraining power is ambiguous. This is understandable, as the two FoMs used in Section~\ref{sec:dissect} are based on partial derivatives of the data vector, not the data vector per se. Furthermore, in the context of a $3\!\times\!2$pt analysis, the signal-to-noise is usually dominated by galaxy clustering, making it even less indicative of the whole picture. ${\rm SNR}^2$ may be a more useful metric for cosmic shear alone, but cosmic shear makes a subdominant contribution to our FoMs.

\subsection{Conversion from $A_{\rm s}$ to $\sigma_8$} \label{ss:convert}

{\sc CoCoA} takes $A_{\rm s}$ as an input parameter and yields $\sigma_8$ as an output value. For convenience, we compute partial derivatives with respect to $A_{\rm s}$ or with $A_{\rm s}$ kept fixed, and then convert them to those with respect to $\sigma_8$ or with $\sigma_8$. Such conversion is done using the mathematical equalities (parameters to the right of the vertical line are kept fixed)
\begin{equation}
    \left. \frac{\partial {\boldsymbol m}}{\partial \sigma_8} \right|_{\boldsymbol \theta'} = \left. \frac{\partial {\boldsymbol m}}{\partial A_{\rm s}} \right|_{\boldsymbol \theta'} \left( \left. \frac{\partial \sigma_8}{\partial A_{\rm s}} \right|_{\boldsymbol \theta'} \right)^{-1},
    \label{eq:As2sigma8_1}
\end{equation}
where ${\boldsymbol \theta'}$ denotes the collection of parameters other than $A_{\rm s}$ or $\sigma_8$, and
\begin{equation}
    \left. \frac{\partial {\boldsymbol m}}{\partial \theta_\alpha} \right|_{\sigma_8, {\boldsymbol \theta''}} = \left. \frac{\partial {\boldsymbol m}}{\partial \theta_\alpha} \right|_{A_{\rm s}, {\boldsymbol \theta''}} - \left. \frac{\partial {\boldsymbol m}}{\partial \sigma_8} \right|_{\boldsymbol \theta'} \left. \frac{\partial \sigma_8}{\partial \theta_\alpha} \right|_{A_{\rm s}, {\boldsymbol \theta''}},
    \label{eq:As2sigma8_2}
\end{equation}
where ${\boldsymbol \theta''}$ denotes the collection of parameters other than $A_{\rm s}$ or $\sigma_8$ and $\theta_\alpha$. Since $\sigma_8$ does not depend on nuisance parameters, Equation~(\ref{eq:As2sigma8_2}) only needs to be applied to cosmological parameters (other than $A_{\rm s}$).

\subsection{Validating Fisher Calculations} \label{ss:logp}

\begin{figure*}
    \plotone{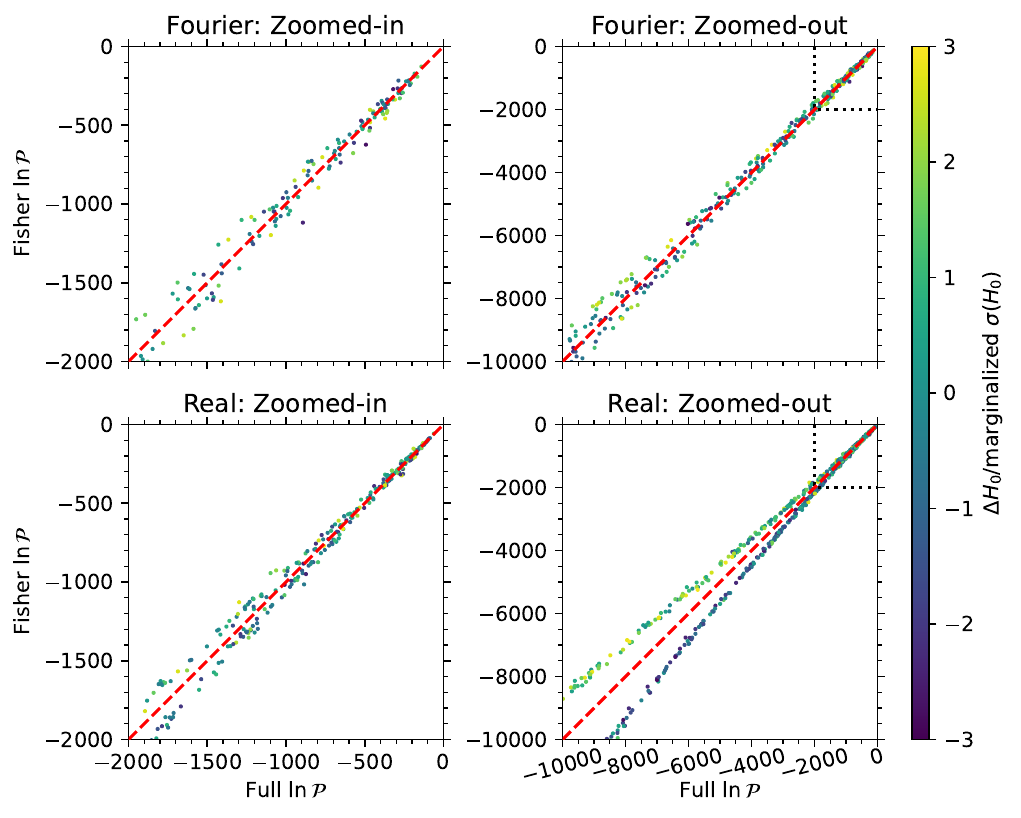}
    \caption{\label{fig:logp_check}Validating logarithmic posterior probabilities computed from Fisher analysis ($y$-axes) against full calculations ($x$-axes). Each data point corresponds to a random realization of the uniform distribution between ${\hat \theta}_\alpha \pm 3\sigma_\alpha$, where ${\hat \theta}_\alpha$ is the MAP parameter value and $\sigma_\alpha$ is the 1D marginalized error bar. The two columns present results in Fourier and real spaces, respectively. The left column shows the $\ln {\cal P}$ range of interest for $1\sigma$ marginalized errors in 1D or 2D subspaces, while the right column shows a larger range where deviations due to non-Gaussianity are more significant. In each panel, the data points are color-coded by the discrepancy between the random and MAP $H_0$ values, and perfect agreement is shown as a diagonal red dashed line.}
\end{figure*}

Figure~\ref{fig:logp_check} demonstrates the good agreement between Fisher and full calculations in terms of $\ln {\cal P}$ values. We create random realizations of parameters ${\boldsymbol \theta}$ by drawing a value of each parameter $\theta_\alpha$ from a uniform distribution between ${\hat \theta}_\alpha \pm 3\sigma_\alpha$, where ${\hat \theta}_\alpha$ is the MAP parameter value and $\sigma_\alpha$ is the 1D marginalized error bar. We then compute $\ln {\cal P} ({\boldsymbol \theta})$ from a full {\sc CoCoA} calculation and from the Gaussian approximation implicit in Fisher analysis,
\begin{equation}
    \ln {\cal P}_{\rm Fisher} ({\boldsymbol \theta}) = \ln {\cal P} ({\boldsymbol {\hat \theta}}_{\cal P}) - \frac12 ({\boldsymbol \theta} - {\boldsymbol \theta}_{\cal P})^{\rm T} {\bf \Sigma}_{\cal P}^{-1} ({\boldsymbol \theta} - {\boldsymbol \theta}_{\cal P}).
    \label{eq:logp_fisher}
\end{equation}
The first element of ${\boldsymbol \theta}$ is $A_{\rm s}$ for full calculations and $\sigma_8$ for Fisher calculations; the other elements are the same.

For $31$ parameters, a $99.7\%$ confidence interval corresponds to $\Delta \ln {\cal P} \equiv \ln {\cal P} ({\boldsymbol \theta}) - \ln {\cal P} ({\boldsymbol {\hat \theta}}_{\cal P}) \simeq -28.7$. Thus, we expect that the MCMC contours shown in Figures~\ref{fig:corsub_fourier} and ~\ref{fig:corsub_real} are 2D projections of points with $\Delta \ln {\cal P} \gtrsim -30$. However, varying a single parameter {\it in isolation} by its {\it marginalized} $1\sigma$ uncertainty can produce a $|\Delta \ln {\cal P}|$ that is much larger because other parameters are not varied to compensate. (If all other parameters were perfectly known, the $1\sigma$ uncertainty would be much smaller.) In our case, because the baseline DC1 data vector is noiseless, $\ln {\cal P} ({\boldsymbol {\hat \theta}}_{\cal P}) \approx 0$ ($-4.59$ in Fourier space and $-2.19$ in real space), and $\Delta \ln {\cal P} \approx \ln {\cal P}$. We find that varying individual parameters one at a time over their $1\sigma$ ranges gives $\ln {\cal P} > -{\cal O} (10^3)$ for $H_0$, $\ln {\cal P} > -{\cal O} (10^2)$ for other cosmological parameters, and $\ln {\cal P} > -{\cal O} (1)$ for nuisance parameters which is not a surprise since the nuisance parameters are prior-dominated. The $|\ln {\cal P}|$ values in Figure~\ref{fig:logp_check} are larger because we are varying all parameters simultaneously and independently. 

Over $\ln {\cal P} > -2000$, much larger than the value corresponding to typical confidence intervals, Figure~\ref{fig:logp_check} shows good agreement between the full and Fisher calculations of $\ln {\cal P}$. At larger $|\ln {\cal P}|$ we see a bifurcation for values of $H_0$ above and below the MAP value, particularly in real space. For single-parameter variations of $H_0$, the $\ln {\cal P}$--$H_0$ curve (not shown here) deviates asymmetrically from a parabola, and a parabolic approximation (e.g., Fisher analysis) underestimates $\ln {\cal P}$ when $H_0$ is smaller and overestimates it when $H_0$ is larger, explaining the bifurcation in Figure~\ref{fig:logp_check}. Such asymmetric deviations can be seen from 1D marginalized distributions based on MCMC results (purple solid curves in diagonal panels of our corner plots, like Figures~\ref{fig:corsub_fourier} and \ref{fig:corsub_real}). A smaller but similar effect appears in the $\ln {\cal P}$--$\Omega_{\rm b}$ curve. Since $H_0$ manifests a much more significant non-Gaussianity effect than any other parameter given the domains we are considering here, the bifurcation can be primarily ascribed to $H_0$, with scatter due to errors in other parameters.

The good agreement in Figure~\ref{fig:logp_check} suggests that the moderate differences between Fisher and MCMC contours in Figures~\ref{fig:corsub_fourier} and \ref{fig:corsub_real} are caused mainly by the bounded cosmological priors in the MCMC analysis vs. unbounded priors in the Fisher analysis, rather than by a breakdown of the Fisher approximation itself. However, further investigation is warranted to understand how far out in confidence levels the Fisher approximation remains accurate. We also note that both the MCMC and Fisher analyses assume that the likelihood of the data is described by a multi-variate Gaussian with the specified covariance matrix, and that this approximation might become inaccurate in some regimes \citep{2009MNRAS.395.2065T, 2020MNRAS.499.2977L}.

\section{Density of the Lens Sample} \label{app:qtrlens}

\begin{table*}[]
    \centering
    \begin{tabular}{cc|ccccccc}
    \hline
        FoM (Space) & Choice & $3\!\times\!2$pt & GGL+Clus & Shear Only & Shear+GGL & GGL Only & Shear+Clus & Clus Only \\
    \hline
        \multirow{5}{5em}{$1/\Delta^2(\sigma_8)$$\times 10^{-3}$ (Fourier)}
        & W & $57.03\%$ & $67.62\%$ & $100.00\%$ & $87.43\%$ & $81.96\%$ & $76.88\%$ & $80.01\%$ \\
        & B & $60.99\%$ & $60.01\%$ & $100.00\%$ & $89.25\%$ & $87.18\%$ & $56.99\%$ & $54.31\%$ \\
        & C & $54.82\%$ & $64.87\%$ & $100.00\%$ & $87.41\%$ & $85.29\%$ & $68.58\%$ & $66.65\%$ \\
        & BC & $64.75\%$ & $64.31\%$ & $100.00\%$ & $91.11\%$ & $87.38\%$ & $62.33\%$ & $50.94\%$ \\
        & N & $61.27\%$ & $63.67\%$ & $100.00\%$ & $89.52\%$ & $86.77\%$ & $65.82\%$ & $50.42\%$ \\
    \hline
        \multirow{5}{5em}{$1/\Delta^2(\sigma_8)$$\times 10^{-3}$ (Real)}
        & W & $64.74\%$ & $73.93\%$ & $100.00\%$ & $54.05\%$ & $81.56\%$ & $75.20\%$ & $84.50\%$ \\
        & B & $71.14\%$ & $79.29\%$ & $100.00\%$ & $66.52\%$ & $84.32\%$ & $86.35\%$ & $64.59\%$ \\
        & C & $95.47\%$ & $77.44\%$ & $100.00\%$ & $67.26\%$ & $86.81\%$ & $89.87\%$ & $70.42\%$ \\
        & BC & $79.71\%$ & $73.83\%$ & $100.00\%$ & $82.62\%$ & $86.19\%$ & $86.54\%$ & $64.15\%$ \\
        & N & $94.92\%$ & $68.19\%$ & $100.00\%$ & $77.36\%$ & $86.10\%$ & $86.92\%$ & $64.09\%$ \\
    \hline
        \multirow{5}{5em}{$\det^{-1/2}({\rm Cov}$ $(\sigma_8, \Omega_{\rm m}))$$\times 10^{-3}$ (Fourier)}
        & W & $64.18\%$ & $67.88\%$ & $100.00\%$ & $87.79\%$ & $84.69\%$ & $74.40\%$ & $69.52\%$ \\
        & B & $75.61\%$ & $74.22\%$ & $100.00\%$ & $93.31\%$ & $88.09\%$ & $73.62\%$ & $59.35\%$ \\
        & C & $55.94\%$ & $62.60\%$ & $100.00\%$ & $88.73\%$ & $87.74\%$ & $64.32\%$ & $60.05\%$ \\
        & BC & $66.37\%$ & $64.46\%$ & $100.00\%$ & $93.81\%$ & $88.50\%$ & $64.44\%$ & $51.07\%$ \\
        & N & $48.88\%$ & $65.76\%$ & $100.00\%$ & $91.74\%$ & $88.03\%$ & $71.00\%$ & $50.70\%$ \\
    \hline
        \multirow{5}{5em}{$\det^{-1/2}({\rm Cov}$ $(\sigma_8, \Omega_{\rm m}))$$\times 10^{-3}$ (Real)}
        & W & $65.34\%$ & $74.49\%$ & $100.00\%$ & $71.45\%$ & $84.62\%$ & $83.48\%$ & $76.78\%$ \\
        & B & $79.11\%$ & $68.20\%$ & $100.00\%$ & $80.59\%$ & $86.79\%$ & $91.44\%$ & $70.93\%$ \\
        & C & $85.60\%$ & $78.07\%$ & $100.00\%$ & $82.41\%$ & $87.75\%$ & $87.78\%$ & $71.46\%$ \\
        & BC & $77.69\%$ & $73.09\%$ & $100.00\%$ & $87.15\%$ & $87.81\%$ & $82.67\%$ & $66.17\%$ \\
        & N & $91.49\%$ & $66.34\%$ & $100.00\%$ & $75.49\%$ & $87.78\%$ & $83.55\%$ & $66.09\%$ \\
    \hline
    \end{tabular}
    \caption{\label{tab:qtrlens}Ratios between FoMs with the density of the lens sample reduced by a factor of $4$ (assuming the redshift distributions are not affected) and FoMs under the ``lens $=$ source'' assumption (Table~\ref{tab:probe}).}
\end{table*}

To investigate the impact of the ``lens $=$ source'' assumption, we conduct a case study with the density of the lens sample reduced by a factor of $4$. For simplicity, we assume that the redshift distributions are not affected; we think the results are qualitatively valid, and leave more rigorous quantitative analysis (which changes the redshift distributions in accordance with the lens density) to future work. From Table~\ref{tab:qtrlens}, we see that the FoMs are unaffected for Shear Only (as expected), reduced by a fraction of about $10$--$20\%$ for GGL and GGL+Shear, and reduced by a fraction of up to $40$--$50\%$ for all subsets involving galaxy clustering (including $3\!\times\!2$pt). We therefore tentatively conclude that the cosmological yields of the Roman weak gravitational lensing program are sensitive to the density of the lens sample, and further development of the measurement pipeline should aim at achieving ``lens $=$ source'' among other goals.

\section{Cosmology-Dependent Covariance} \label{app:varcov}

\begin{table*}[]
    \centering
    \begin{tabular}{cc|ccc|ccc}
    \hline
        Param. & Prior $\sigma$ & Mar. $\sigma$ & Mar. $\sigma'$ & Frac. Diff. & Unmar. $\sigma$ & Unmar. $\sigma'$ & Frac. Diff. \\
    \hline
        $\Omega_{\rm m}$ & --- & $0.000872$ & $0.000862$ & $-1.1458\%$ & $0.000588$ & $0.000587$ & $-0.1783\%$ \\
        $\Omega_{\rm b}$ & --- & $0.002702$ & $0.002531$ & $-6.3227\%$ & $0.000631$ & $0.000627$ & $-0.6231\%$ \\
        $h_0$ & --- & $0.016621$ & $0.015709$ & $-5.4894\%$ & $0.003041$ & $0.003031$ & $-0.3467\%$ \\
        $\sigma_8$ & --- & $0.000563$ & $0.000560$ & $-0.5262\%$ & $0.000421$ & $0.000421$ & $-0.0555\%$ \\
        $n_{\rm s}$ & --- & $0.004608$ & $0.004502$ & $-2.3132\%$ & $0.001882$ & $0.001879$ & $-0.1605\%$ \\
    \hline
        $\Omega_{\rm m}$ & $0.006000 \times 2$ & $0.000792$ & $0.000790$ & $-0.2688\%$ & $0.000587$ & $0.000586$ & $-0.1779\%$ \\
        $\Omega_{\rm b}$ & $0.000305 \times 2$ & $0.000571$ & $0.000569$ & $-0.3086\%$ & $0.000439$ & $0.000438$ & $-0.3029\%$ \\
        $h_0$ & $0.004000 \times 2$ & $0.005541$ & $0.005519$ & $-0.4080\%$ & $0.002843$ & $0.002834$ & $-0.3031\%$ \\
        $\sigma_8$ & $0.006000 \times 2$ & $0.000522$ & $0.000521$ & $-0.2154\%$ & $0.000421$ & $0.000421$ & $-0.0554\%$ \\
        $n_{\rm s}$ & $0.004000 \times 2$ & $0.002995$ & $0.002990$ & $-0.1642\%$ & $0.001832$ & $0.001829$ & $-0.1521\%$ \\
    \hline
    \end{tabular}
    \caption{\label{tab:varcov}Impact of cosmology-dependent covariance. Note that these results assume a larger survey area (the sum of the HLIS Medium and Wide Tiers) than what is considered in the main text, so the values should not be directly compared. ``Mar.'' (``Unmar.'') $\sigma$ denotes the fully marginalized (unmarginalized) 1$\sigma$ uncertainties in cosmological parameters according to the first term in Equation~(\ref{eq:varcov}). The primed uncertainties ($\sigma'$) include the second term (cosmology dependence of the covariance matrix) as well; the fractional differences (``Frac. Diff.'') are defined as $\sigma' / \sigma - 1$. In the first (horizontal) block, infinitely wide priors are assumed for all parameters. In the second block, we still assume infinitely wide priors on nuisance parameters, but Gaussian priors with finite widths are assumed for the cosmological parameters (all of them simultaneously). The widths are twice the Particle Data Group parameter constraints.\footnote{\url{https://pdg.lbl.gov/2023/reviews/rpp2023-rev-cosmological-parameters.pdf}}}
\end{table*}

It is well known that, to incorporate the cosmology dependence of the covariance matrix, a second term needs to be added to Equation~(\ref{eq:fisher}), e.g., Equation~(10.40) in \citet{2023cctp.book.....H}
\begin{equation}
    F_{\alpha \beta} = {\boldsymbol m}_{,\alpha}^{\rm T} {\bf C}^{-1} {\boldsymbol m}_{,\beta} + \frac12 {\rm Tr} [{\bf C}^{-1} {\bf C}_{,\alpha} {\bf C}^{-1} {\bf C}_{,\beta}].
    \label{eq:varcov}
\end{equation}
\citet{2013A&A...551A..88C} argued that including the second term would violate the Cram\'er-Rao inequality, but this statement may or may not apply to real-world covariance matrices with noise and a non-Gaussian component. Therefore, we compute numerical derivatives of the covariance matrix in real space and use Equation~(\ref{eq:varcov}) to estimate the contribution from the second term. As shown in Table~\ref{tab:varcov}, the changes in constraining power are negligible for a $3\!\times\!2$pt analysis. Nonetheless, it is important to note that, when the covariance matrix and the data vector are correlated, the central values of inferred parameters are also affected by this correlation \citep[e.g.,][]{2024PhRvD.110l3517A}. We leave further investigation of how a cosmology-dependent covariance matrices may change our answer for future work.

\bibliography{main}{}
\bibliographystyle{aasjournalv7}

\end{document}